\documentclass[10pt,doublecolumn]{IEEEtran}
\usepackage{amsmath,amsthm,amsfonts,amssymb,bm,mathrsfs}
\usepackage{graphicx,subfigure}
\usepackage[cmintegrals]{newtxmath}
\usepackage{amssymb}
\usepackage{cite}
\usepackage{setspace}
\usepackage{caption}
\usepackage{floatrow}
\usepackage{algorithm}
\usepackage{algorithmic}
\usepackage{float}
\ifCLASSINFOpdf
\else
\fi
\begin{document}
	
%
\title{Study on Base Station Topology in Cellular Networks: Take Advantage of Alpha Shapes, Betti Numbers, and Euler Characteristics}

\author{Ying Chen, \IEEEauthorblockN{Rongpeng Li, Zhifeng Zhao, and Honggang Zhang}\\
	\thanks{Y. Chen, R. Li, Z. Zhao, and H. Zhang are with College of Information Science and Electronic Engineering, Zhejiang University. Email: \{21631088chen\_ying, lirongpeng, zhaozf, honggangzhang\}@zju.edu.cn}
	\thanks{This work was supported in part by National Key R$ \& $D Program of China (No. 2018YFB0803702), National Natural Science Foundation of China (No. 61701439, 61731002), Zhejiang Key Research and Development Plan (No. 2018C03056), the National Postdoctoral Program for Innovative Talents of China (No. BX201600133), and the Project funded by China Postdoctoral Science Foundation (No. 2017M610369).}}


%


\maketitle

\begin{abstract}
 Faced with the ever-increasing trend of the cellular network scale, how to quantitatively evaluate the effectiveness of the large-scale deployment of base stations (BSs) has become a challenging topic. To this end, a deeper understanding of the cellular network topology is of fundamental significance to be achieved. In this paper, $ \alpha $-Shape, a powerful algebraic geometric tool, is integrated into the analysis of real BS location data for six Asian countries and six European countries, respectively. Firstly, the BS spatial deployments of both Asian and European countries express fractal features based on two different testifying metrics, namely the Betti numbers and the Hurst coefficients. Secondly, it is found out that the log-normal distribution presents the best match to the cellular network topology when the practical BS deployment is characterized by the Euler characteristics.
\end{abstract}


\section{Introduction}
Driven by the explosive installment of base stations (BSs) globally recently, the issue of the optimal BS deployment, i.e., a collection of BSs distributed on the two-dimensional plane \cite{Kibi2016Modelling}, of the cellular network has attracted tremendous concern from both academic and industrial communities. An inefficient way of deployments usually leads to disappointing network performances, such as poor capacity, low spectrum efficiency, waste of energy consumption, and intolerable delay \cite{Ericson2011Total,Lin2009Optimizing}. One promising attempt to optimize network deployments is the ultra-dense cellular network by shrinking the size of individual BS's coverage \cite{Ericson2011Total}. Despite the significantly gained momentum, ultra-dense cellular network scheme is still far away from the perfect solution to network deployments because of the unpleasant consequences it brings about, including the escalation of energy consumption, inter-cell interference, void cell, tough power control and so on \cite{Peng2015Optimal}.

\subsection{Related Works}
In order to meet various performance requirements, substantial efforts have been made on how to effectively address the cellular network deployment issues. \cite{Ericson2011Total} investigated the influence of cell size on the network energy consumption for the purpose of maximizing energy efficiency. In \cite{Lin2009Optimizing}, an optimal approach was presented to prolong the lifetime of wireless sensor networks (WSNs) by deploying BSs appropriately. Stochastic geometry models were built in \cite{Kibi2016Modelling} for BS deployments in shared cellular networks, and the proposed models were observed to be suitable for different countries. \cite{Liu2012Femtocell} proposed to jointly optimize the placement of BSs with other factors like power control. \cite{Wu2012Energy} studied the BS deployment scheme based on the overall traffic variation process rather than the peak traffic load, which can successfully bring energy consumption reduction and performance improvement. \cite{Coskun2015A} developed a greedy framework to arrange the BSs in heterogeneous cellular networks, so as to improve the energy efficiency. In the strive for meeting coverage demands and traffic requirements, the BSs were automatically planned and distributed in \cite{Istv2010An} in mobile networks. From an energy reduction perspective, \cite{Mahmud2011Efficient} proposed a heuristic method to lay out the BSs in the WSNs.

Undoubtedly, the knowledge of topology of the cellular networks is extremely beneficial for the guidance on BS deployments. Since the locations of BSs are of paramount importance in determining the network topology, a range of researches on the spatial distribution of BSs have been carried out in recent decades. In \cite{Zhou2015On}, the severe divergence of Poisson distribution from practical BS density distribution was found out, and $ \alpha $-stable distribution, one of the heavy-tailed distributions, was shown to match the real distribution more properly. By separating the BSs into diverse subsets, a comprehensive study on the cellular network's spatial structure was presented in \cite{Zhou2016Large}, and the clustering feature of BS density distribution was also revealed. In \cite{Li2016On}, further study about the strong linear dependence was exhibited between BS deployment and spatial traffic distribution. Taking different scenarios into account, \cite{Chiaraviglio2016What} claimed that the log-normal and Weibull distributions were the more precise ones to fit the real BS deployments in rural and coastal scenarios, respectively. \cite{Zhao2017Temporal} verified the validity of $ \alpha $-stable distribution for the temporal traffic series in addition to the spatial density of BSs.

Understanding the network topology can prominently facilitate the design of efficient cellular networks, and pervasive applications depending upon the network topology have been found in the wireless communication field. For instance, the knowledge of network topology can be exploited to perform automatic computations for topology control in WSNs \cite{Athanassopoulos2012Cellular}, investigate the pros and cons of BS switching on/off scheme \cite{Kwon2015Random}, develop energy-saving mechanism for green cellular networks \cite{Xiang2015Topology}, improve the spectral efficiency and reduce the bit error rate (BER) in space-time network coding (STNC) scheme \cite{Torrea2017Topology}, manage the interference and allocate spectrum resource for a femtocell-based cellular network \cite{Dharmaraj2013Distributed} and so on.

\subsection{Our Contributions}
Most of the previous literatures analyzed the cellular network deployments from the perspective of BS spatial density distribution by using straightforward simulation methods. As one of the unprecedented researches, this paper introduces a powerful algebraic geometric tool, namely $ \alpha $-Shapes \cite{Weygaert2011Alpha}, into the fundamental analysis of real BS location data for twelve countries around the world, and generalizes essential topological characteristics through the mass data sets from the perspective of topological invariants, i.e., Betti numbers and Euler characteristics \cite{Weygaert2011Alpha}. Briefly speaking, $ \alpha $-Shapes are geometric manifolds constructed from a specific point set so that they are closely related to topological nature of the point set. Moreover, the Betti numbers and the Euler characteristics can be used to describe the topological information in the $ \alpha $-Shapes, thus also tightly associated with the topological features of the point set. Specifically, our works aim to answer this kind of question: is there any essentially identical features hidden in the network topology regardless of the geographical differences? In other words, it is an interesting and profound issue to find out the topological features in the cellular network topology, and these outcomes can be an invaluable tool for fruitful guidelines on the design of cellular networks.

In this regard, the main contributions of this paper are as follows:
\begin{itemize}
	\item First of all, the fractal phenomenon is revealed in the BS spatial deployments for both Asian and European countries in terms of the Betti numbers;
	\item Secondly, the fractal features in the topology of BSs in cellular networks are also confirmed by taking advantage of the Hurst coefficients;
	\item Thirdly, it is verified that log-normal distribution provides the most conforming fitness with the cellular network topology when the practical BS deployments are characterized by the Euler characteristics.
\end{itemize}

This paper is organized as follows: The vital algebraic topological invariants and tools, including $ \alpha $-Shapes, Betti numbers and Euler characteristics, are introduced in Section II. The detailed description for the cellular network data is given in Section III. The fractal features of BS deployments in Asian and European countries are clarified in Section IV. The identical log-normal distribution of the Euler characteristics for the above countries is illustrated in Section V. Finally, conclusions are drawn in Section VI.

\section{Fundamental Tools for Topology Discovery}
In this section, the principal concepts of the algebraic topological tools used for topology discovery in our works are presented. Various information in all aspects about these tools will be provided, including the basic definitions, the relevant applications, the constructions and so on. The relationships between them will also be expounded.

\subsection{Alpha Shapes ($ \alpha $-Shapes)}

As one of the most fundamental notions from the domain of computational topology in algebraic geometry \cite{Wagner2012Computational,Zomorodian2005Topology}, $ \alpha $-Shapes were first introduced by Edelsbrunner in 1973 \cite{Knuth1973The}. $ \alpha $-Shapes have been extensively applied to a plenty of vital domains. Typical examples of the applications based on $ \alpha $-Shapes modeling include: protein-DNA interactions can be precisely predicted \cite{Zhou2010A}, molecular shapes can be detailedly described \cite{Wilson2009Alpha}, properties of macro-molecular can be analytically exacted \cite{Liang1998Analytical} in the field of biology; interactions between the origin and destinations can be summarized and visualized in geography \cite{Mu2011A}; network edges can be detected locally in ad hoc networks \cite{Fayed2009Localised}; surface morphology of products can be analyzed in manufacturing. Applications in other fields include shape similarity comparison of 3D models \cite{Ohbuchi2003Shape}, medical image analysis and so forth.

In general, unlike common tools in algebraic topology field, which have to resort to some kind of user-defined smoothing functions or thresholds for the analysis of a discrete point set, $ \alpha $-Shapes depend entirely on the point distribution itself, and focus on the features determined by the set of points exclusively \cite{Weygaert2011Alpha}. In this regard, $ \alpha $-Shapes are significantly superior and are regarded as the optimal technique for our study for the same reason.

In an informal definition, $ \alpha $-Shapes can be considered as the intuitive notion of the shape of a specific discrete point set \cite{Edelsbrunner2010Alpha}. For example, without loss of generality, a point set $ \mathbf{S} $ and one of its $\alpha $-Shapes are illustrated in Fig. 1. As shown in Fig. 1, the $\alpha $-Shape is in accordance with our intuition about the shape of $ \mathbf{S} $. And the processes of construction will be elaborated later in this subsection.
\begin{figure}[htbp]
	\centering
	\includegraphics[scale=0.15]{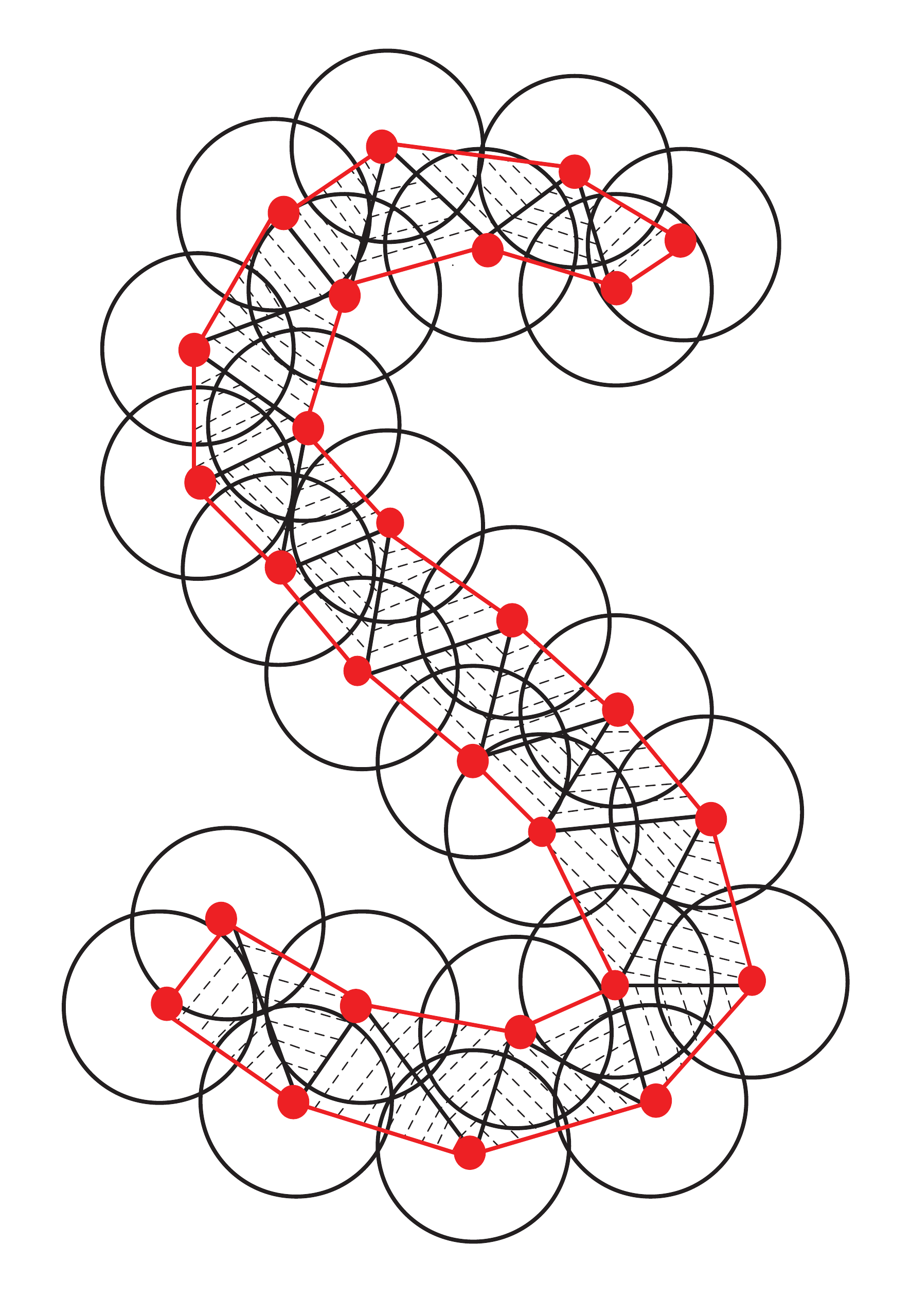}
	\caption{A point set $ \mathbf{S} $ and one of its $\alpha $-Shapes. The point set $ \mathbf{S} $ is represented by the red nodes in the two-dimensional plane, and the $ \alpha $-Shape of $ \mathbf{S} $ is denoted by the red outline of the area with black dotted shadows. In addition, the black circles and edges between nodes include the intermediate results during the construction of $ \alpha $-Shapes.}
\end{figure}

Two topological concepts closely related to the construction of $ \alpha $-Shapes are the Voronoi tessellation and the Delaunay triangulation. Let a two-dimensional point set $ \mathbf{A} $, a subset of the two-dimensional real number point set $ \mathbf{R^{2}} $, consist of node $ a_{i}\ (i=1,2,...,n) $, where $ n $ is the total number of $ \mathbf{A} $. Then a Voronoi cell $ V_{i} $ refers to the set of points which are closer to $ a_{i} $ than any other nodes in $ \mathbf{A} $, as expressed in Eq. \eqref{7}, and the Voronoi diagram of $ \mathbf{A} $ is composed of the set of Voronoi cells \cite{Edelsbrunner2010Alpha}.
\begin{equation}\label{7}
{V_i} = \{ x \in \mathbf{R^{2}}\left| {{\rm{ }}\left\| {x - {a_i}} \right\|} \right. \le \left\| {x - {a_j}} \right\|,{\rm{ }}\forall j = 1,2,...n,{\rm{ }}j \ne i\} 
\end{equation}

The Delaunay triangulation is coupled with the Voronoi tessellation. An edge connecting two nodes belongs to the Delaunay triangulation if and only if the two corresponding Voronoi cells share a common side. Similarly, a triangle connecting three nodes belongs to the Delaunay triangulation if and only if the three corresponding Voronoi cells share a common corner \cite{Edelsbrunner2010Alpha}. The Voronoi tessellation and the Delaunay triangulation of $ \mathbf{A} $ are depicted as Fig. 2.
\begin{figure}[htbp]
	\centering
	\includegraphics[scale=0.2]{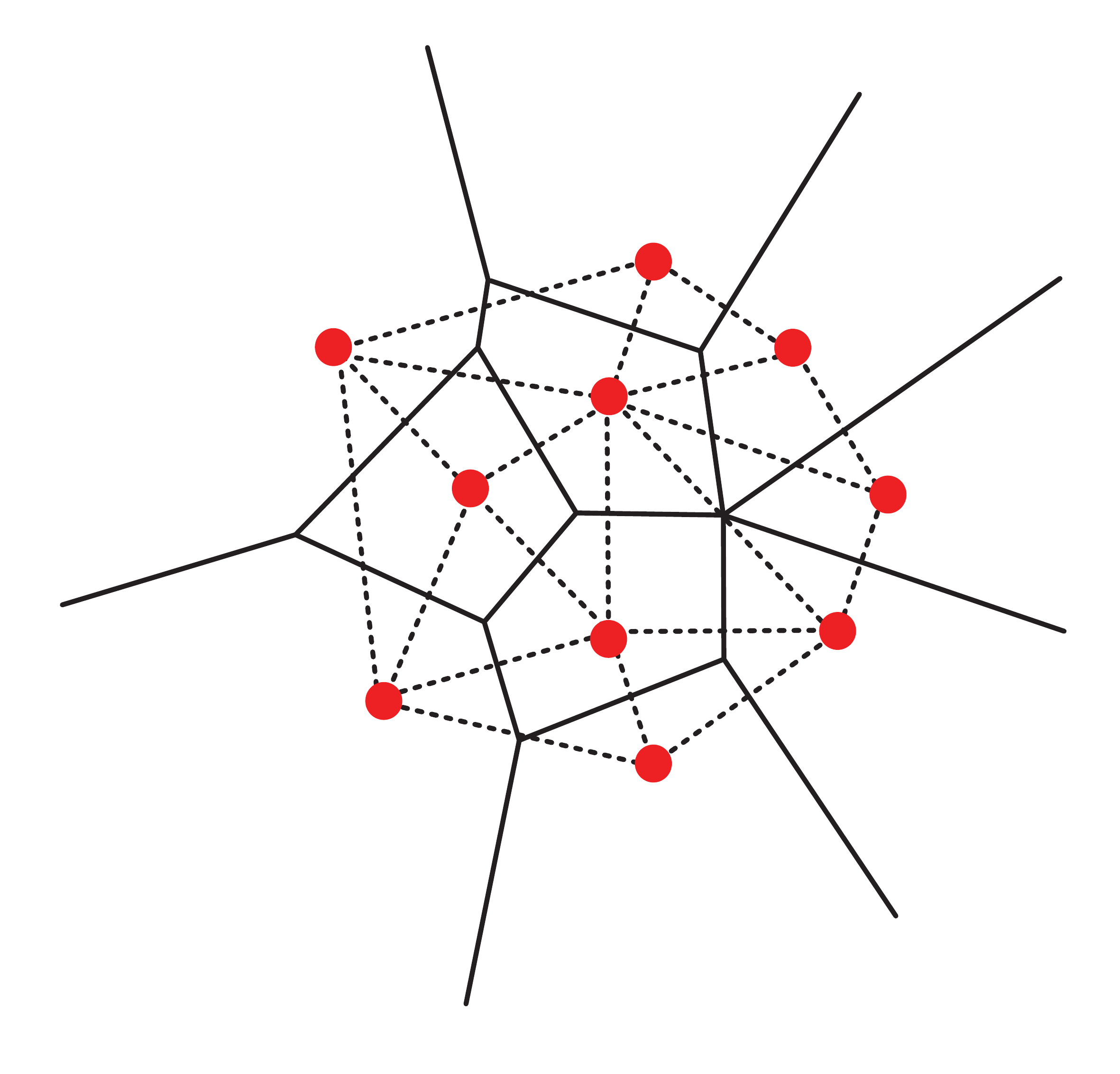}
	\caption{The Voronoi tessellation and the Delaunay triangulation of $ \mathbf{A} $, where the red nodes belong to the point set $ \mathbf{A} $, the dotted lines refer to the Delaunay triangulation, and the solid lines indicate the edges of the Voronoi tessellation.}
\end{figure}

Fig. 3 represents the process of one $ \alpha $-Shape construction given the scale parameter $ \alpha $. Specifically, here follows the outline of building up $ \alpha $-Shapes. Firstly, the Voronoi tessellation and the Delaunay triangulation of the point set $ \mathbf{A} $ is obtained in Fig. 3(a). Secondly, taking the scale parameter $ \alpha $ as the radius, circles centered by every point in $ \mathbf{A} $ are drawn. Then the set of circles forms the area $ C(\alpha) $ as shown by the pink area in Fig. 3(b), in which the straight lines indicate the edges of the Voronoi tessellation. Thirdly, the alpha complex consists of the simplexes, namely, the vertexes, edges and triangles, which are entirely included in the area $ C(\alpha) $, as displayed by the red part in Fig. 3(c). Finally, one $ \alpha $-Shape is obtained in Fig. 3(d).
\begin{figure}[htbp]
	\centering
	\includegraphics[scale=0.15]{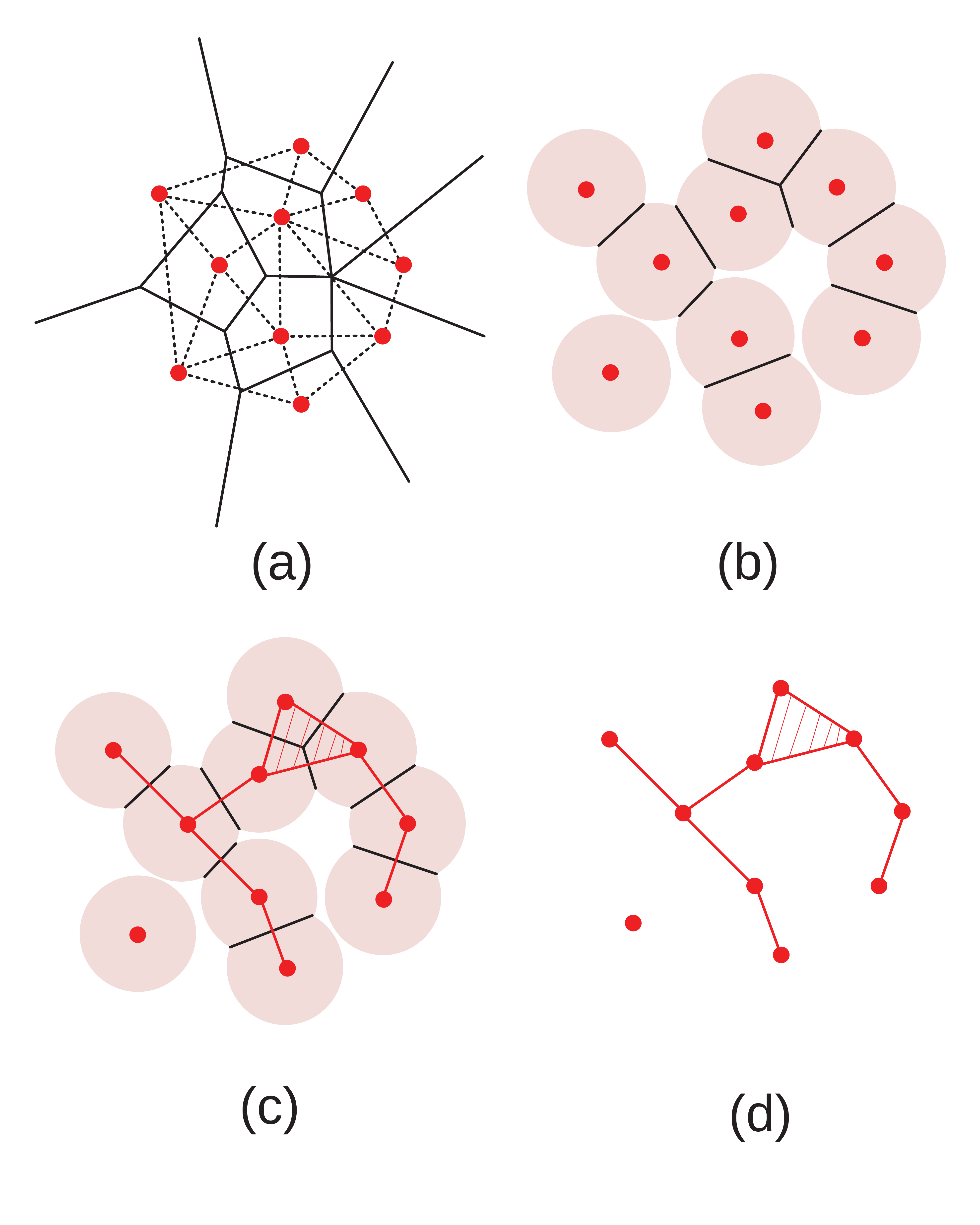}
	\caption{The illustration of $ \alpha $-Shapes construction.}
\end{figure}

The features of $ \alpha $-Shapes demonstrate themselves when $ \alpha $ increases gradually from zero to infinity, as shown in Fig. 4. The $ \alpha $-Shape is equivalent to the point set itself when $ \alpha $ equals to zero. Then a new Delaunay simplex is added into the $ \alpha $-Shapes when $ \alpha $ exceeds some threshold, which means the $ \alpha $-Shapes vary by the values of $ \alpha $ in a discrete manner. So alpha complex is the same as the Delaunay triangulation of $ \mathbf{A} $ when $ \alpha $ reaches some maximum value $ \alpha_{max} $, and the $ \alpha $-Shape keeps being the convex hull of $ \mathbf{A} $ afterwards. Therefore, the number of $ \alpha $-Shapes is actually finite although the scale parameter $ \alpha $ varies from $ 0 $ to $ +\infty $.

\begin{figure}[htbp]
	\centering
	\includegraphics[scale=0.16]{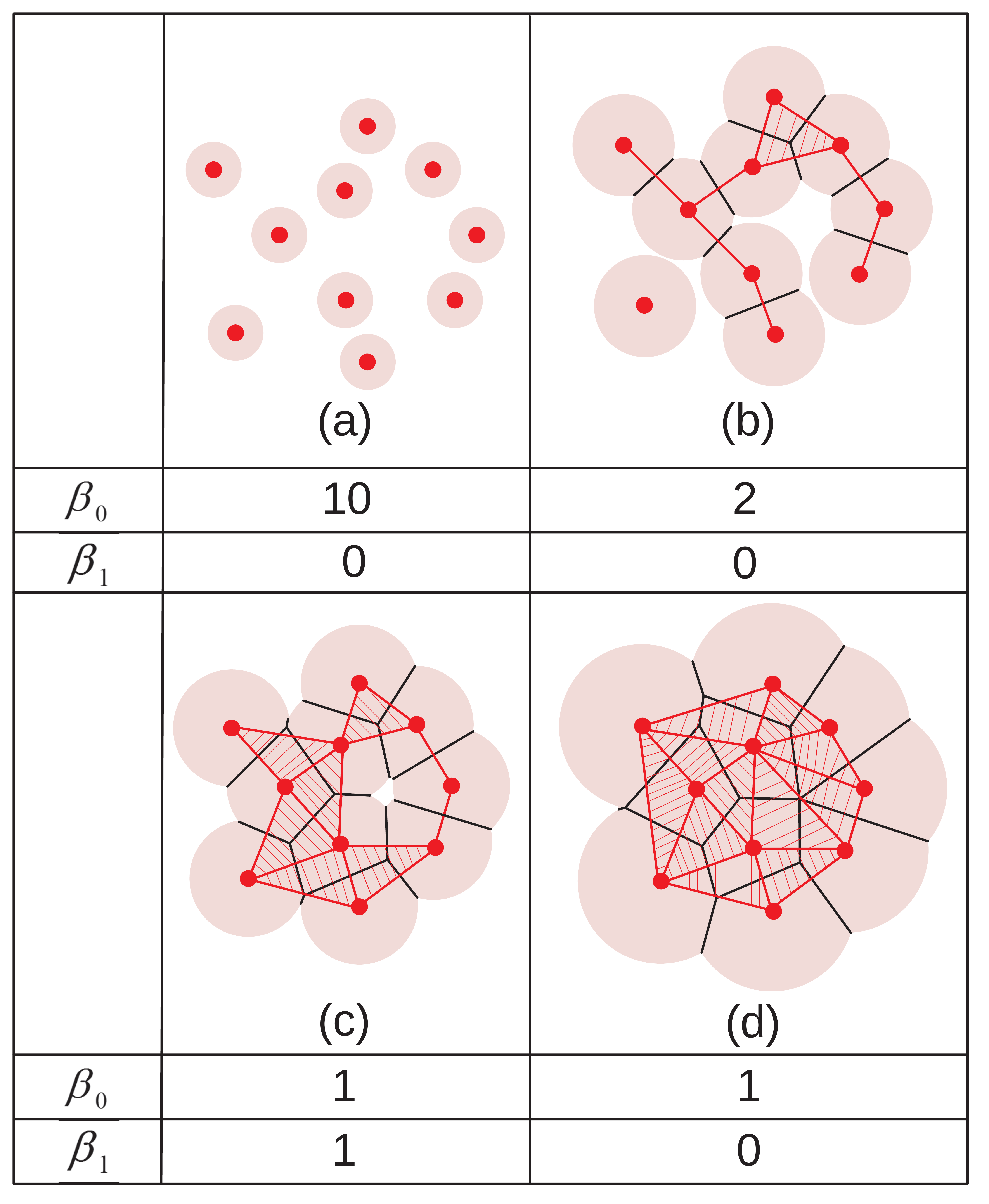}
	\caption{The features of $ \alpha $-Shapes as $ \alpha $, i.e., the radius of the pink circles in each subgraph, increases gradually from zero to infinity.}
\end{figure}

In summary, $ \alpha $-Shapes are the intuitive description of point sets, the subsets of the Delaunay triangulation, and the generation of the convex hull.

\subsection{Betti Numbers}
The basic notion of homology needs to be stated first for a better elaboration of the Betti numbers. Homology describes the connectivity of space via the characterization of two fundamental morphological elements, namely, holes and boundaries \cite{Weygaert2010Alpha}. In terms of three-dimensional space, a 0-dimensional hole refers to an independent component or a gap between two disconnected components; a 1-dimensional hole means a tunnel which appears when an edge is added to two connected points, and the tunnel can be crossed over in either direction without running into a boundary; a 2-dimensional hole is a cavity or void, which is completely encircled by a 2-dimensional surface. Boundaries are usually described by cycles. A 0-cycle indicates a connected object or a point; a closed loop is identified as a 1-cycle; while a 2-dimensional surface is referred to as a 2-cycle. The $ p $-th homology group $ H_{p} $ is composed of all $ p $-dimensional holes or cycles \cite{Pranav2017The}. 

The $ p $-th Betti number $ \beta_{p} $ can be seen as the number of elements in the $ p $-th homology $ H_p $. As a result, the Betti numbers contain complete topological information about a space. Similar as the $ p $-th hole, $ \beta_{0} $ is the number of independent components of a space; $ \beta_{1} $ means the number of independent tunnels; $ \beta_{2} $ indicates the number of independent enclosed voids \cite{Pranav2017The}.

The Betti numbers can be inferred from $ \alpha $-Shapes in a straightforward way. As illustrated in Fig. 4, the $ \alpha $-Shapes evolve from the point set itself into the Delaunay triangulation gradually as the growth of the scale parameter $ \alpha $. In other words, the set of $ \alpha $-Shapes is actually a filtration of the Delaunay triangulation, where $ \alpha $ plays the role of filtration parameter \cite{Zhou2014Alpha}. Therefore, each Betti number can be expressed as a function of $ \alpha $ since every $ \alpha $-Shape corresponds to certain values of the Betti numbers as listed in Fig. 4 for the enhancement of understanding. 

\subsection{Euler Characteristics}
The Euler characteristic is one of the most principle concepts in analyzing the topology of a space, especially in capturing the global features and statistical characteristics.  

Let $ \mathbf{O} $ be a 3-dimentional object whose boundary $ \partial {\rm \mathbf{O}} $ is a 2-dimentional enclosed shell. In the formal definition, the Euler characteristic $ \chi (\partial \mathbf{O}) $ equals to the integrated Gaussian curvature of the surface \cite{Weygaert2011Alpha}:
\begin{equation}\label{8}
\chi (\partial \mathbf{O}) = \frac{1}{{2\pi }}\oint_x {\frac{{dx}}{{{R_1}(x){R_2}(x)}}},
\end{equation}

\noindent where $ R_{1}(x) $ and $ R_{2}(x) $ are two principal radius of the curvature at the point $ x $ of the shell. Surprisingly, the integration above remains invariant under continuous deformation of the surface \cite{Weygaert2011Alpha}, which is the reason why the Euler characteristic can reflect the essential property of a space and is referred to as a topological invariant.

According to the Euler-Poincare Formula, the Euler characteristic can also be obtained by the Betti numbers \cite{Weygaert2011Alpha}. Specifically, the Euler characteristic equals to the alternating sum of the Betti numbers \cite{Weygaert2011Alpha}:
\begin{equation}\label{9}
\chi  = {\beta _0} - {\beta _1} + {\beta _2} - {\beta _3} + ......
\end{equation}

In our works, only two Betti numbers $ \beta_{0} $ and $ \beta_{1} $ are taken into account since the 2-dimensional location data of BSs in the cellular networks are analyzed.

\section{Dataset Description}
In order to achieve precise and effective topological characterizations, our works are based on the analysis of a rich supply of mass data gained from \emph{OpenCellID} community, an open source data platform for providing the locations of BSs in various countries around the world (https://community.opencellid.org/) \cite{Ulm2015Characterization}. The massive BS data of twelve countries, including six representative countries from Asia and Europe respectively, have been extracted from the \emph{OpenCellID} dataset. Each information entry consists of the location of BSs, namely, the latitude and longitude data and other auxiliary information. The basic information about the 12 selected countries is listed in Table I, including the territory area, the total number and the average density of BSs. 
\begin{table}[htbp]
	\centering
	\includegraphics[scale=0.3]{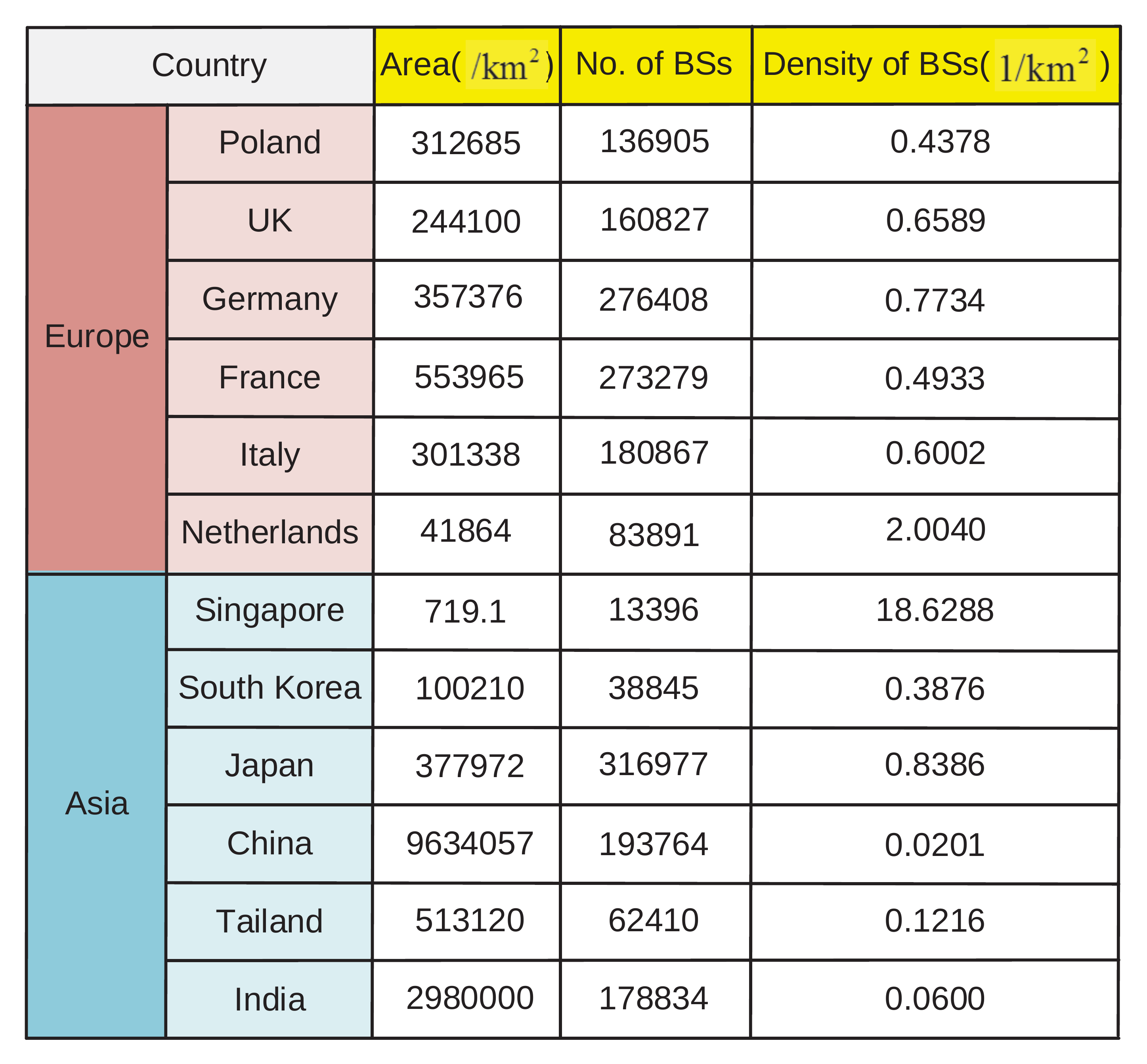}
	\caption{The basic information of 12 selected countries.}
\end{table}

\begin{table}[htb]
	\centering
	\includegraphics[scale=0.45]{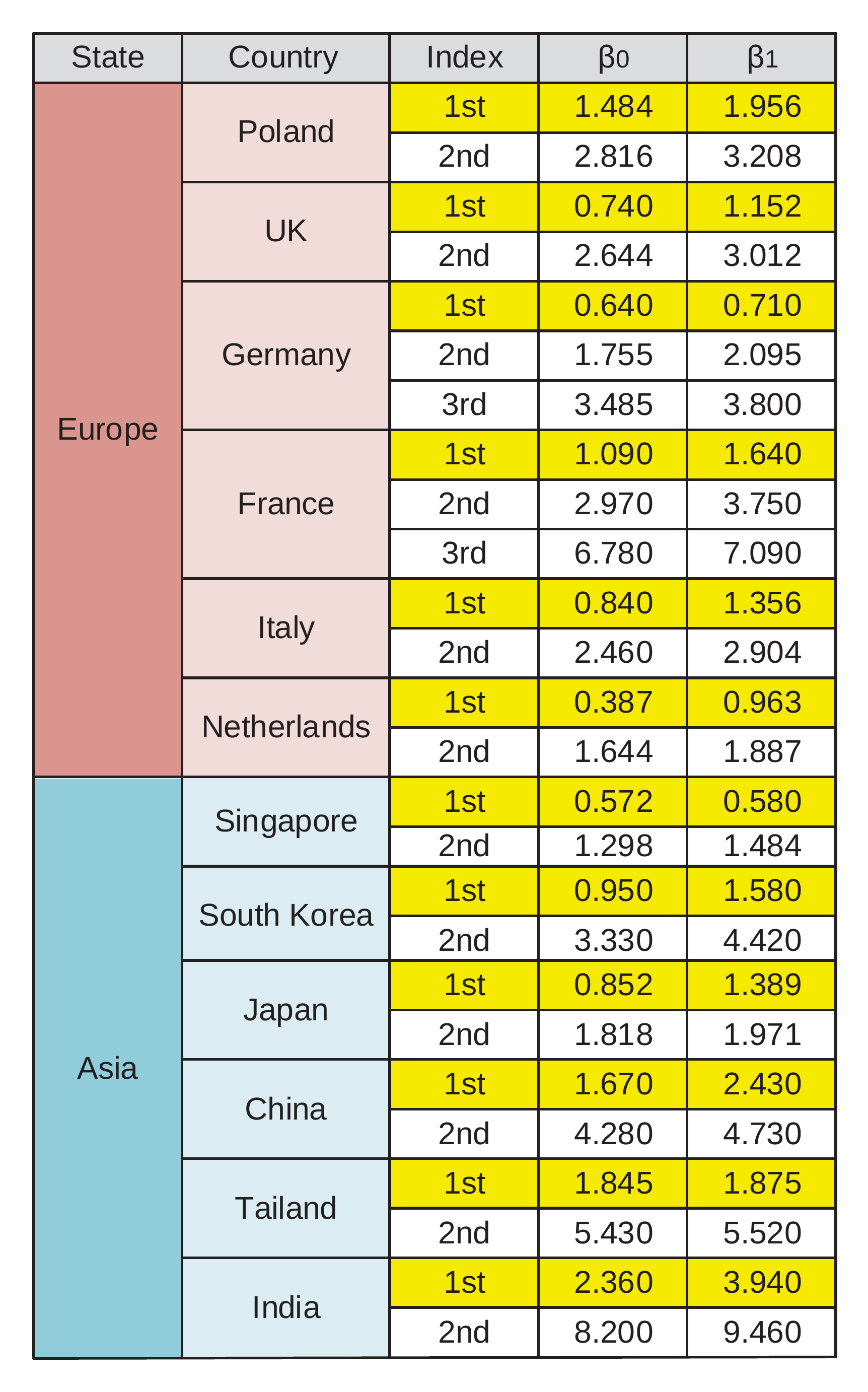}
	\caption{The positions of the ripples and peaks in the Betti curves.}
\end{table}

To express the reliability of the extracted data, the latitude and longitude data are converted into the location data on a 2-dimensional coordination system in the form of $ x $ and $ y $, and the BS deployment diagrams of several representative countries, including three Asian countries and three European ones, are illustrated in Fig. 5. The red boundary lines are extracted from the Google Map, and it is observed that the layout formed by the BSs is basically the same as the real territory configuration since almost all the BS data fall within the boundary line of each country.
\begin{figure*}[htbp]
	\centering
	\subfigure[China$ \qquad\qquad\qquad\qquad\qquad\qquad\qquad $(d) France]{
		\makebox[6cm][c]{
			\includegraphics[scale=0.15]{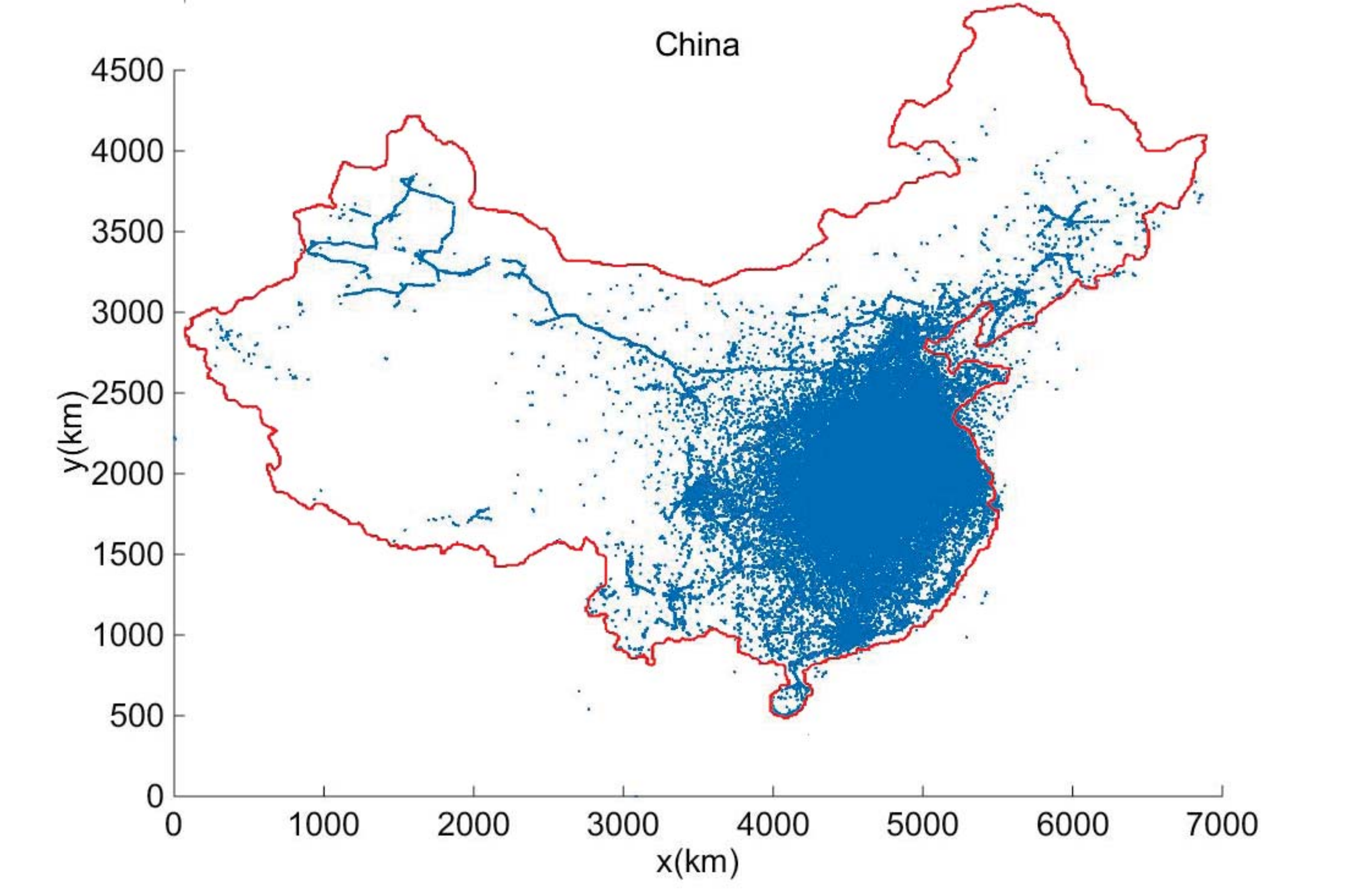}
		}
		\makebox[6cm][c]{
			\includegraphics[scale=0.15]{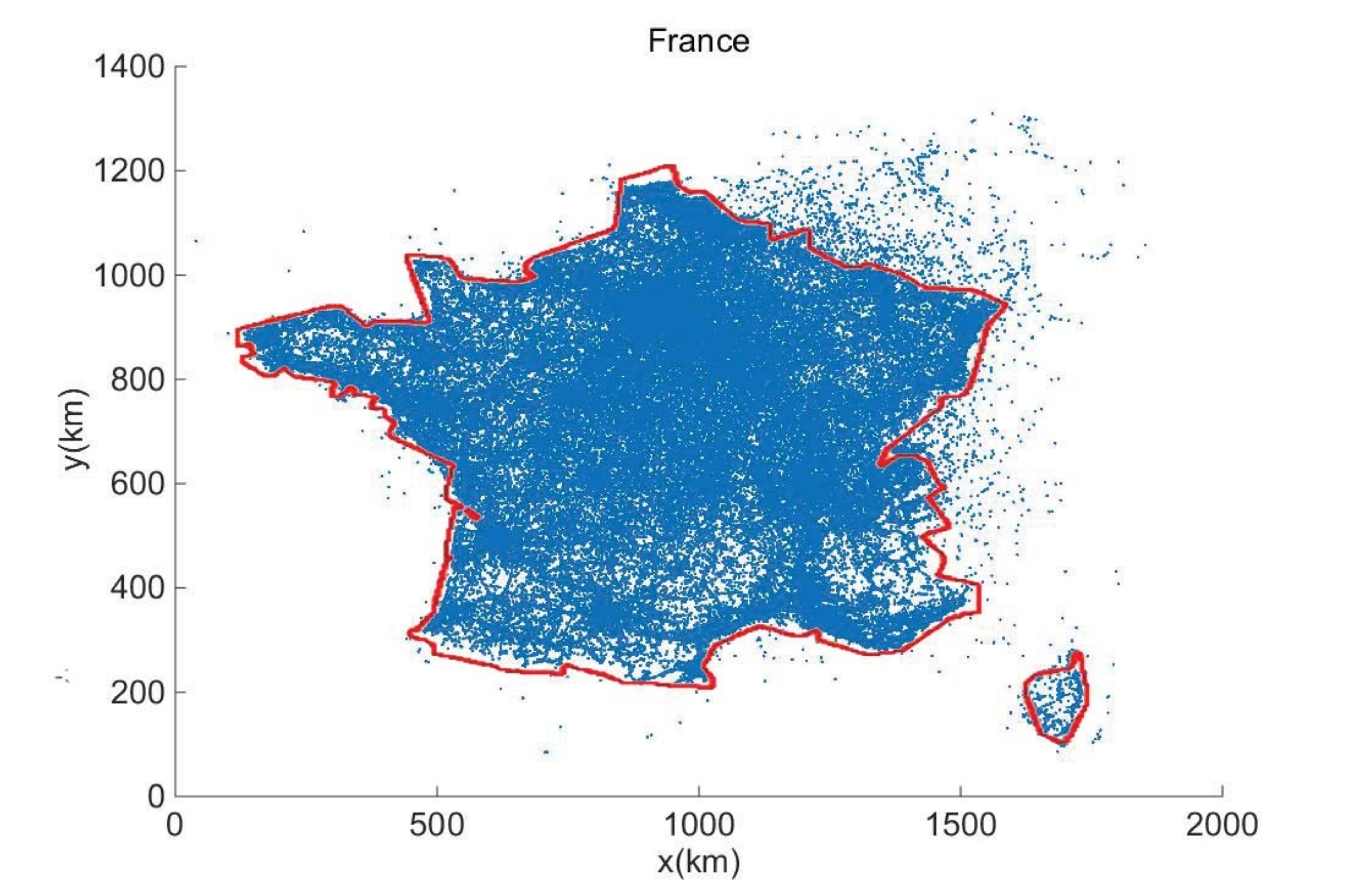}
		}
	}
	
	\subfigure[Japan$ \qquad\qquad\qquad\qquad\qquad\qquad\qquad $(e) Germany]{
		\makebox[6cm][c]{
			\includegraphics[scale=0.15]{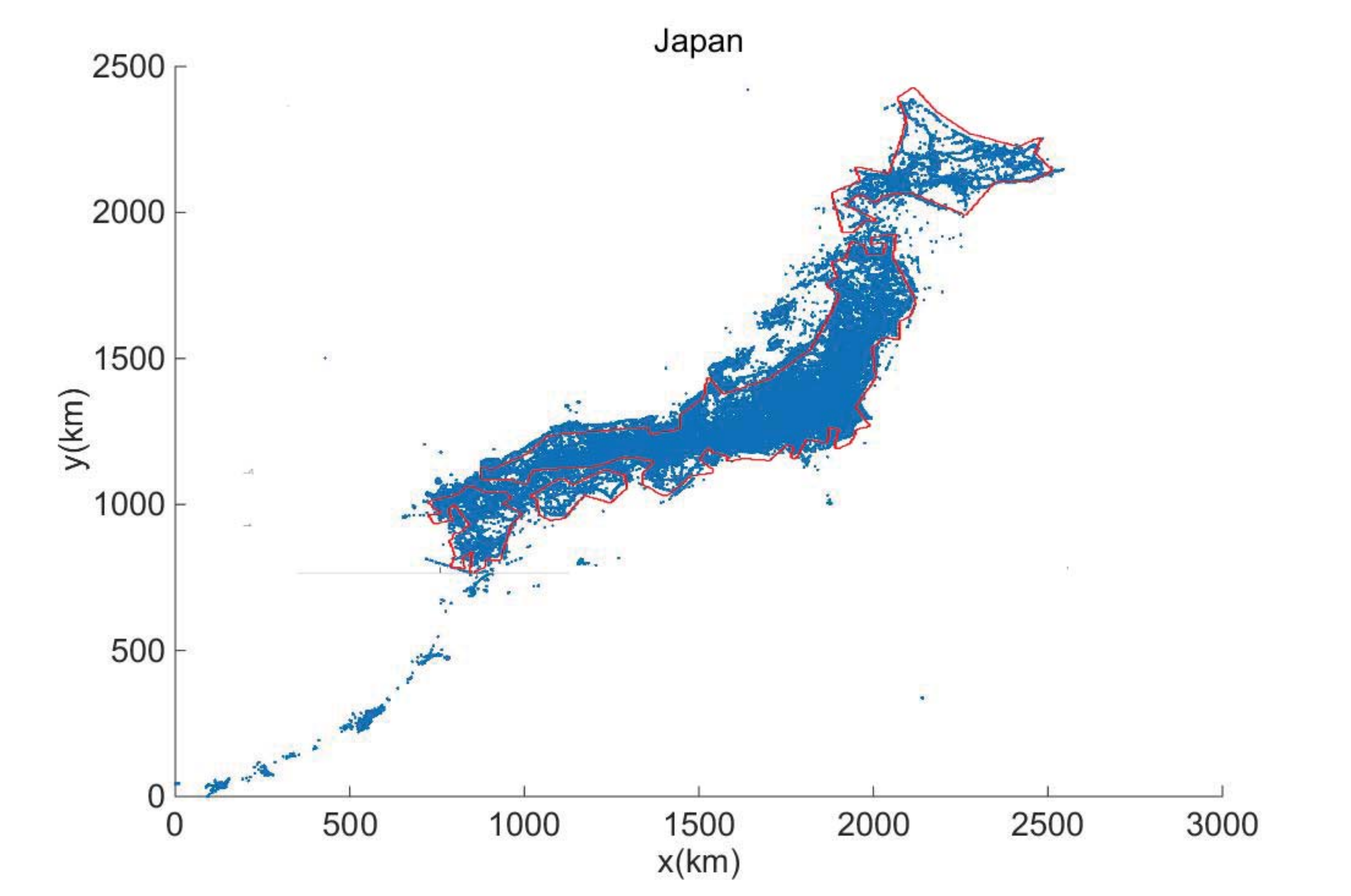}
		}
		\makebox[6cm][c]{
			\includegraphics[scale=0.15]{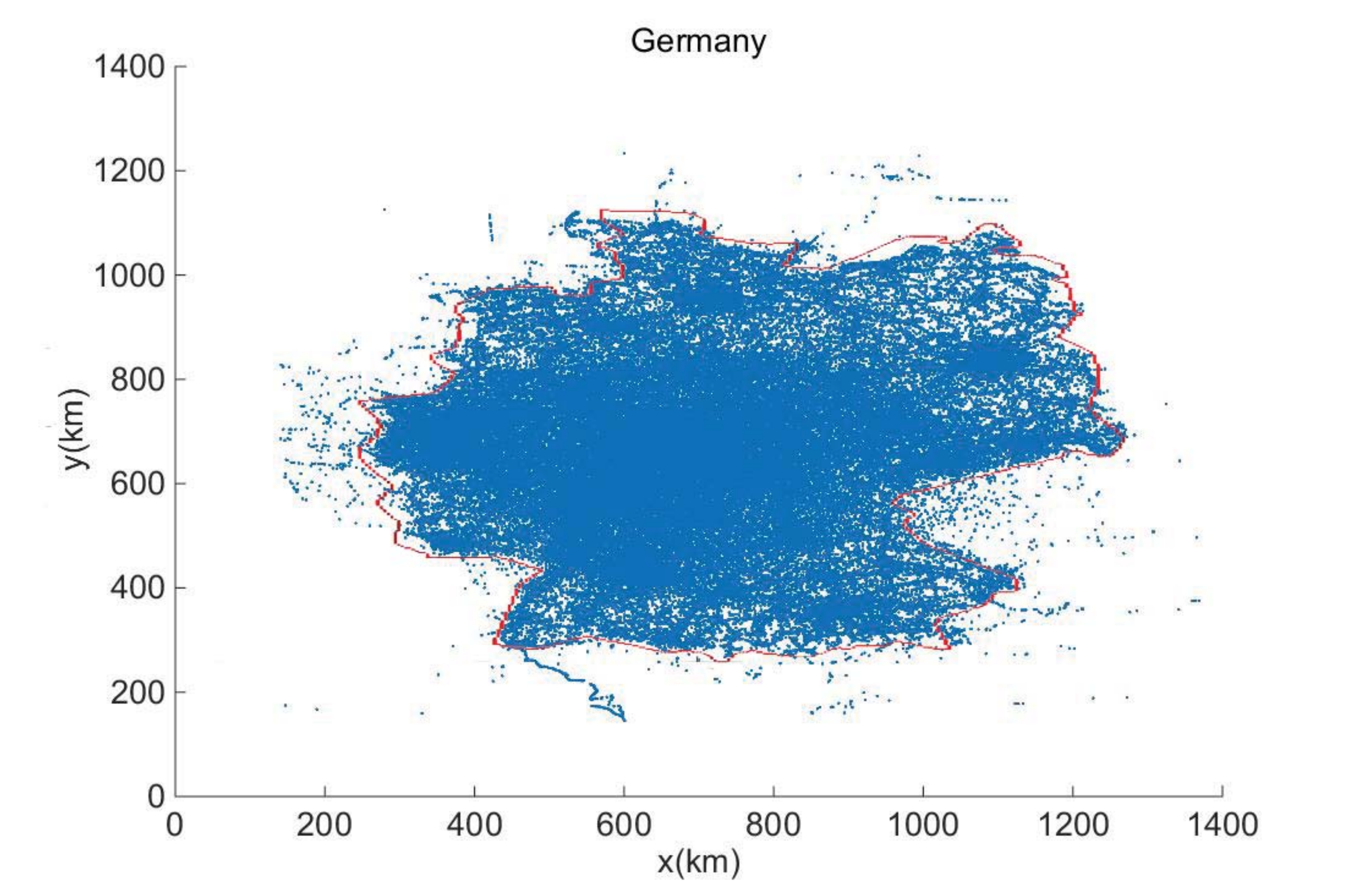}
		}
	}
	
	\subfigure[South Korea$ \qquad\qquad\qquad\qquad\qquad\qquad $(f) Italy]{
		\makebox[6cm][c]{
			\includegraphics[scale=0.15]{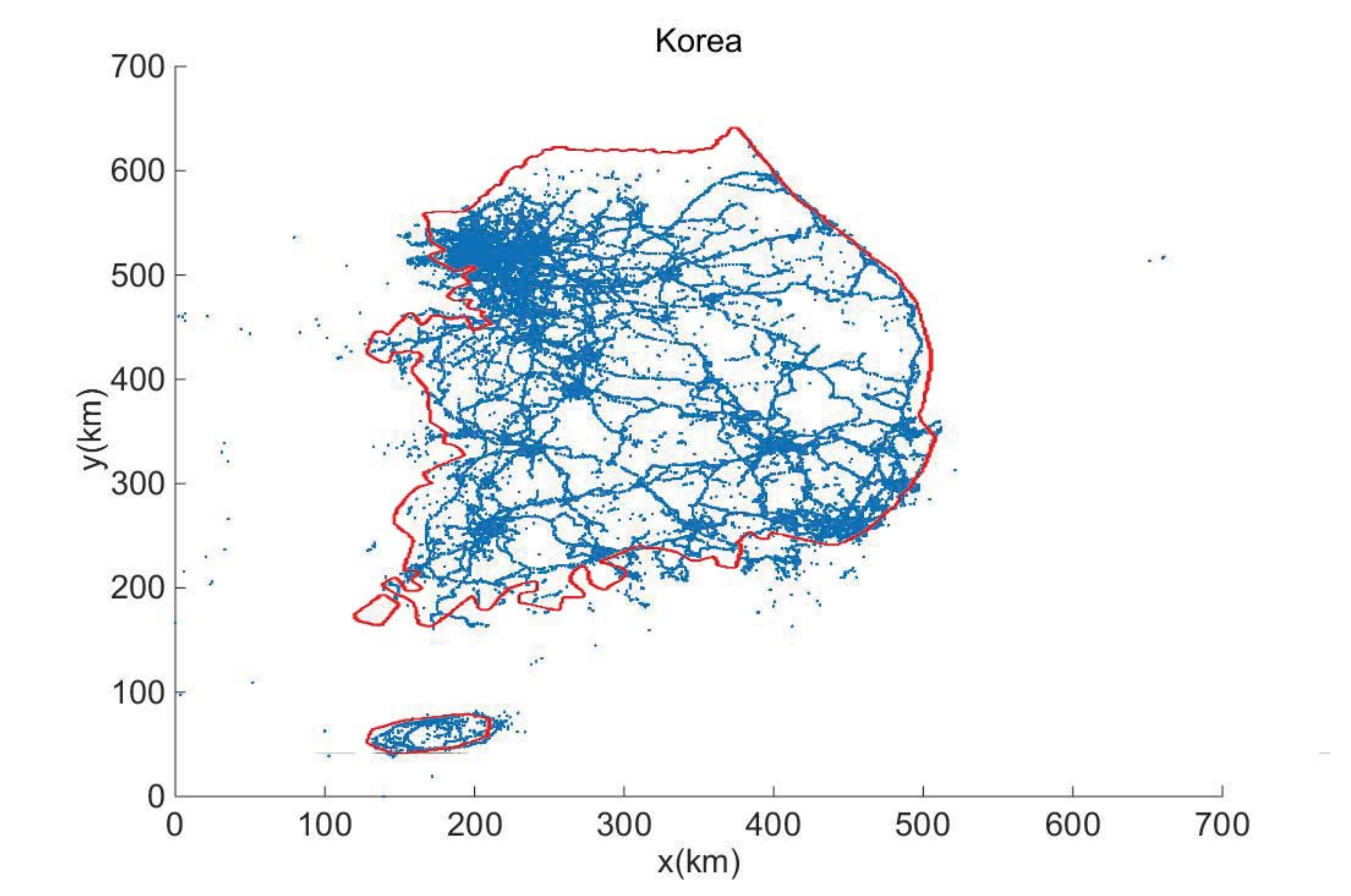}
		}
		\makebox[6cm][c]{
			\includegraphics[scale=0.15]{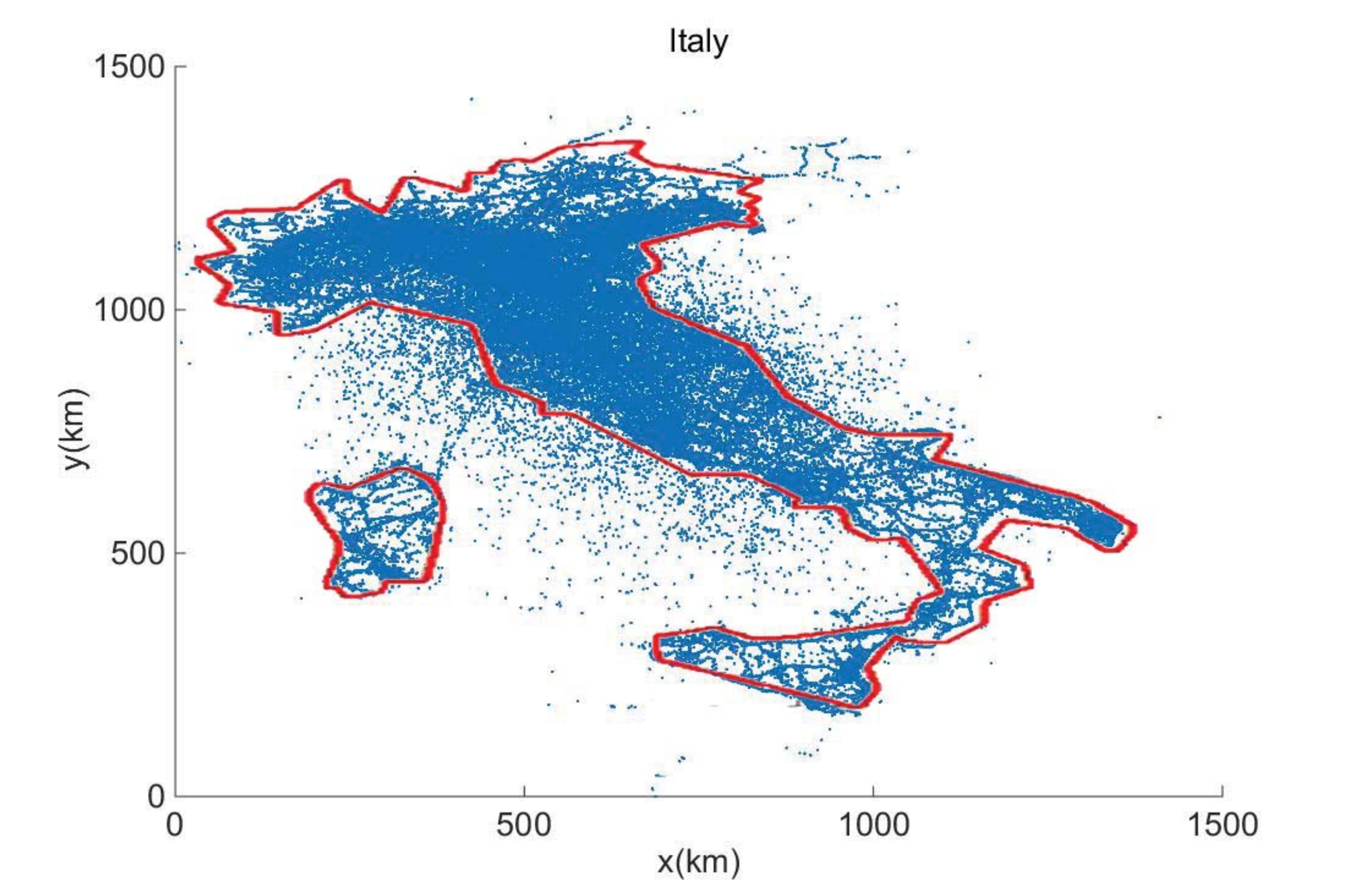}
		}
	}
	
	\caption{The BS deployment diagrams of three (left) Asian and (right) European countries.}
\end{figure*}

\section{Fractal Nature in the Cellular Networks Topology}
As a fundamental feature of networks, fractal phenomenon has been found in a number of wireless networking scenarios \cite{Yuan2017The}. For instance, the design of hand-off scheme for mobile terminals can be inspired by the fractal property \cite{Ge2016Wireless}, and a fractal shape demonstrates itself in the coverage boundary of the wireless cellular network \cite{Hao2017Wireless}. Moreover, a rich number of networks in the real world exhibit the significant fractal characteristics naturally, such as the world-wide web (WWW), yeast interaction, protein homology, and social networks \cite{Strogatz2005Complex,Song2005Self}. Based on the tools of $ \alpha $-Shapes and Betti numbers, this section confirms the fractal nature in the cellular networks from the perspective of the topology of BSs \cite{Pranav2017The}. 

\subsection{Comparison between the Betti Curves of Fractal and Random Point Distributions}
Fig. 6 gives an intuitive comparison between the Betti curves brought by the fractal patterns and the random characteristics.

\begin{figure*}[htbp]
	\centering
	\subfigure[Points deployment diagrams: (left) random; (right) fractal.]{
		\makebox[6cm][c]{
			\includegraphics[scale=0.2]{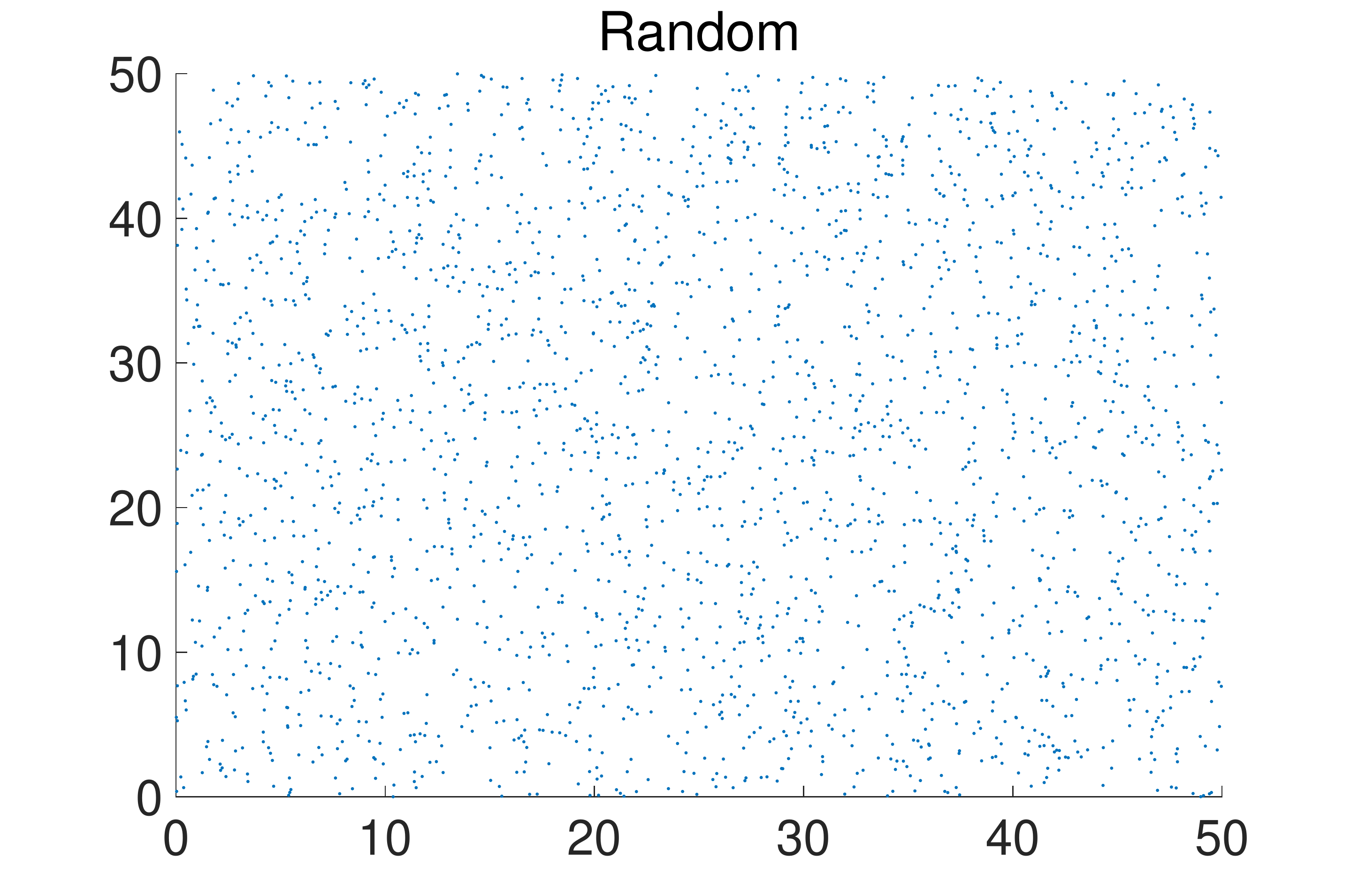}
		}
		\makebox[6cm][c]{
			\includegraphics[scale=0.2]{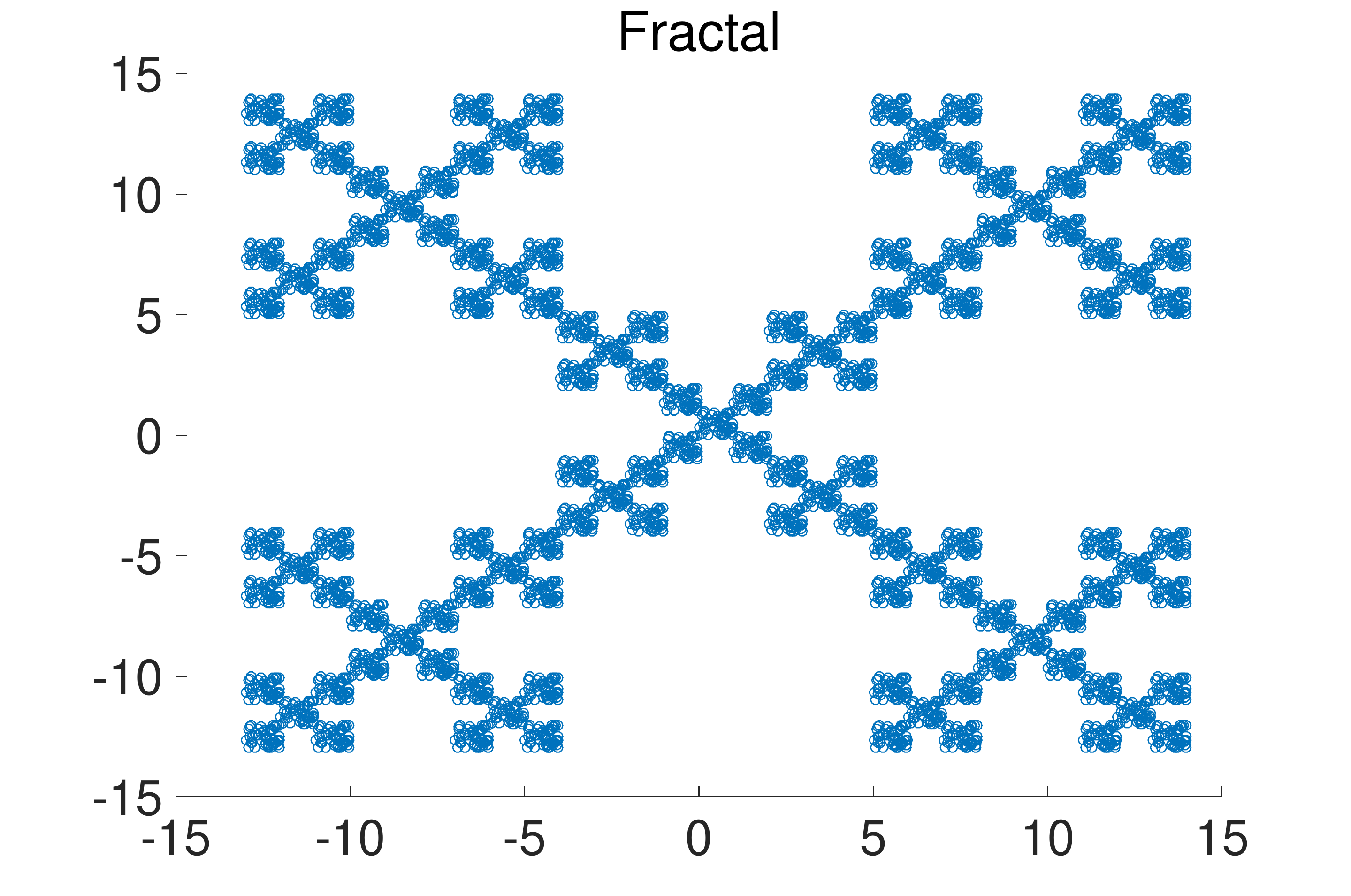}
		}
	}

	\subfigure[Betti curves for random: (left) $ \beta_{0} $; (right) $ \beta_{1} $.]{
		\makebox[6cm][c]{
			\includegraphics[scale=0.2]{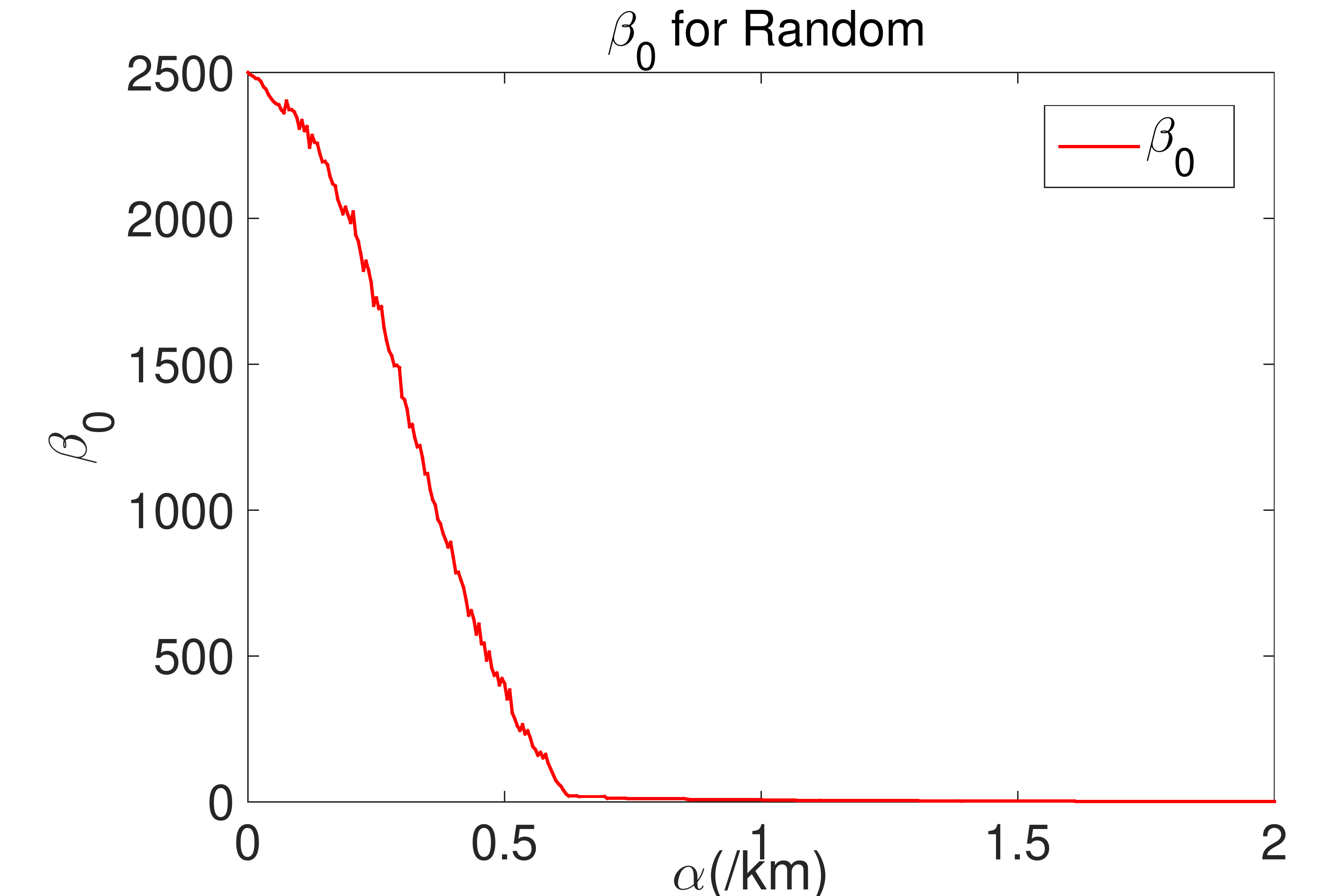}
		}
		\makebox[6cm][c]{
			\includegraphics[scale=0.2]{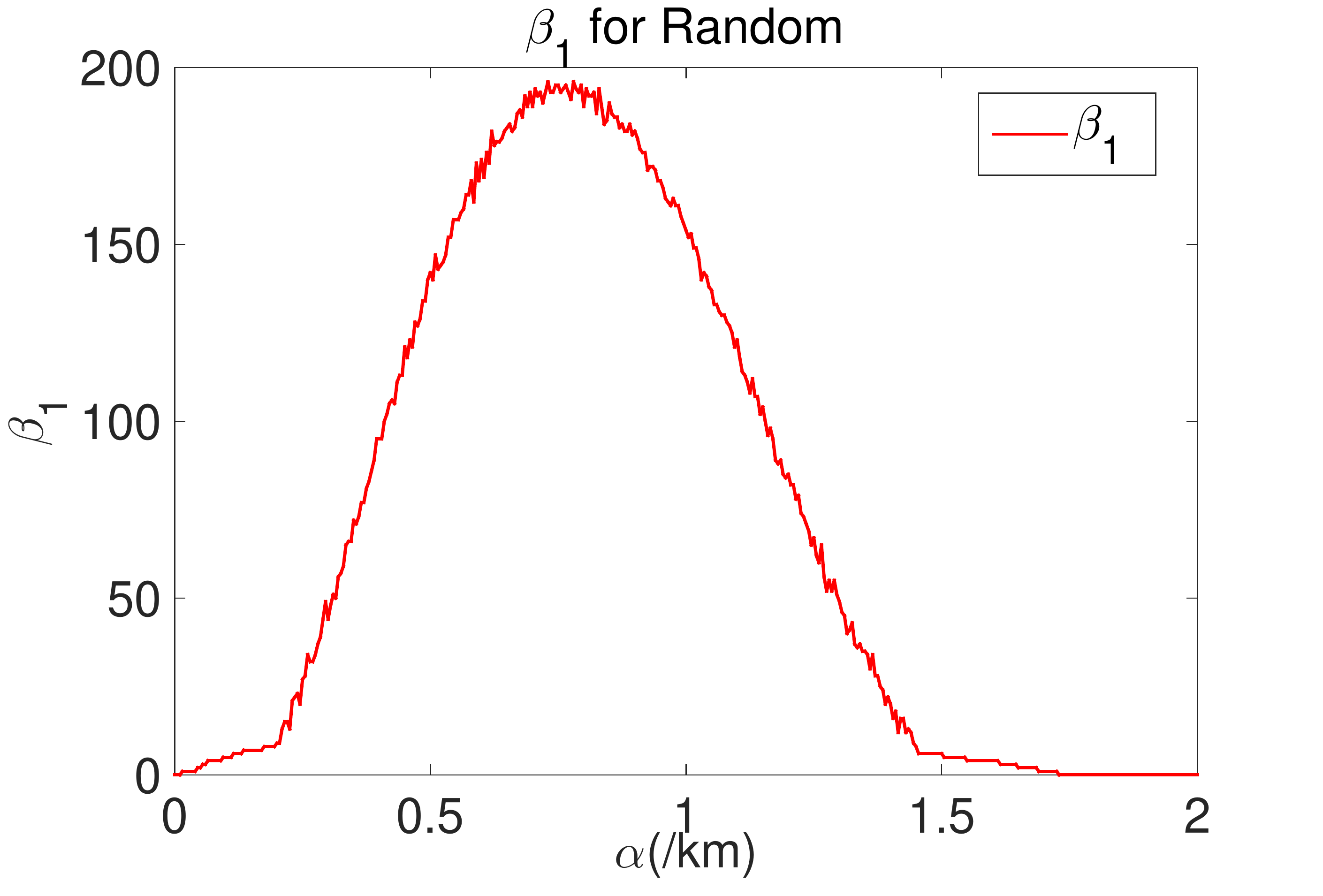}
		}
	}

    \subfigure[Betti curves for fractal: (left) $ \beta_{0} $; (right) $ \beta_{1} $.]{
	\makebox[6cm][c]{
		\includegraphics[scale=0.2]{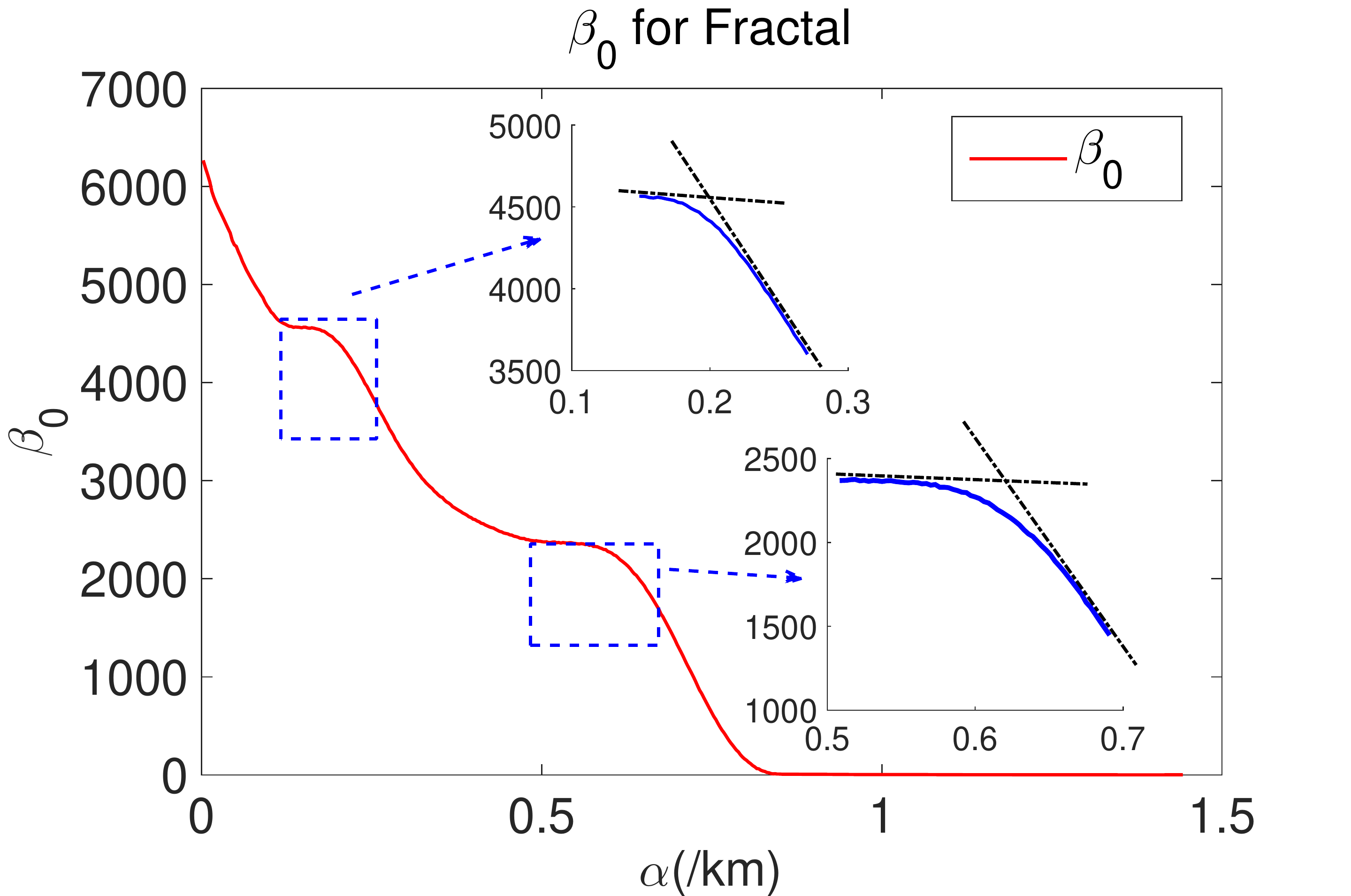}
	}
	\makebox[6cm][c]{
		\includegraphics[scale=0.2]{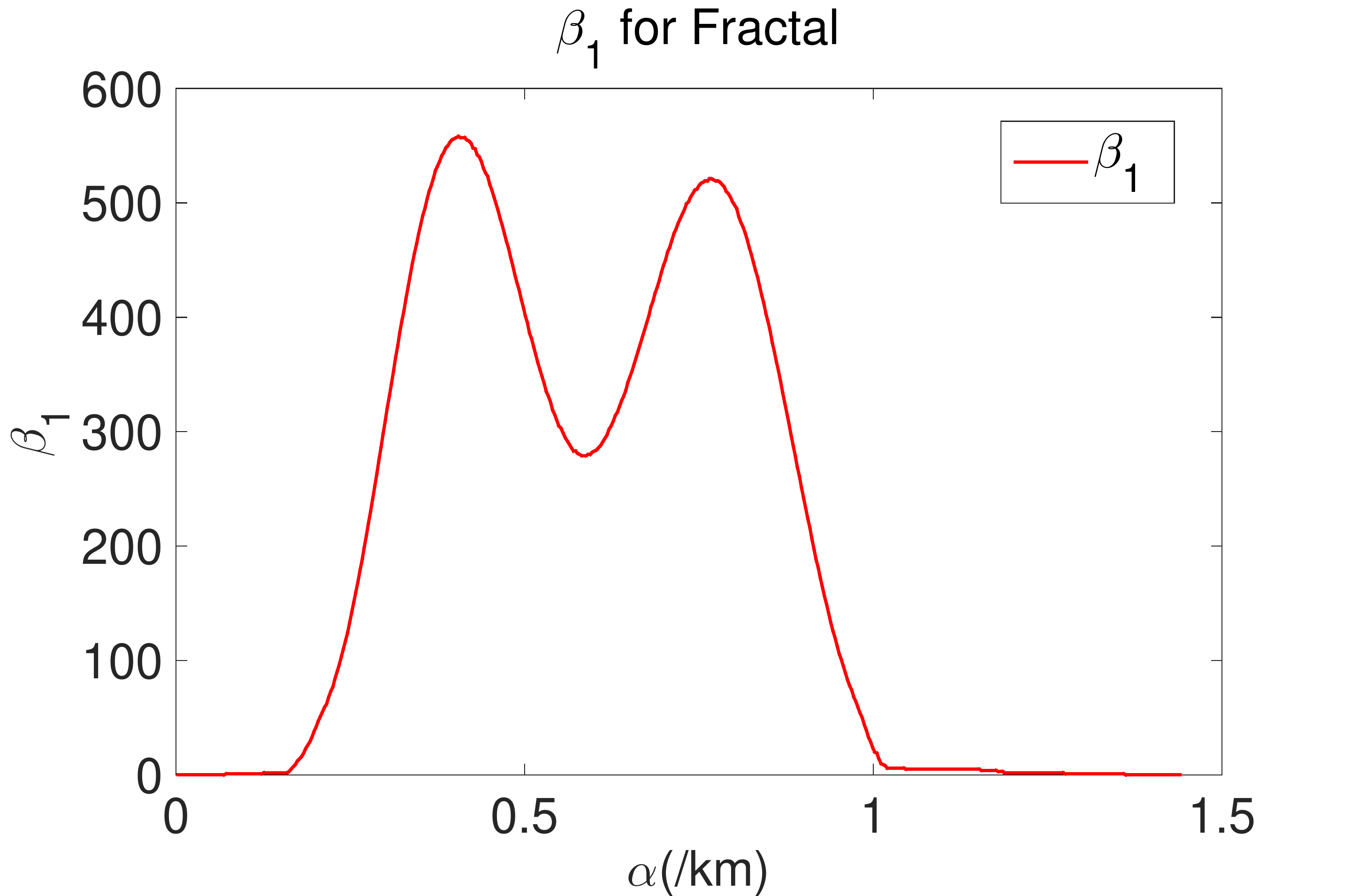}
	}
}
	
	\caption{Comparison between the representative Betti curves of random and fractal point distributions. }
\end{figure*}

Fig. 6(a) expresses the practical point deployments for the fractal and random cases. One of the most significant features of fractal behavior is the self-similarity under any length scale \cite{Benhaiem2018Self}, and self-similarity is added into the point distribution by the hierarchical partitions of the area among the points, while the random point distribution is simply realized by the random choice of the location for each point. 

Fig. 6(b) shows the Betti curves for the random case, while Fig. 6(c) for the fractal case. The $ \beta_{0} $ curve for the random case demonstrates a trend of descending monotonously, and the $ \beta_{1} $ curve presents a single peak formed by the monotone decrease after the monotone rise of $ \beta_{1} $. However, in the fractal case, distinctive from the random case, the $ \beta_{0} $ curve manifests itself by the multiple ripples in the global decline trend, where a ripple is characterized by the rapid slope switch within a narrow range of $\alpha$, as shown by the amplified blue curves and two black straight lines with distinctive slopes in the left window in Fig. 6(c). In the meantime, the $ \beta_{1} $ curve for the fractal case fluctuates in the form of multiple peaks. In other words, the fractal behavior brings the distinguishing features of multiple ripples and peaks into the $ \beta_{0} $ and $ \beta_{1} $ curves, respectively \cite{Pranav2017The}. In addition, the identical number of ripples and peaks is an indication for the hierarchical levels in the fractal distribution \cite{Pranav2017The}. For example, the first two levels, out of the total three levels in the fractal distribution, demonstrate themselves in a visible way by the two ripples or peaks in the Betti curves, while the last level is not observed clearly in the curves because of the trivial number of elements in this level. Moreover, the ripples arrive prior to the corresponding peaks, which is quite reasonable because the size of components is smaller than that of loops. 

\subsection{Fractal Features in Terms of the Betti Numbers}
Within our works, the fractal phenomenon in the cellular networks is discovered based on the modularity patterns (structures hierarchy) of the Betti curves of the practical BS distributions \cite{Pranav2017The}, as verified in Fig. 7 for six European countries and Fig. 8 for six Asian countries, respectively. Regardless of the geographical differences, it is extremely surprising to observe the essentially identical fractal features (hierarchy of structures) expressed by the multiple ripples, which are highlighted in blue segments as shown in the $ \beta_{0} $ curves, as well as the multiple peaks in the $ \beta_{1} $ curves for each of all the aforementioned twelve countries.

\begin{figure*}[htbp]
	\centering
	\subfigure[Poland$ \qquad\qquad\qquad\qquad\qquad\qquad\qquad\qquad\qquad\qquad\qquad\qquad $(d) France]{
		\makebox[4cm][c]{
			\includegraphics[scale=0.15]{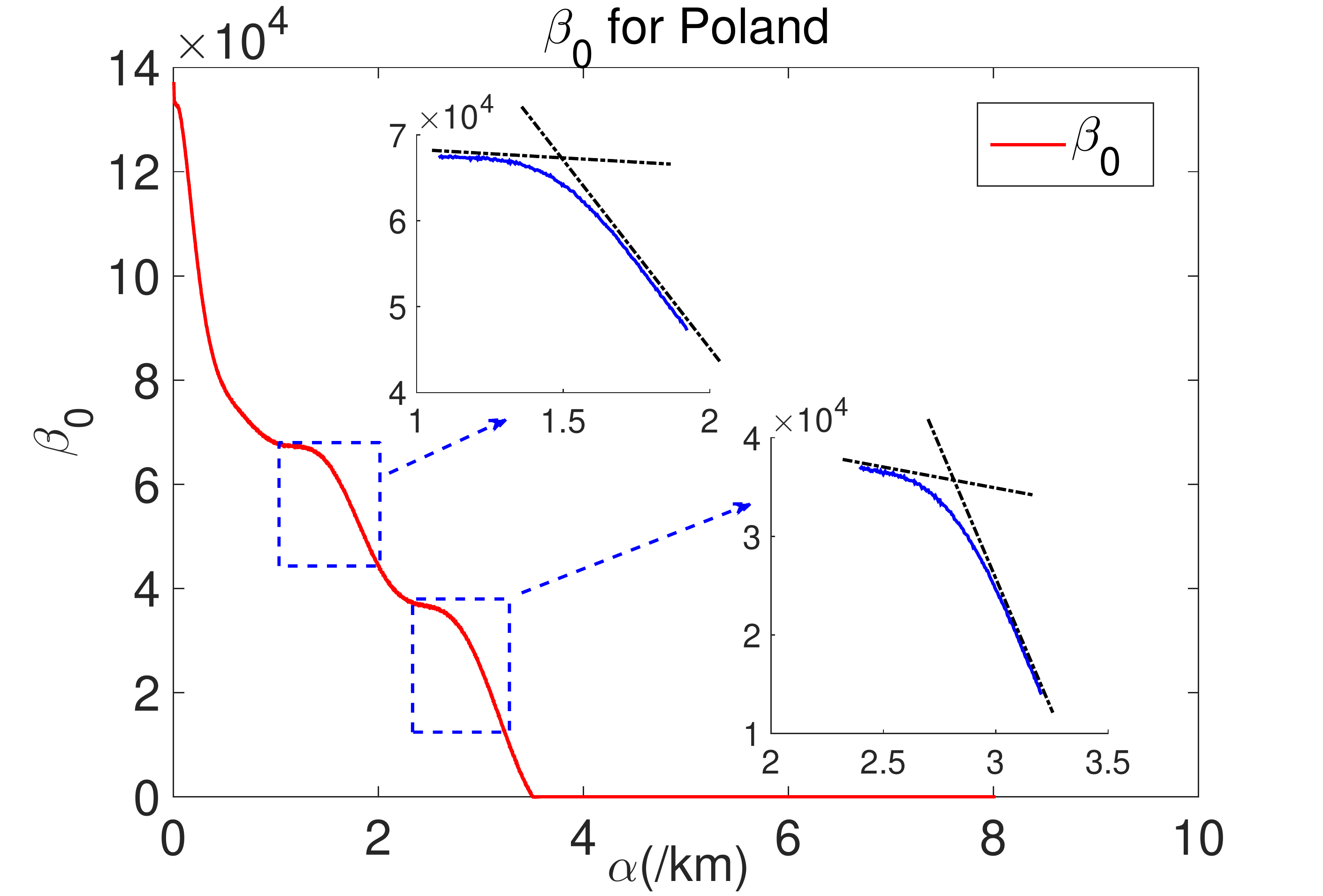}
		}
		\makebox[4cm][c]{
			\includegraphics[scale=0.15]{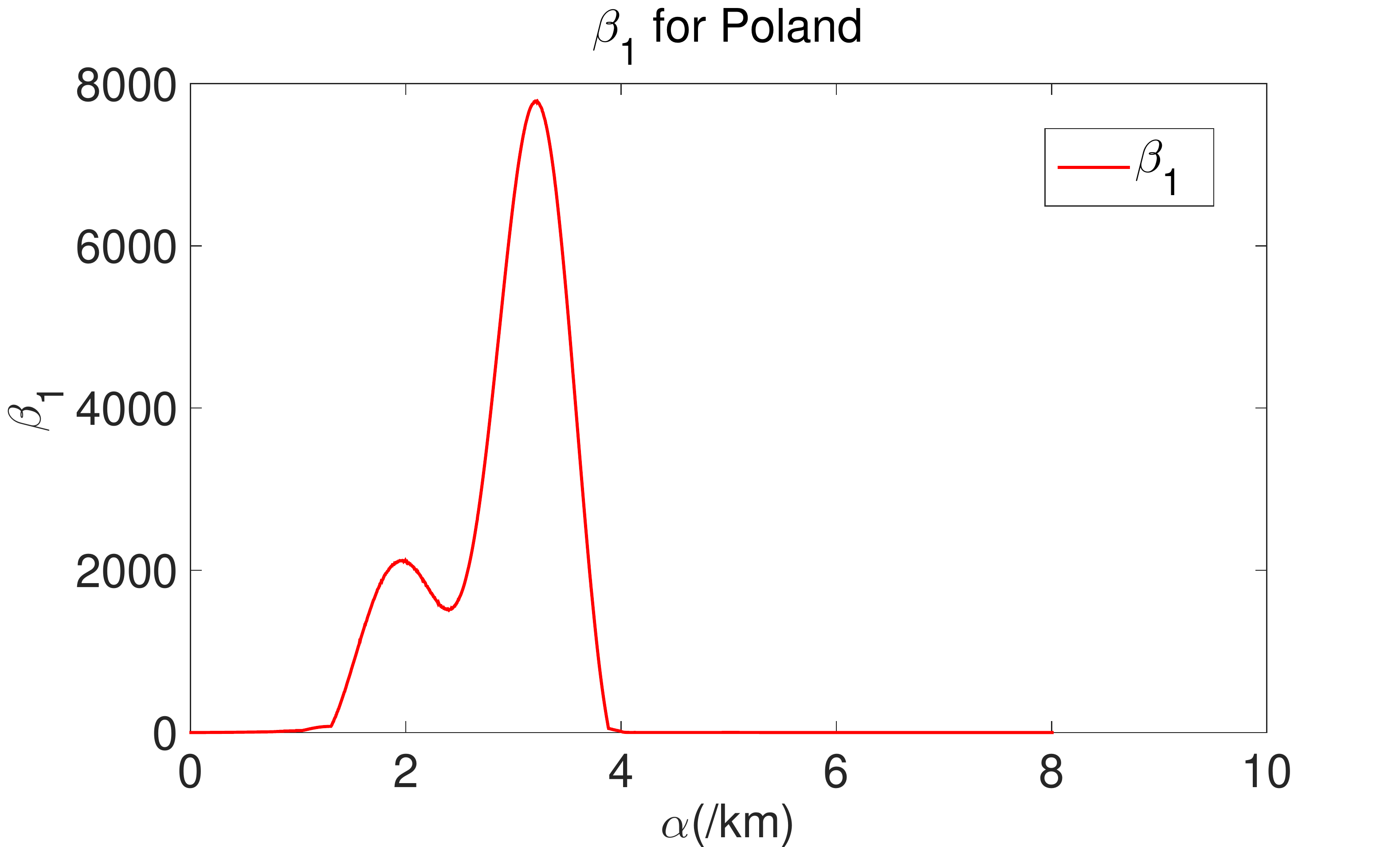}
		}
	   \makebox[4cm][c]{
		\includegraphics[scale=0.15]{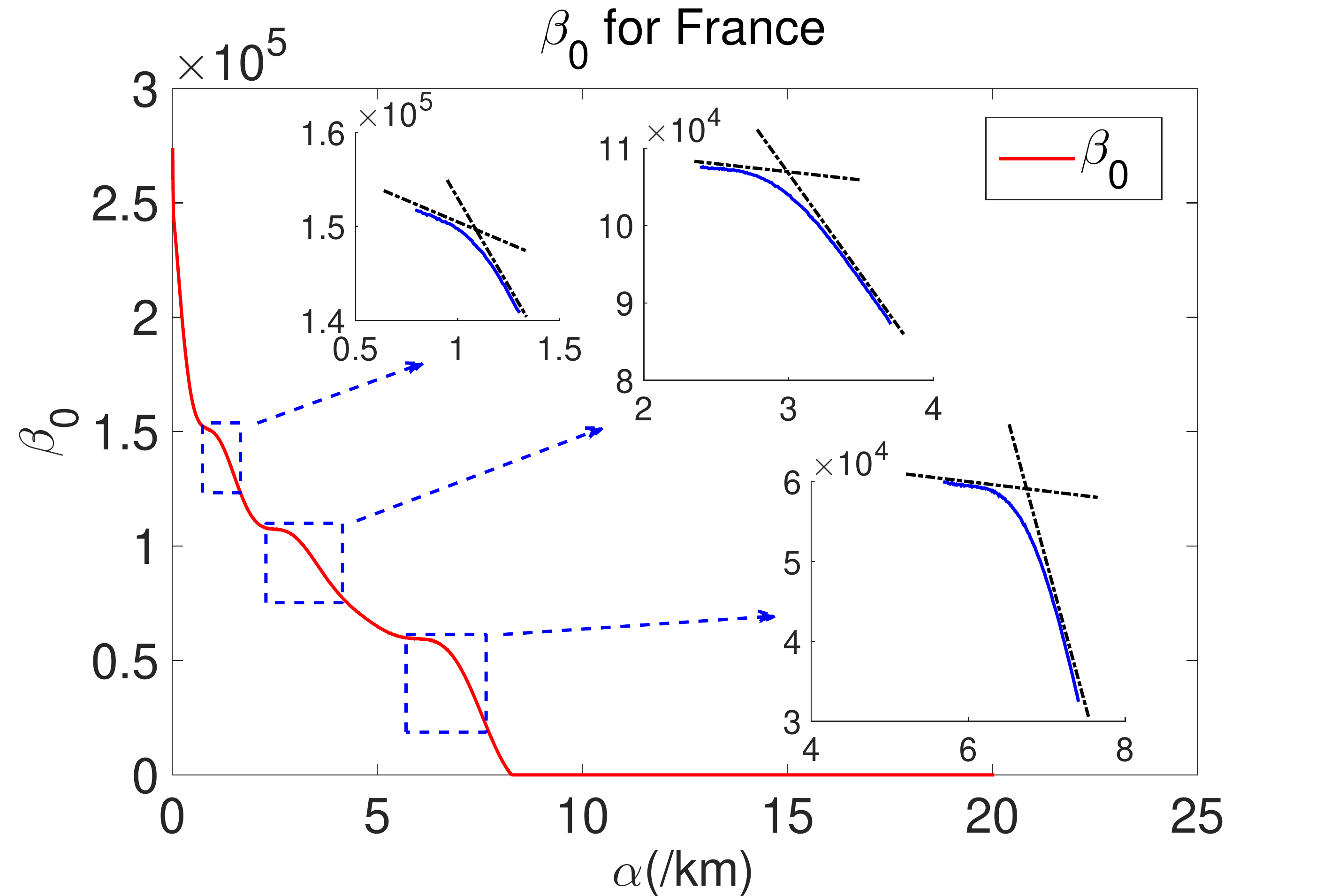}
	}
	   \makebox[4cm][c]{
		\includegraphics[scale=0.15]{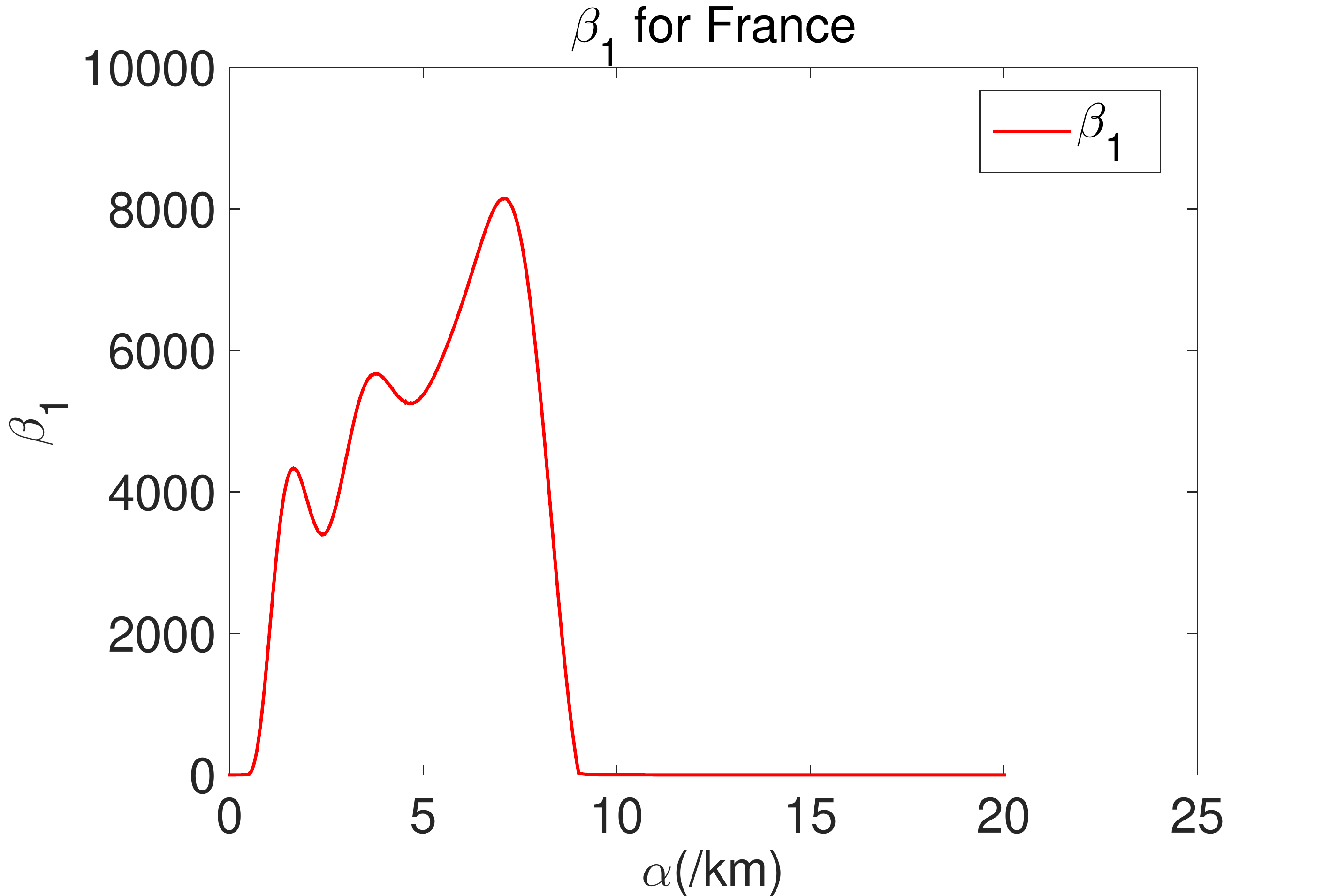}
	}
}
	
	\subfigure[UK$ \qquad\qquad\qquad\qquad\qquad\qquad\qquad\qquad\qquad\qquad\qquad\qquad $(e) Italy]{
		\makebox[4cm][c]{
			\includegraphics[scale=0.15]{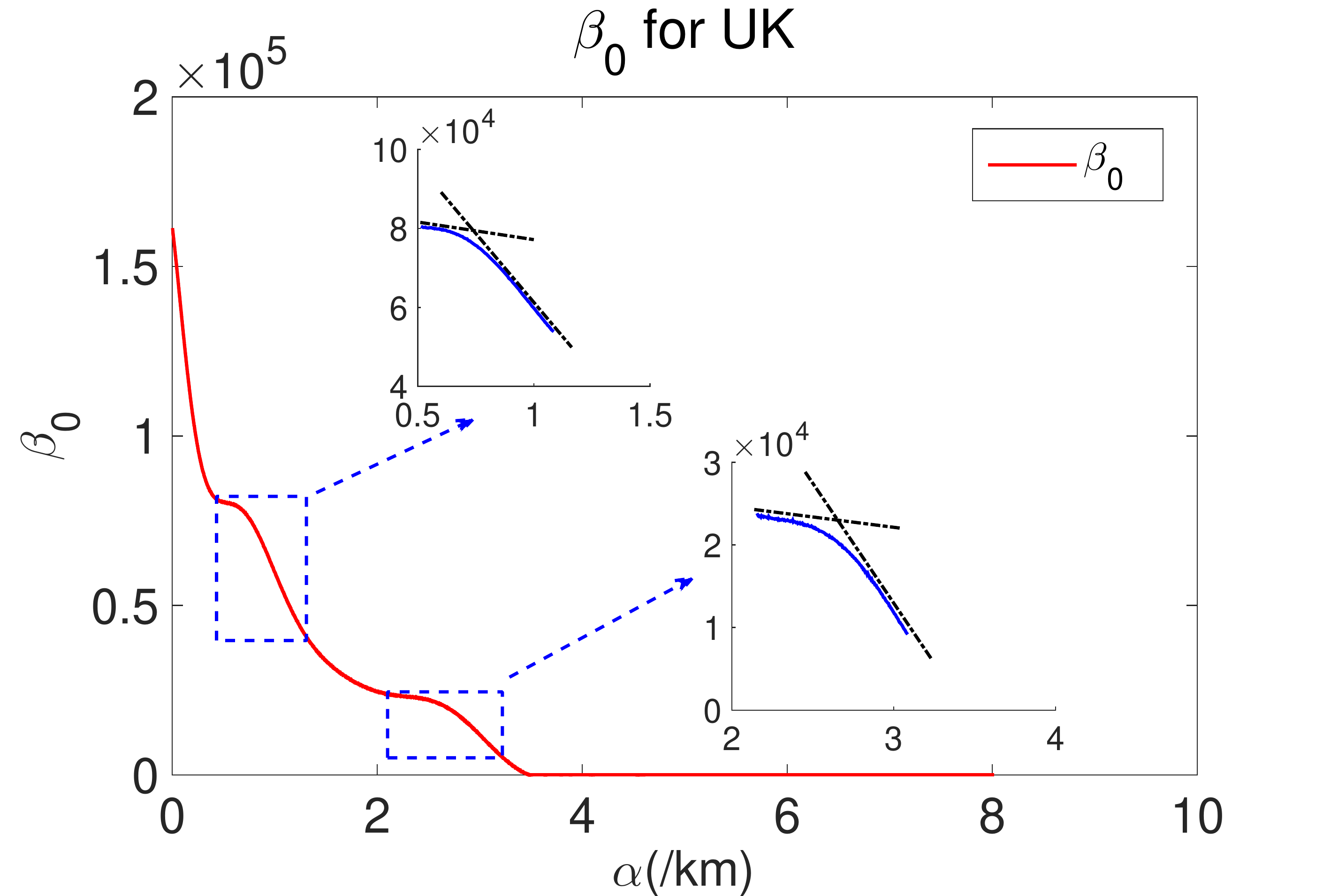}
		}
		\makebox[4cm][c]{
			\includegraphics[scale=0.15]{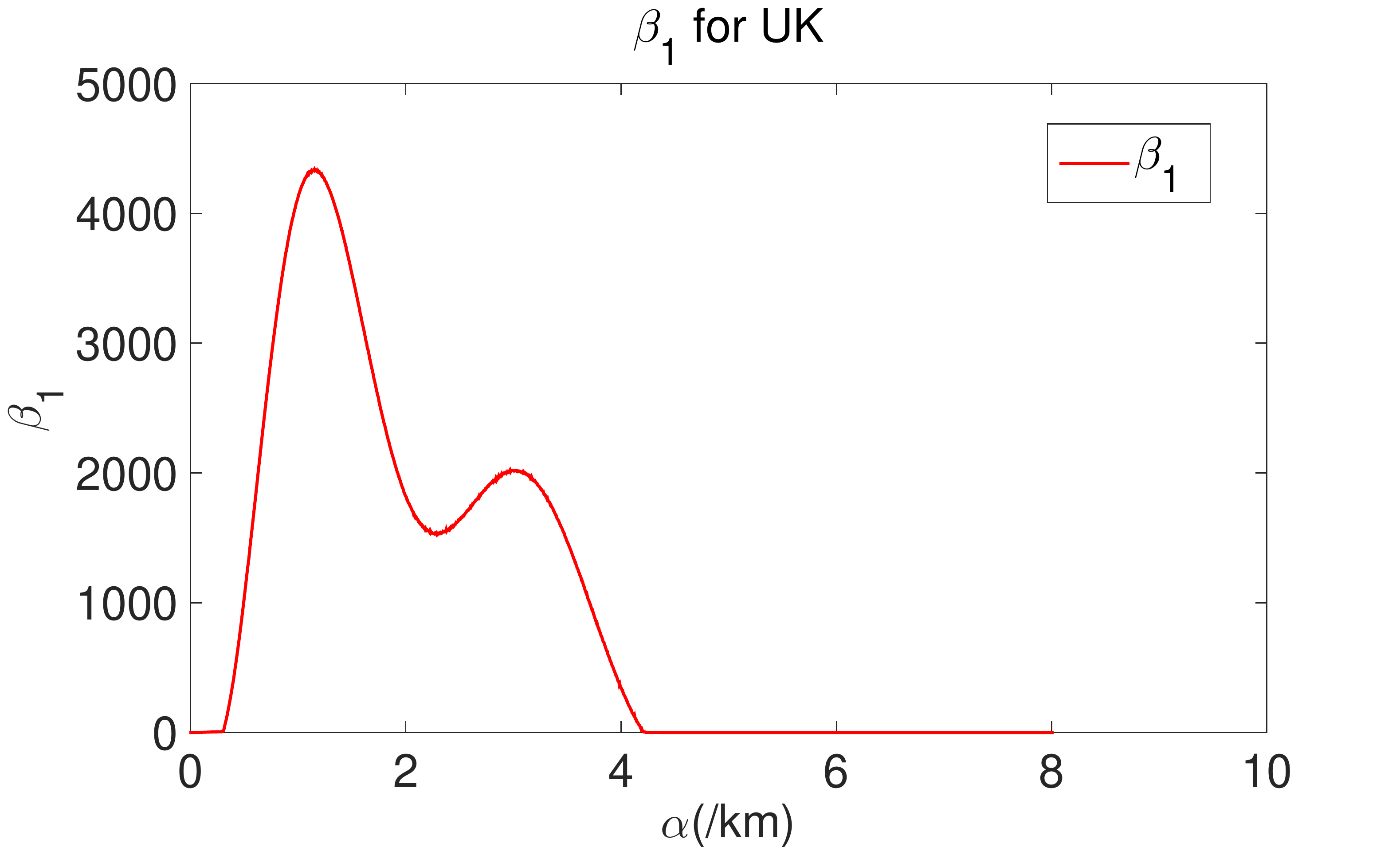}
		}
		\makebox[4cm][c]{
		\includegraphics[scale=0.15]{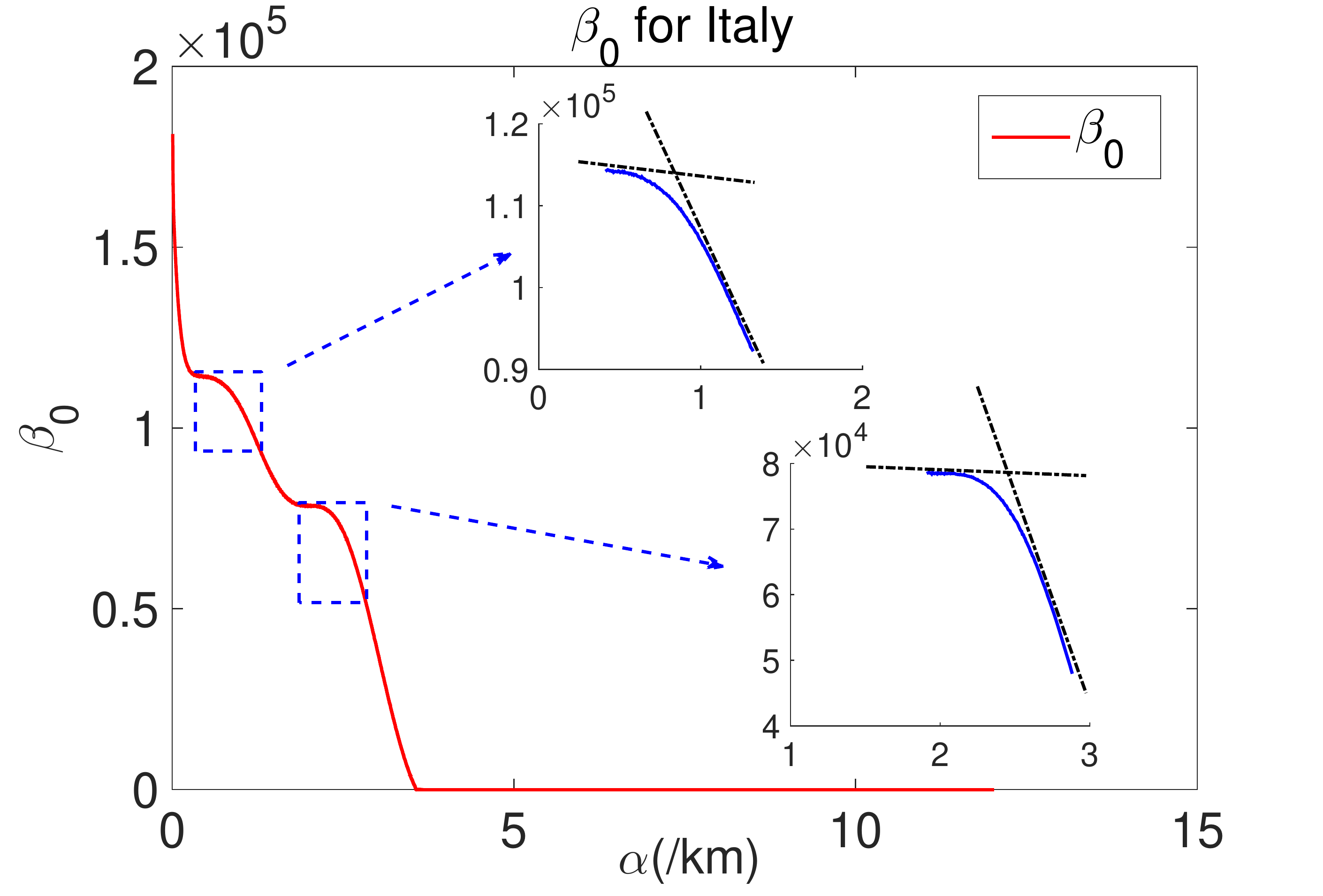}
	}
	    \makebox[4cm][c]{
		\includegraphics[scale=0.15]{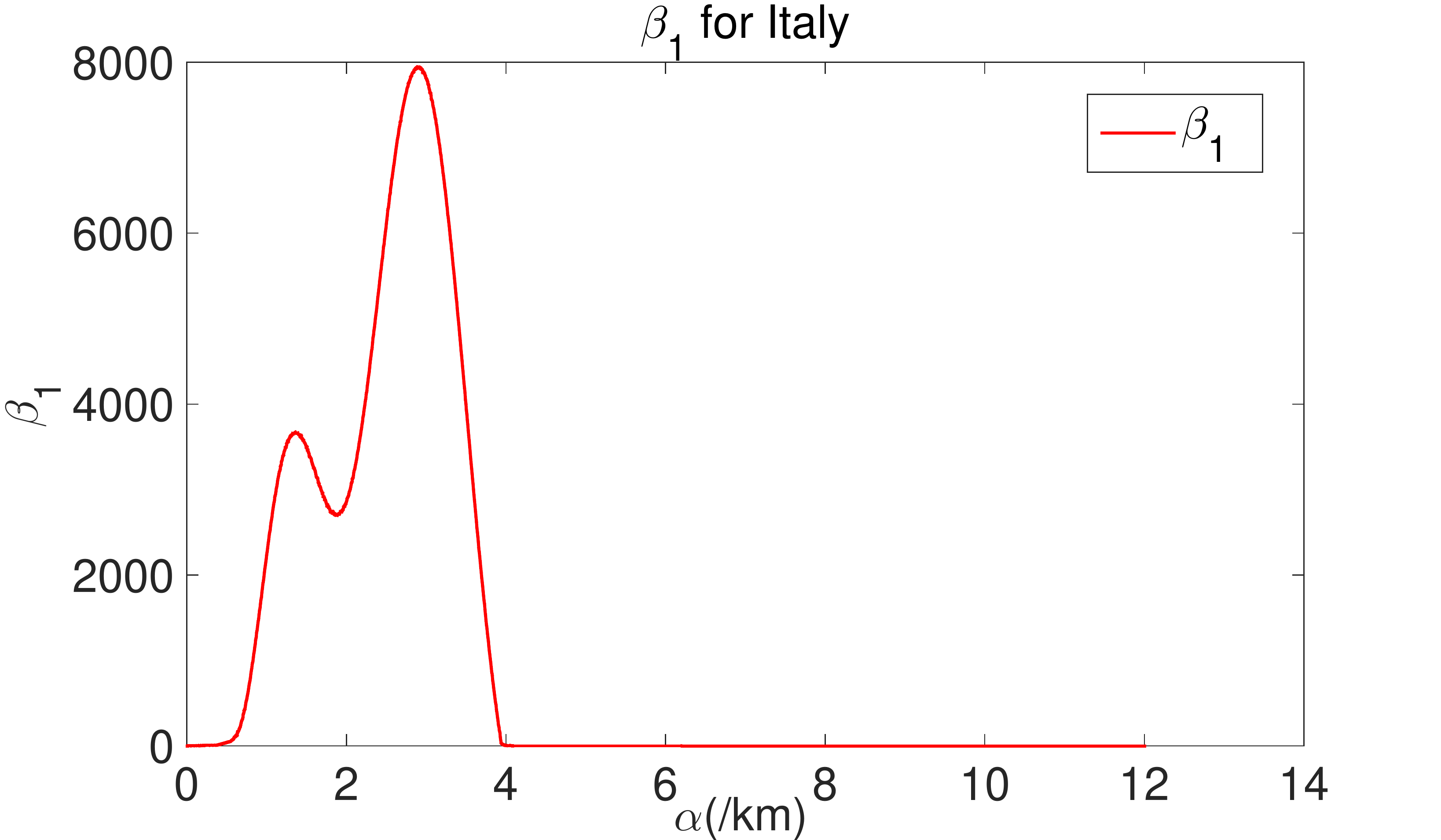}
	}
}
	
	\subfigure[Germany$ \qquad\qquad\qquad\qquad\qquad\qquad\qquad\qquad\qquad\qquad\qquad $(f) Netherlands]{
		\makebox[4cm][c]{
			\includegraphics[scale=0.15]{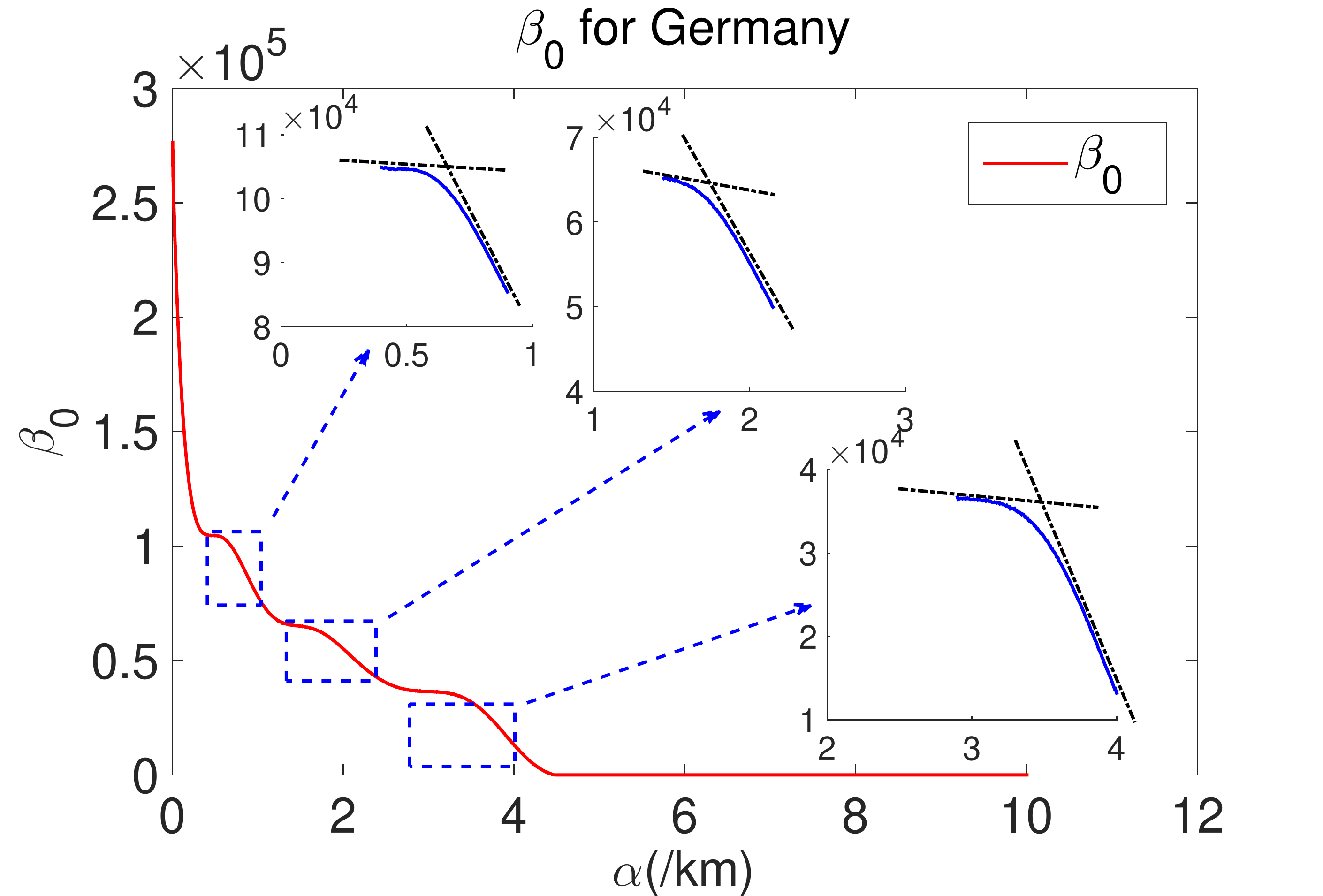}
		}
		\makebox[4cm][c]{
			\includegraphics[scale=0.15]{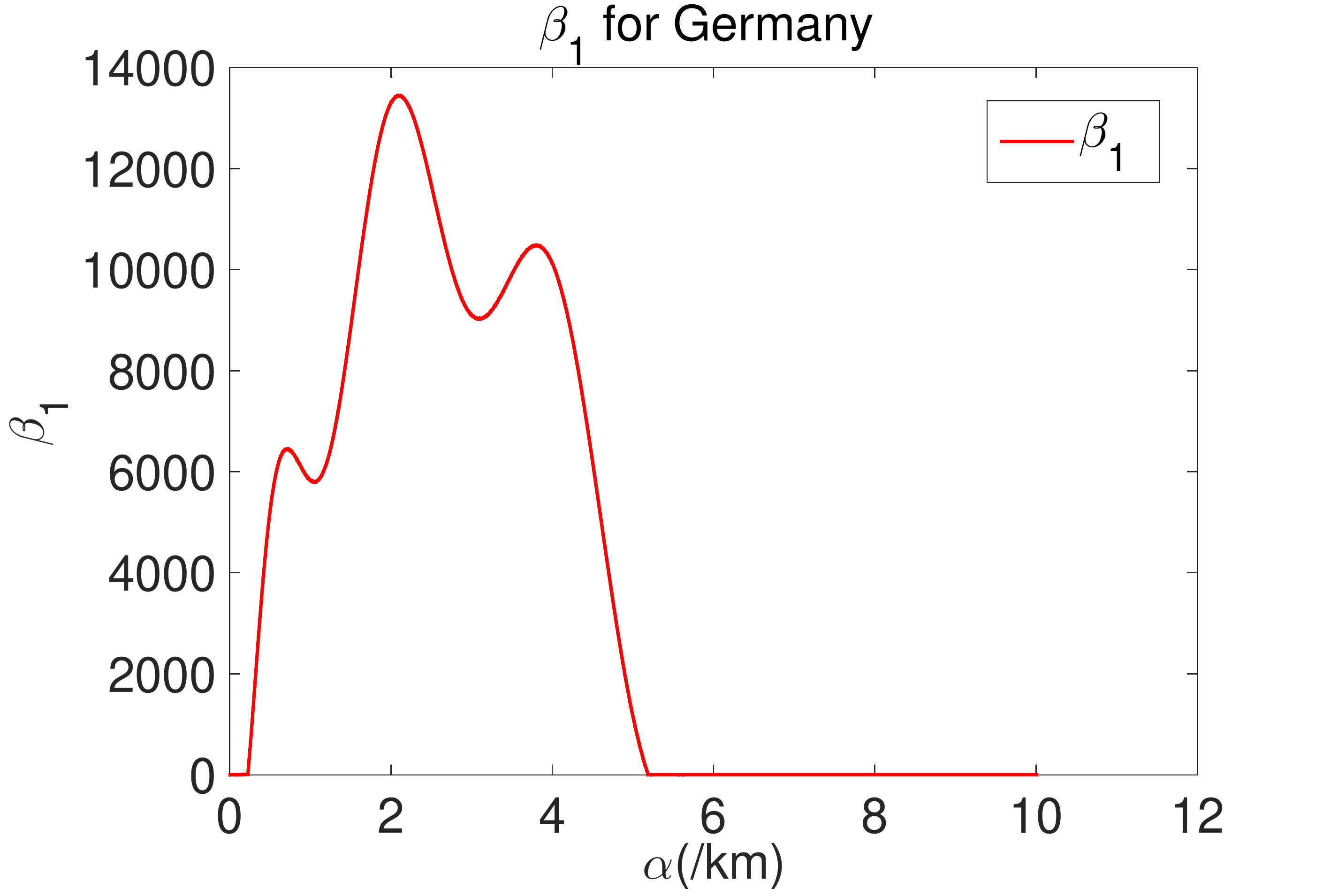}
		}
		\makebox[4cm][c]{
		\includegraphics[scale=0.15]{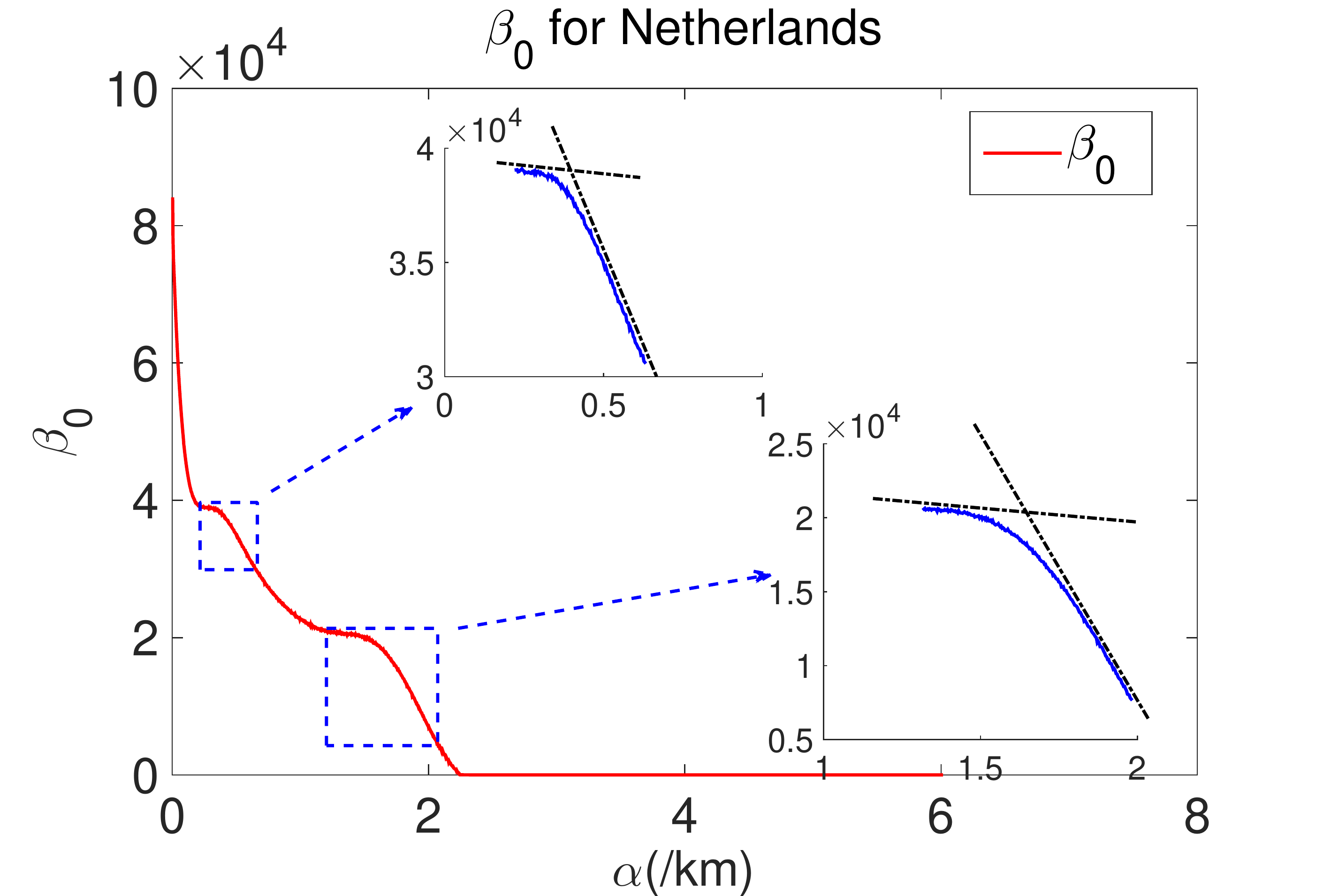}
	}
	   \makebox[4cm][c]{
		\includegraphics[scale=0.15]{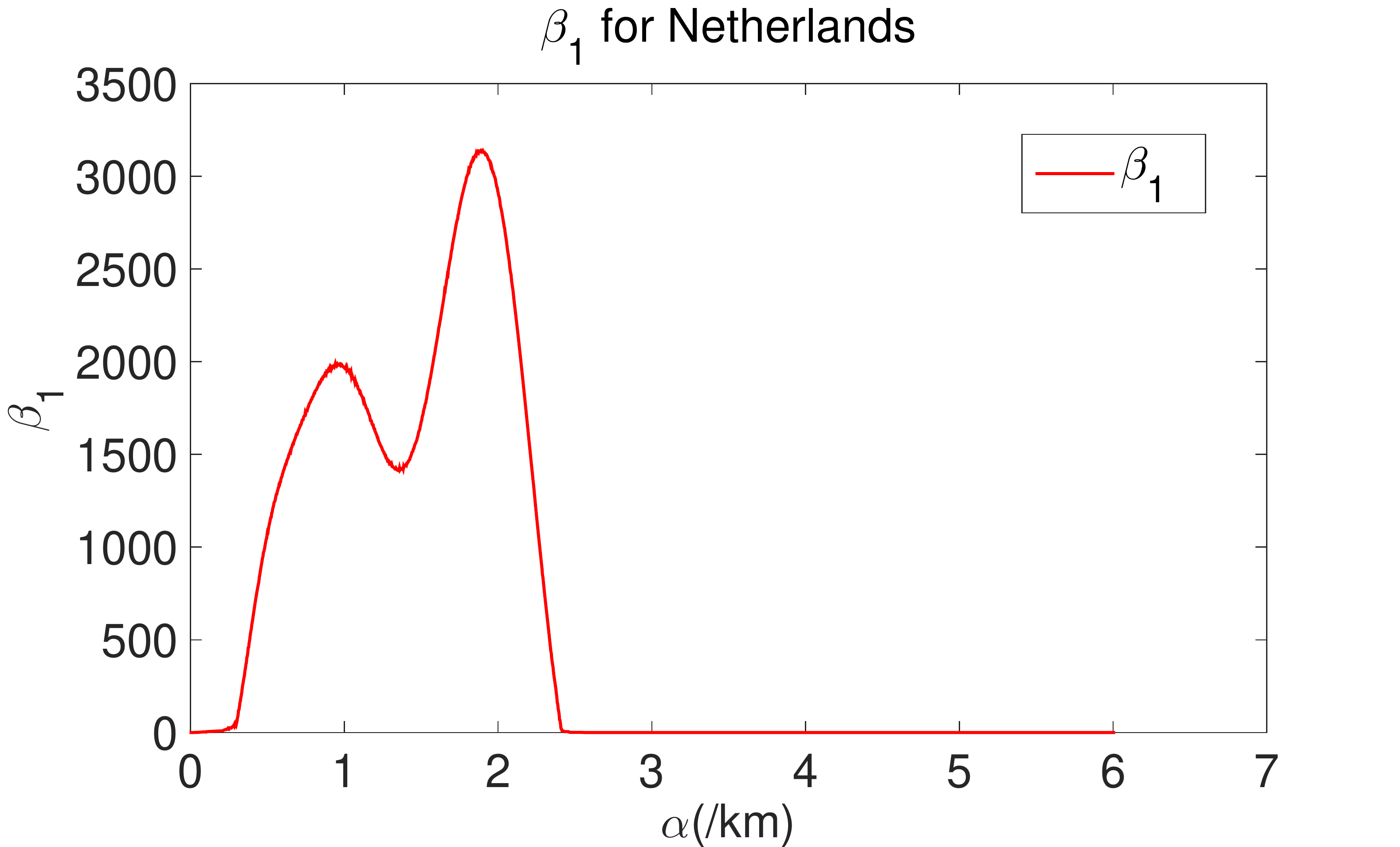}
	}
}	
	
	\caption{The Betti curves of the practical BSs distributions in European countries.}
\end{figure*}

\begin{figure*}[htbp]
	\centering
	\subfigure[Singapore$ \qquad\qquad\qquad\qquad\qquad\qquad\qquad\qquad\qquad\qquad\qquad\qquad $(d) China]{
		\makebox[4cm][c]{
			\includegraphics[scale=0.15]{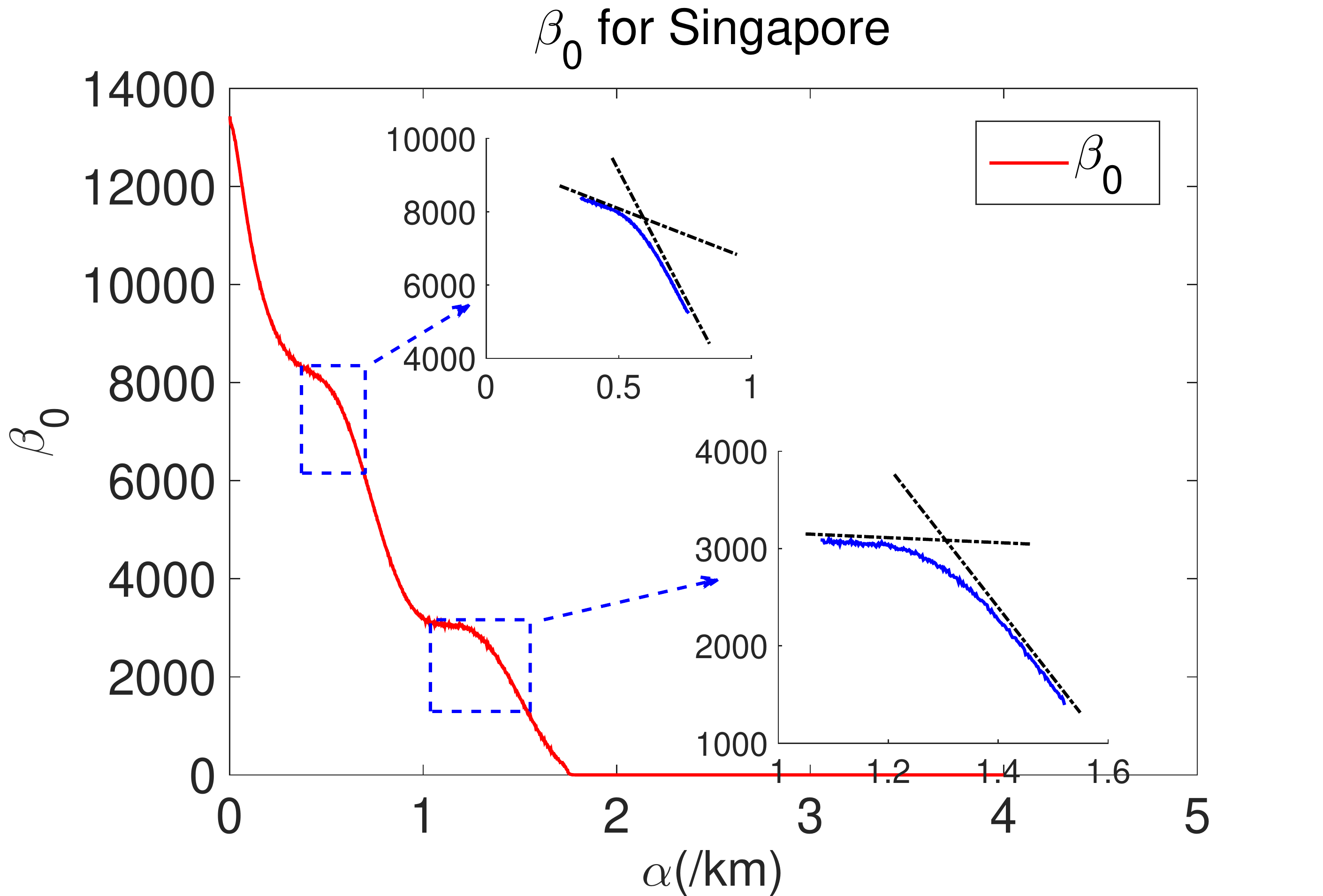}
		}
		\makebox[4cm][c]{
			\includegraphics[scale=0.15]{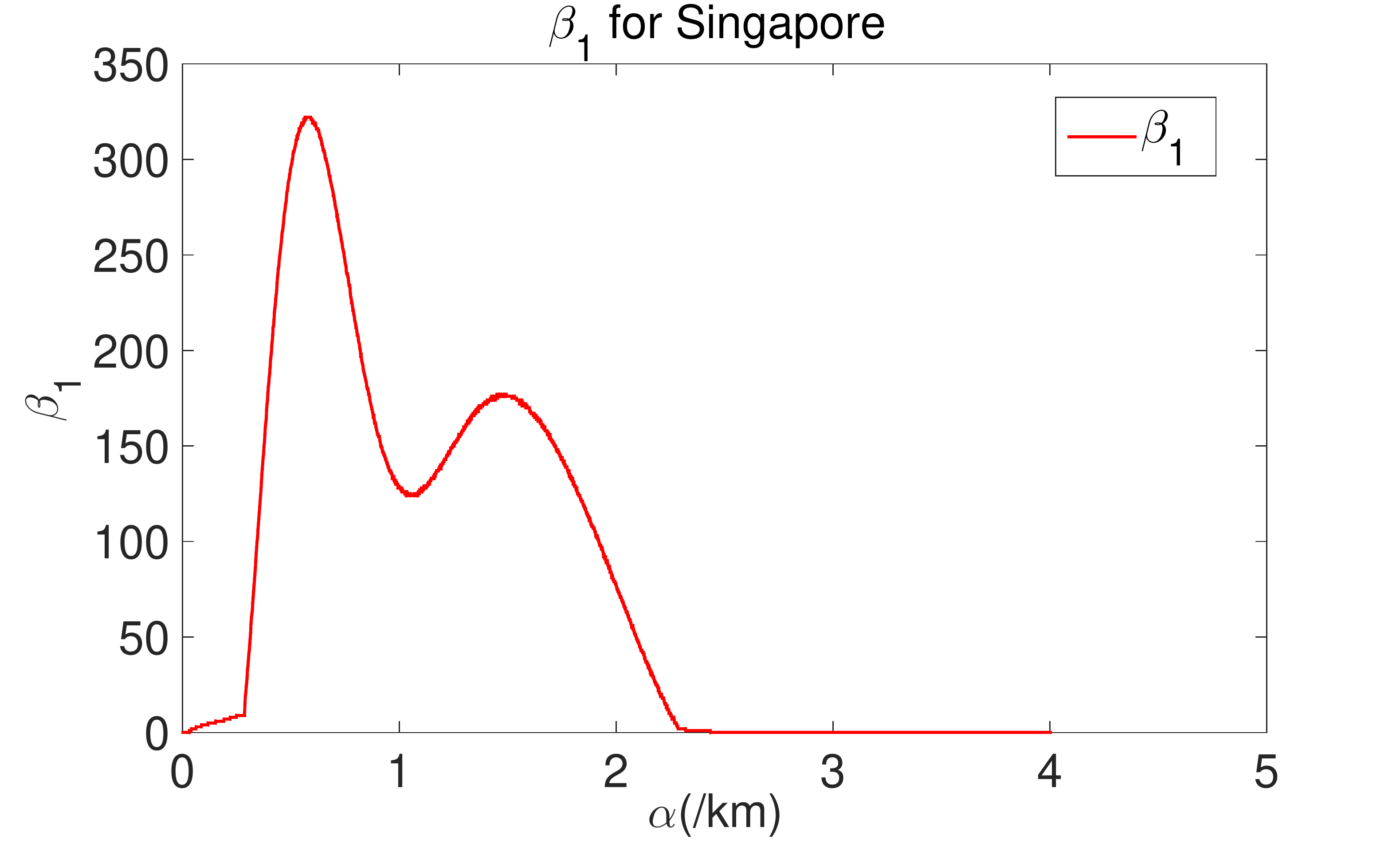}
		}
		\makebox[4cm][c]{
		\includegraphics[scale=0.15]{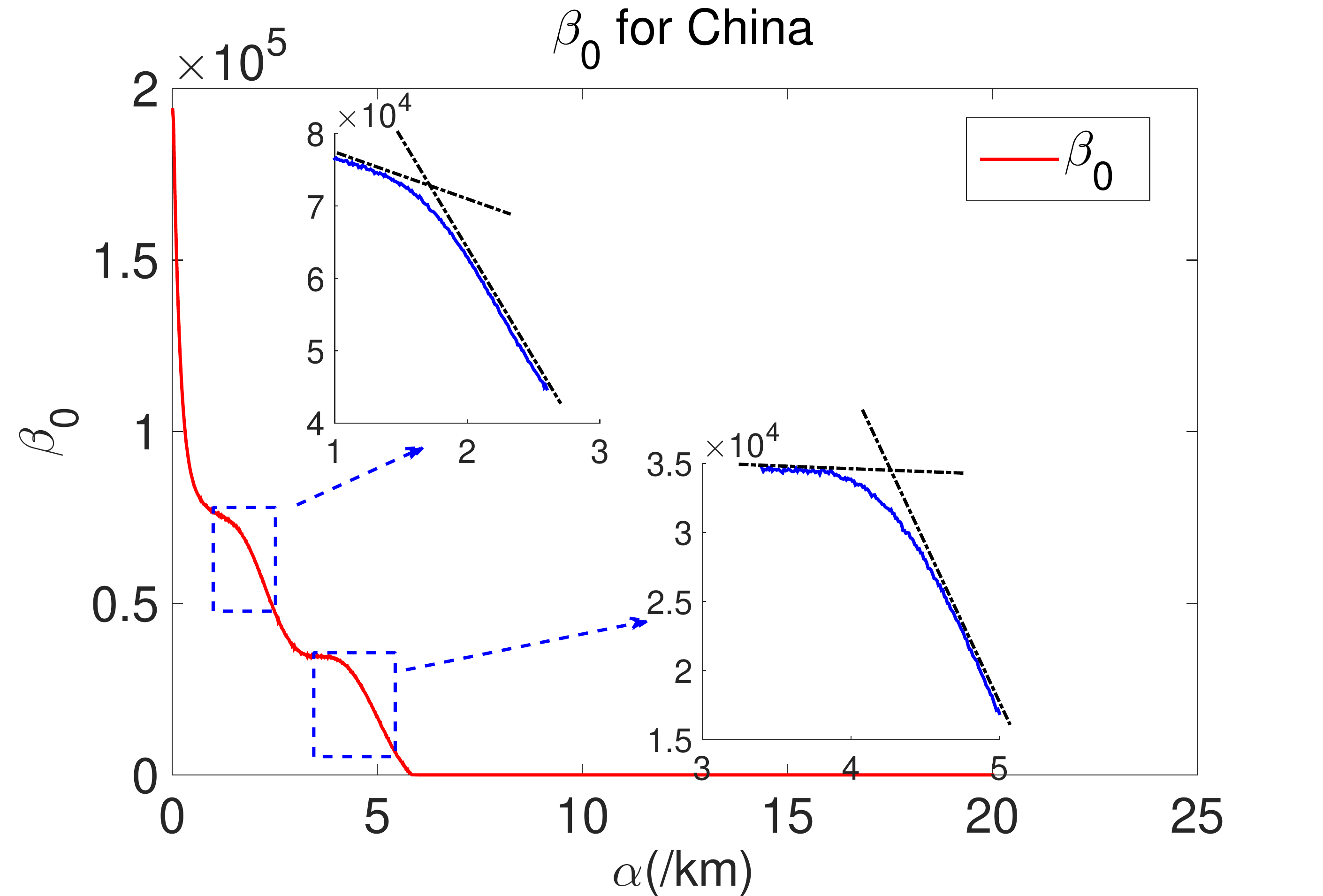}
	}
	\makebox[4cm][c]{
		\includegraphics[scale=0.15]{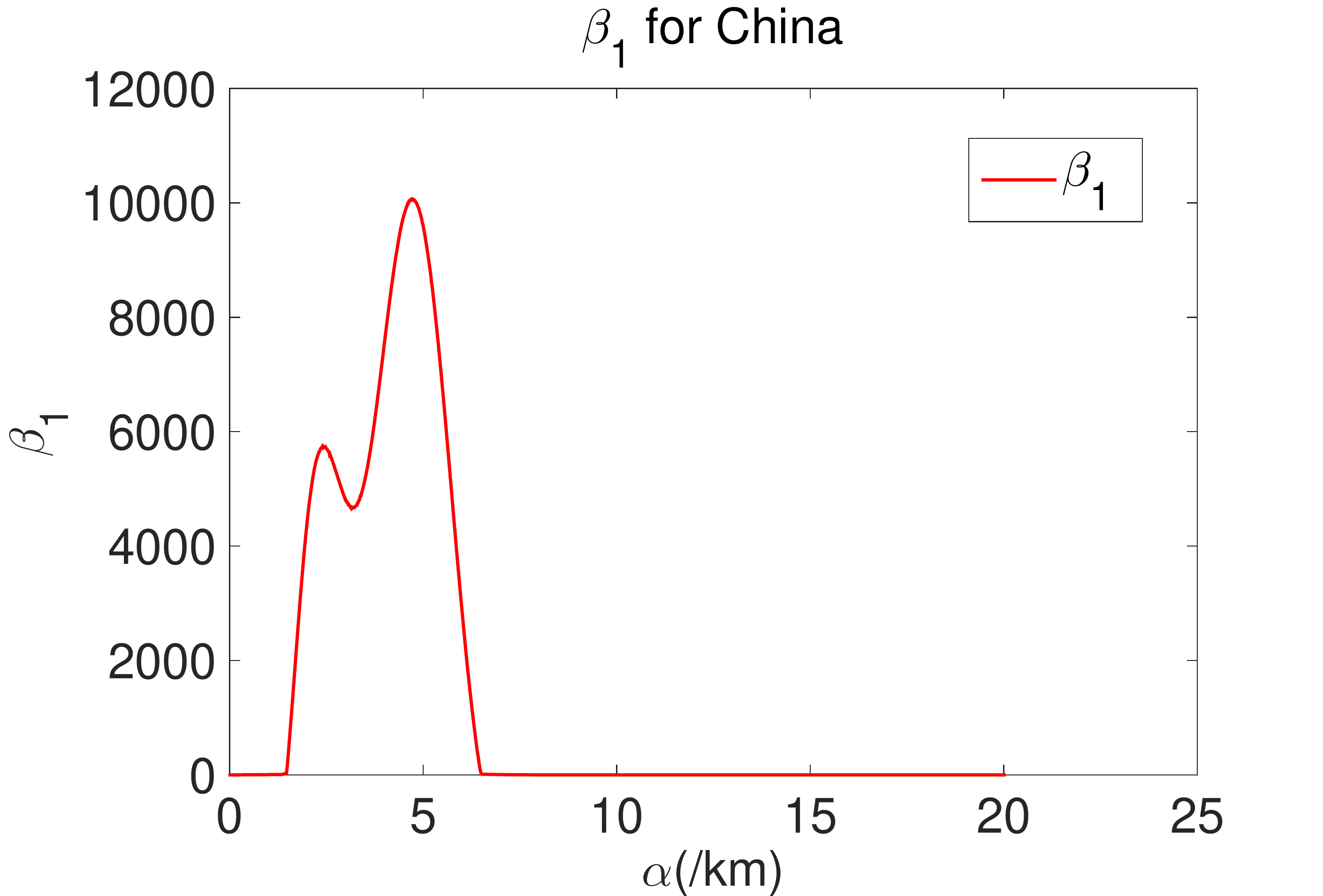}
	}
	}
	
	\subfigure[Korea$ \qquad\qquad\qquad\qquad\qquad\qquad\qquad\qquad\qquad\qquad\qquad\qquad $(e) India]{
		\makebox[4cm][c]{
			\includegraphics[scale=0.15]{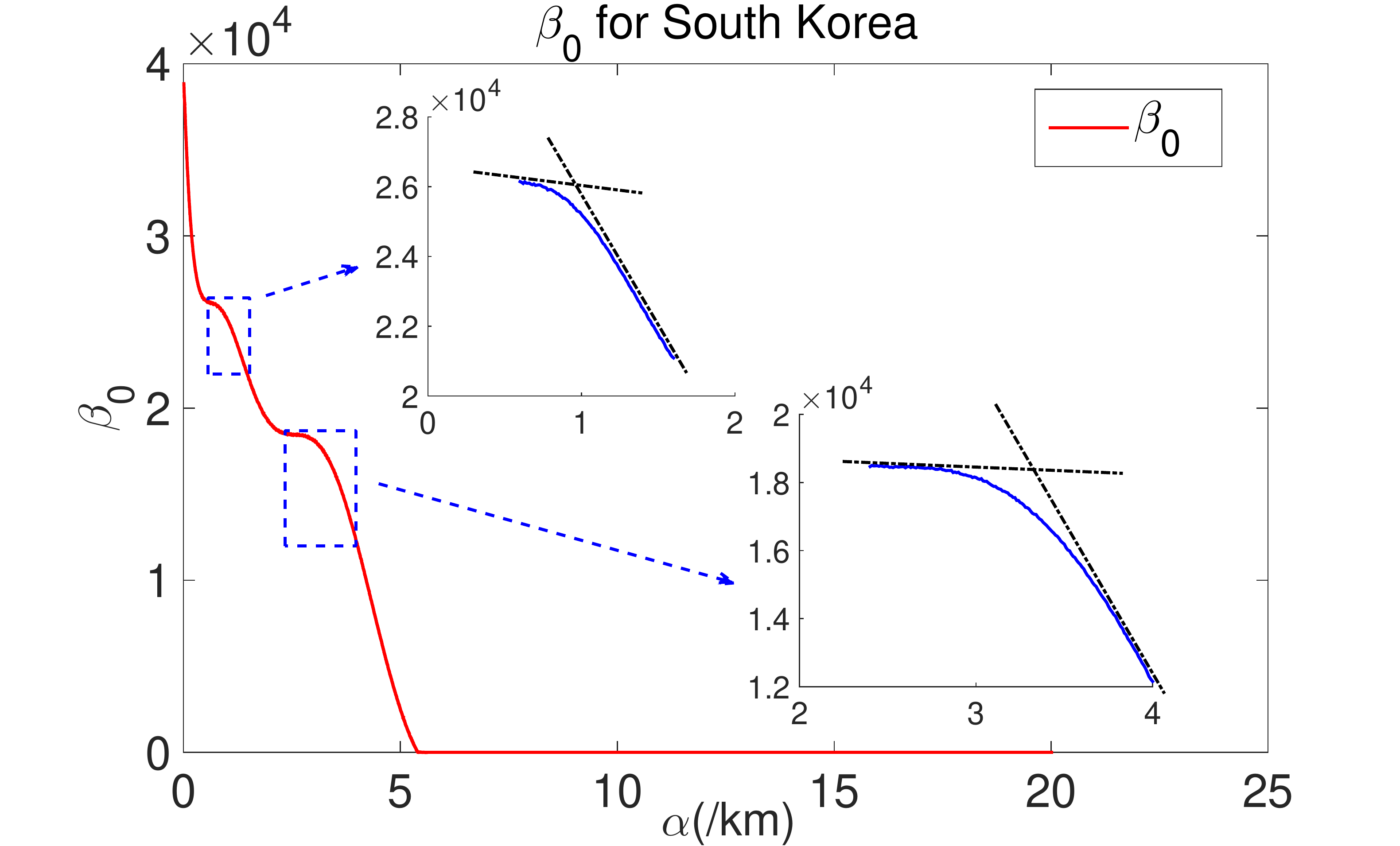}
		}
		\makebox[4cm][c]{
			\includegraphics[scale=0.15]{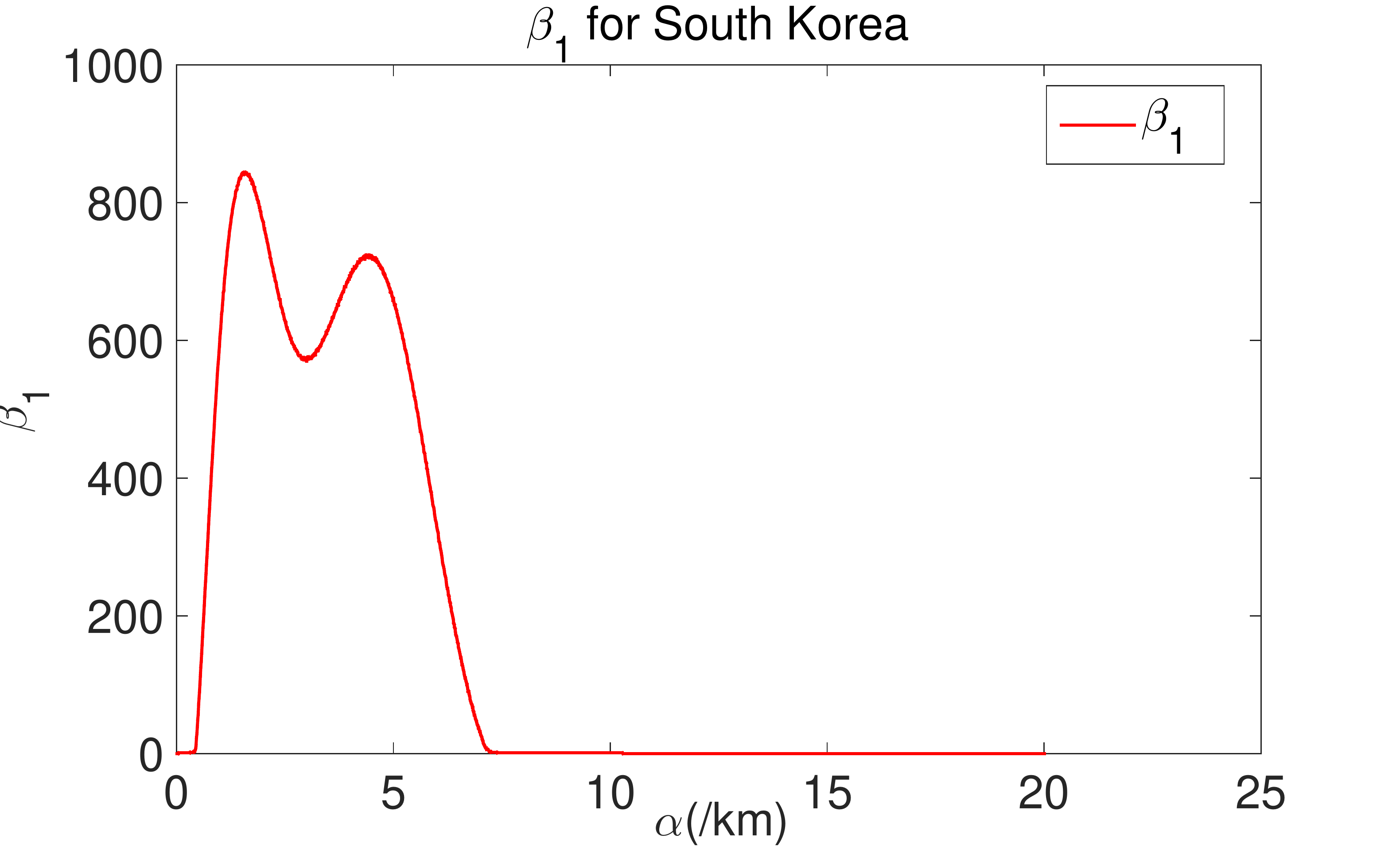}
		}
		\makebox[4cm][c]{
		\includegraphics[scale=0.15]{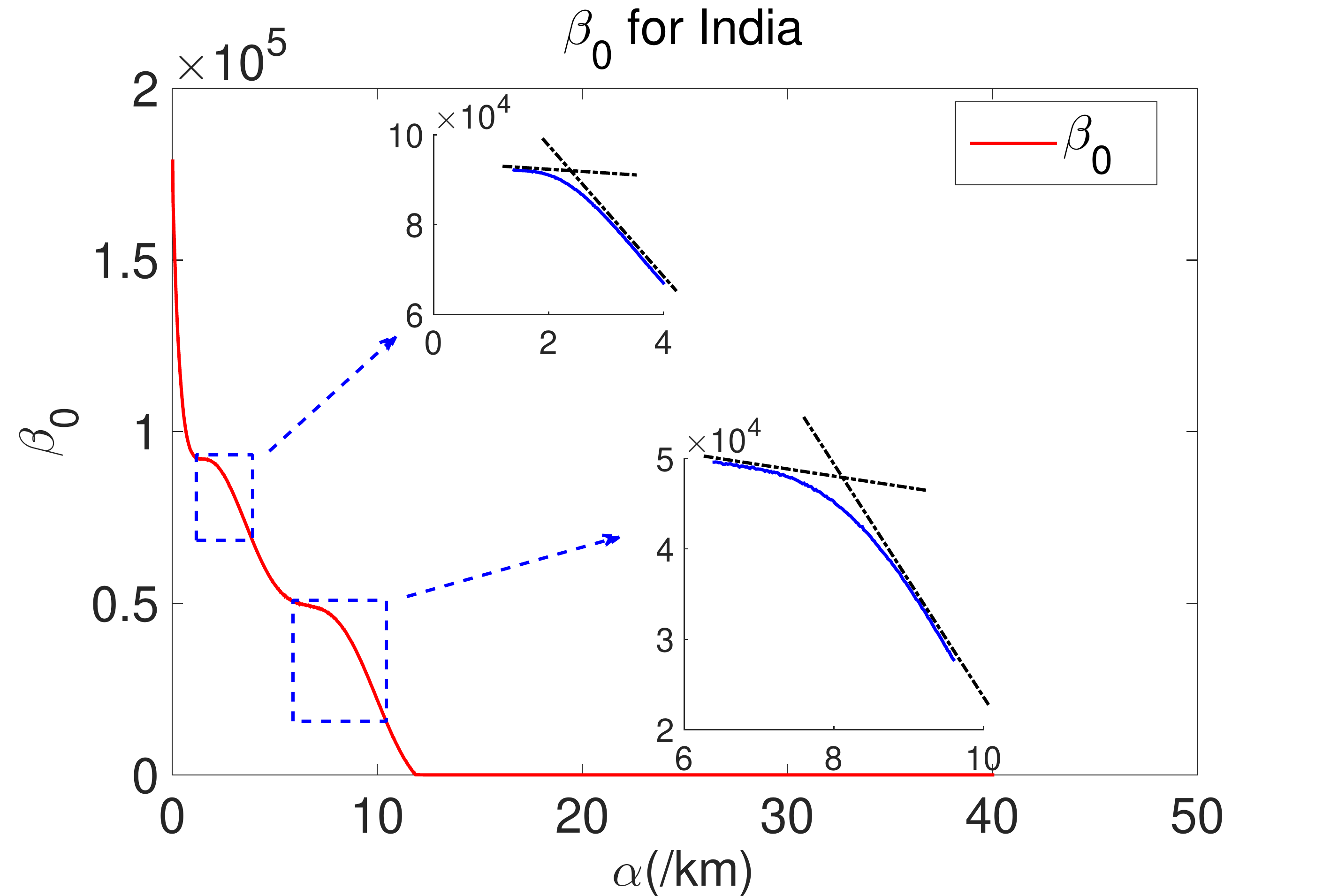}
	}
	   \makebox[4cm][c]{
		\includegraphics[scale=0.15]{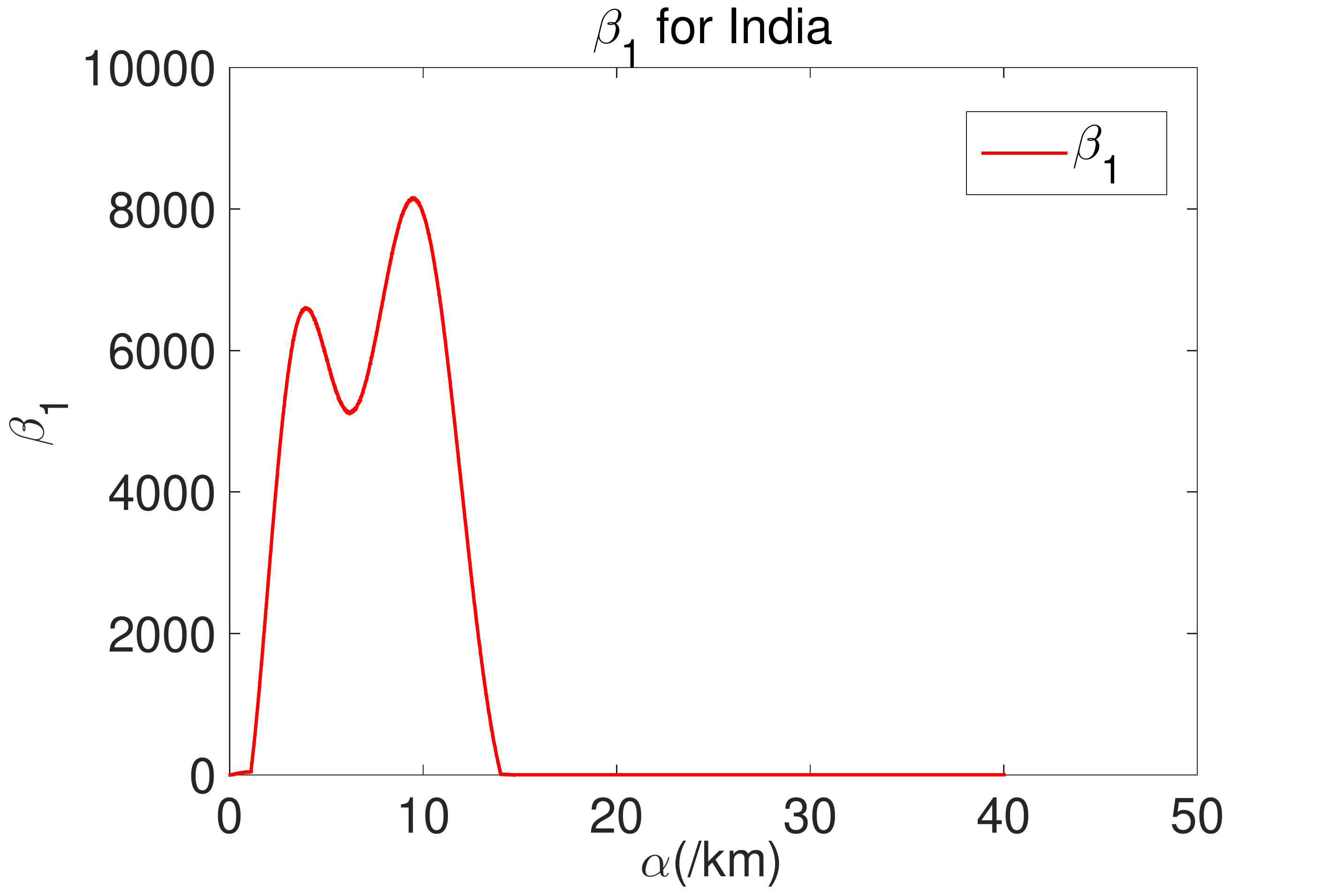}
	}
	}
	
	\subfigure[Japan$ \qquad\qquad\qquad\qquad\qquad\qquad\qquad\qquad\qquad\qquad\qquad\qquad $(f) Tailand]{
		\makebox[4cm][c]{
			\includegraphics[scale=0.15]{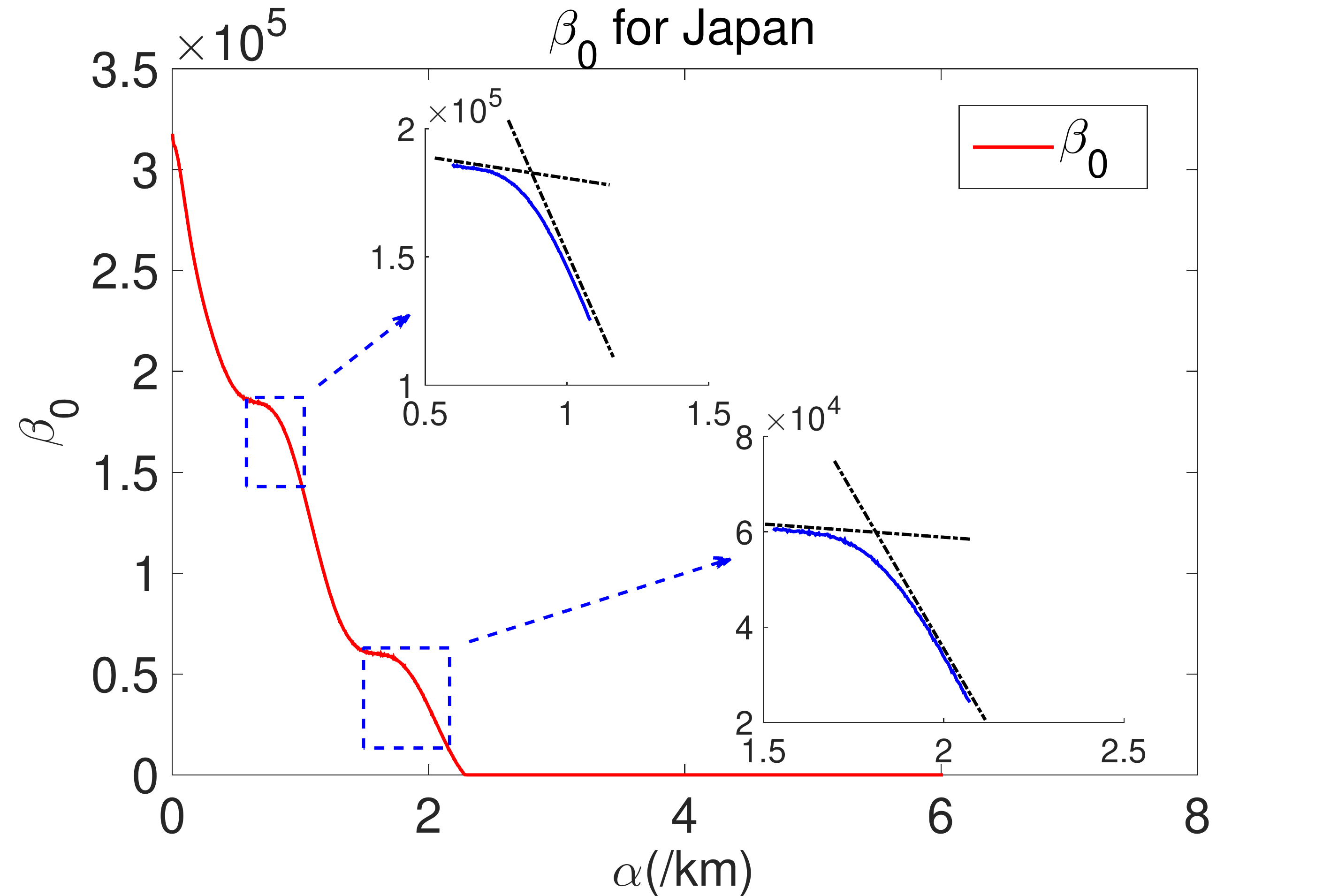}
		}
		\makebox[4cm][c]{
			\includegraphics[scale=0.15]{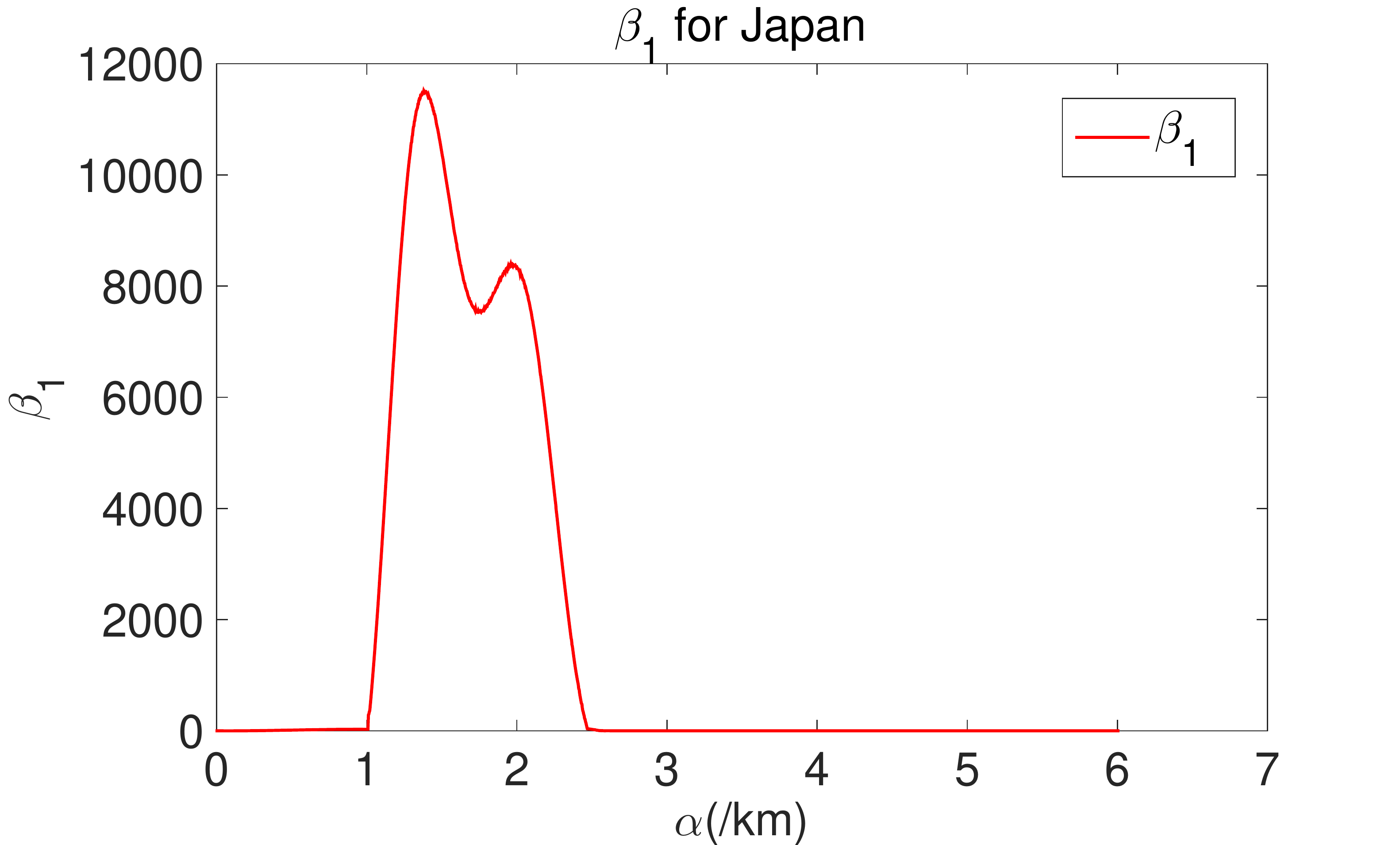}
		}
	   \makebox[4cm][c]{
		\includegraphics[scale=0.15]{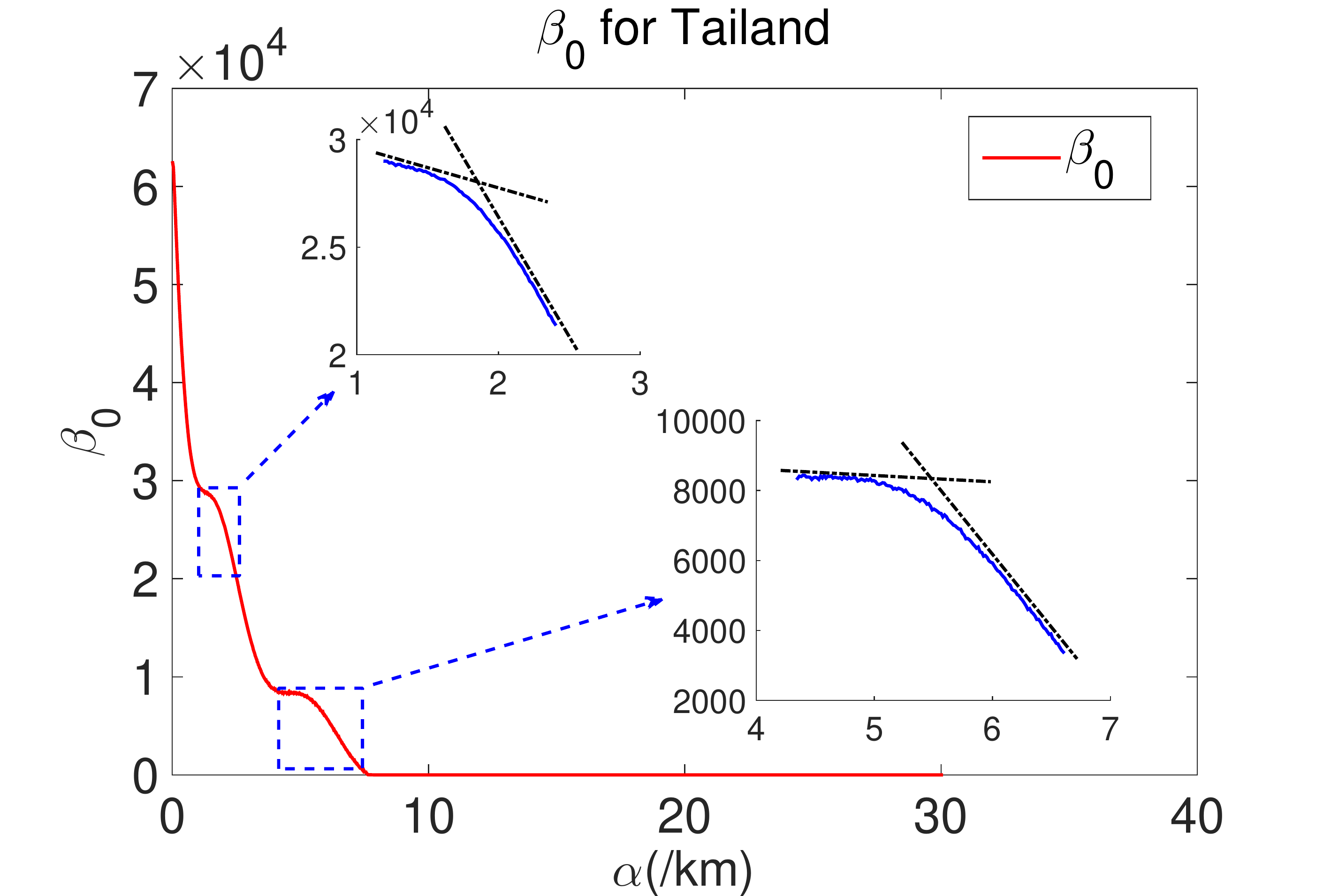}
	}
       \makebox[4cm][c]{
		\includegraphics[scale=0.15]{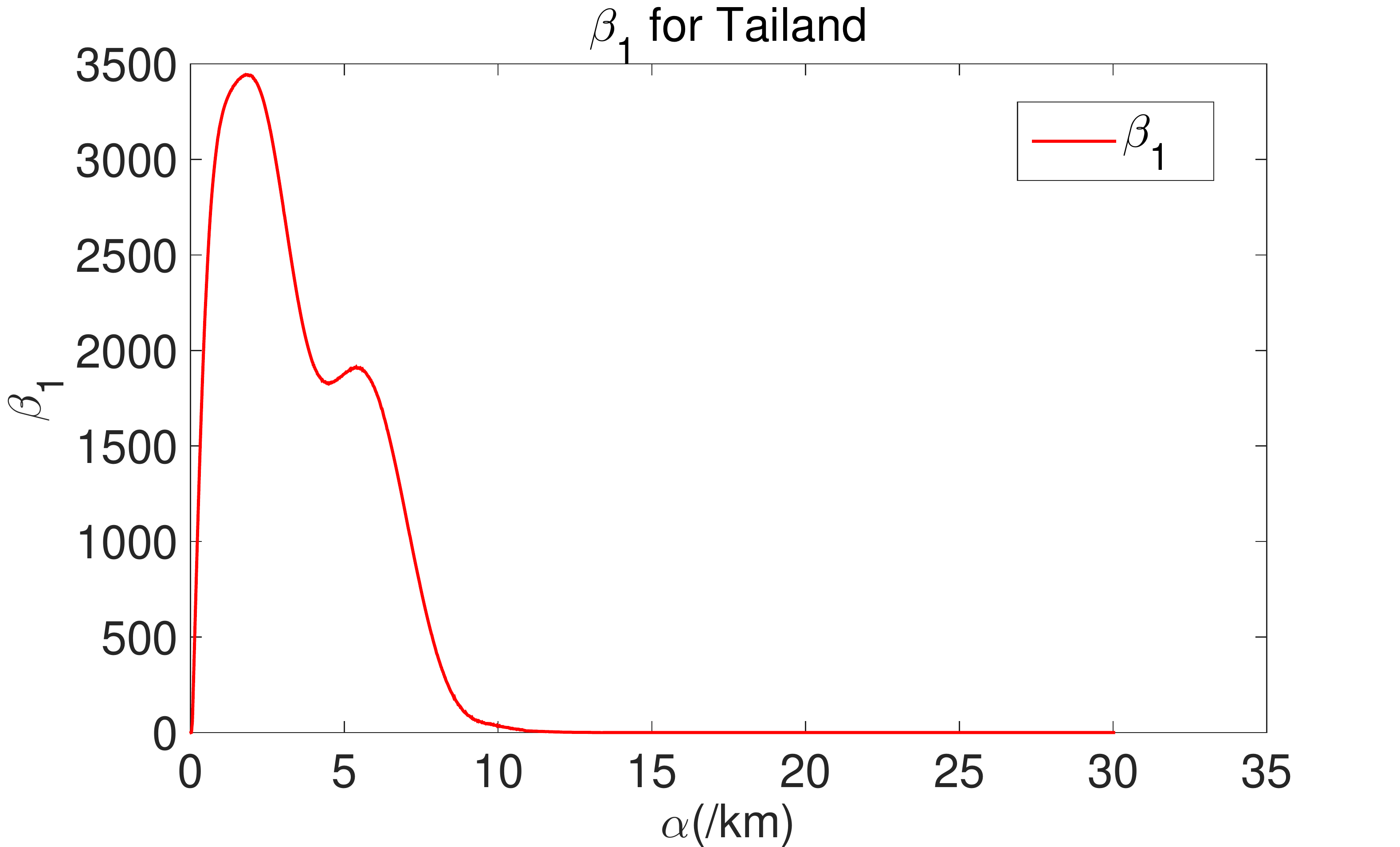}
	}
	}

	\caption{The Betti curves of the practical BSs distributions in Asian countries.}
\end{figure*}

\begin{figure*}[ht]
	\centering
	
	\subfigure[Poland $ \qquad\qquad\qquad\qquad\qquad\qquad $(c) UK$ \qquad\qquad\qquad\qquad\qquad\qquad $ (e) Germany]{
		\makebox[5cm][c]{
			\includegraphics[scale=0.17]{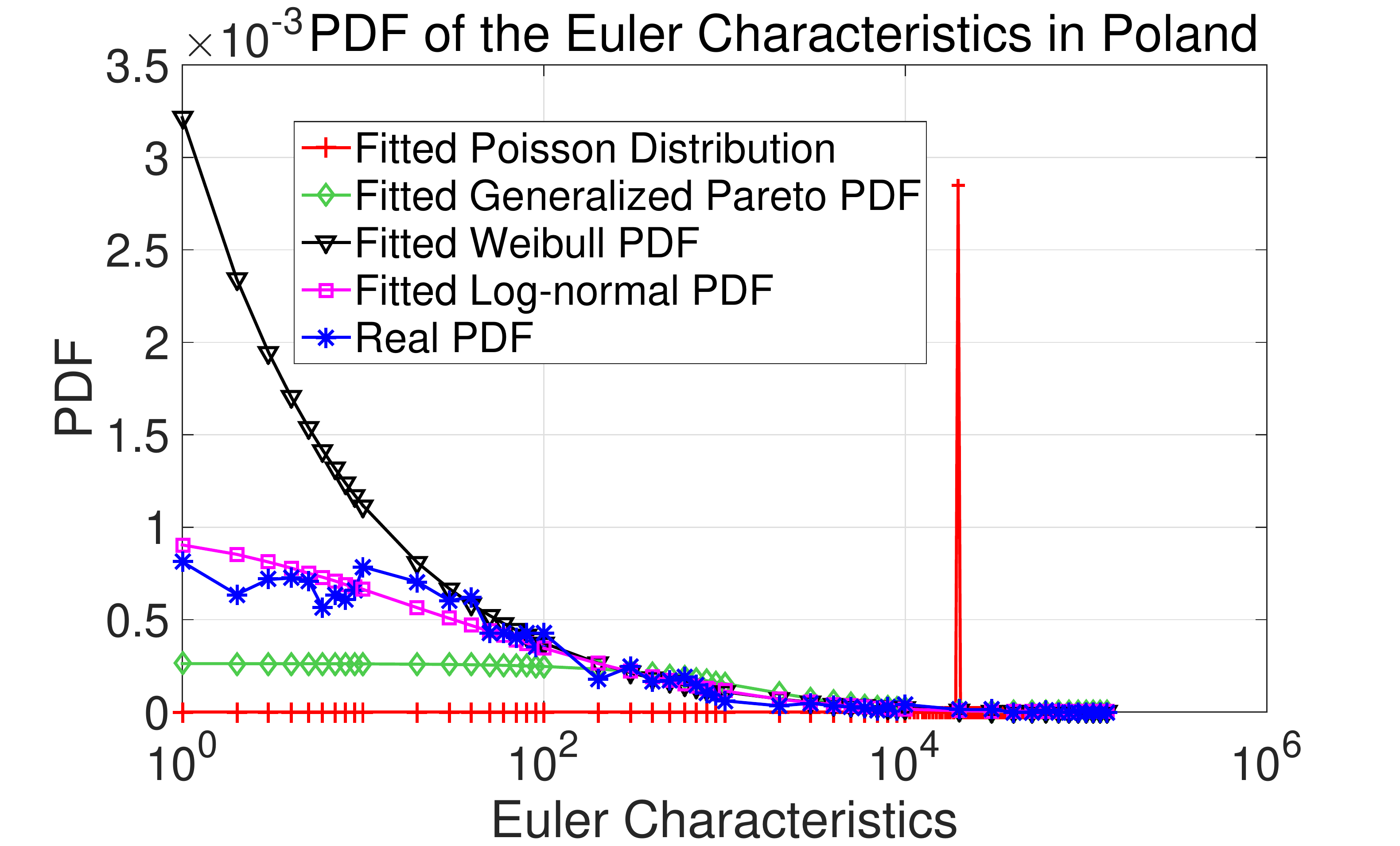}
		}
		
		\makebox[5cm][c]{
			\includegraphics[scale=0.17]{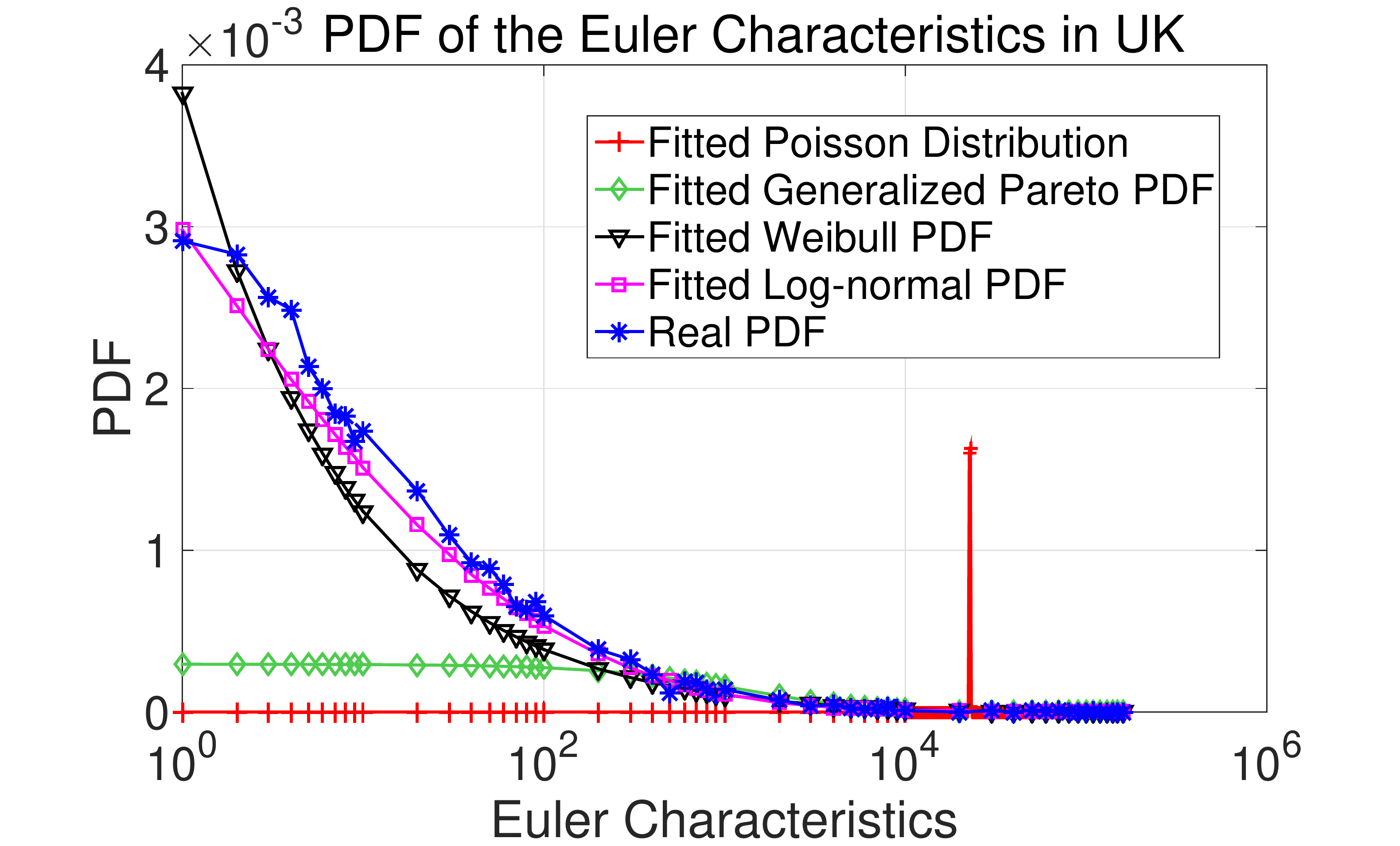}
		}
		
		\makebox[5cm][c]{
			\includegraphics[scale=0.17]{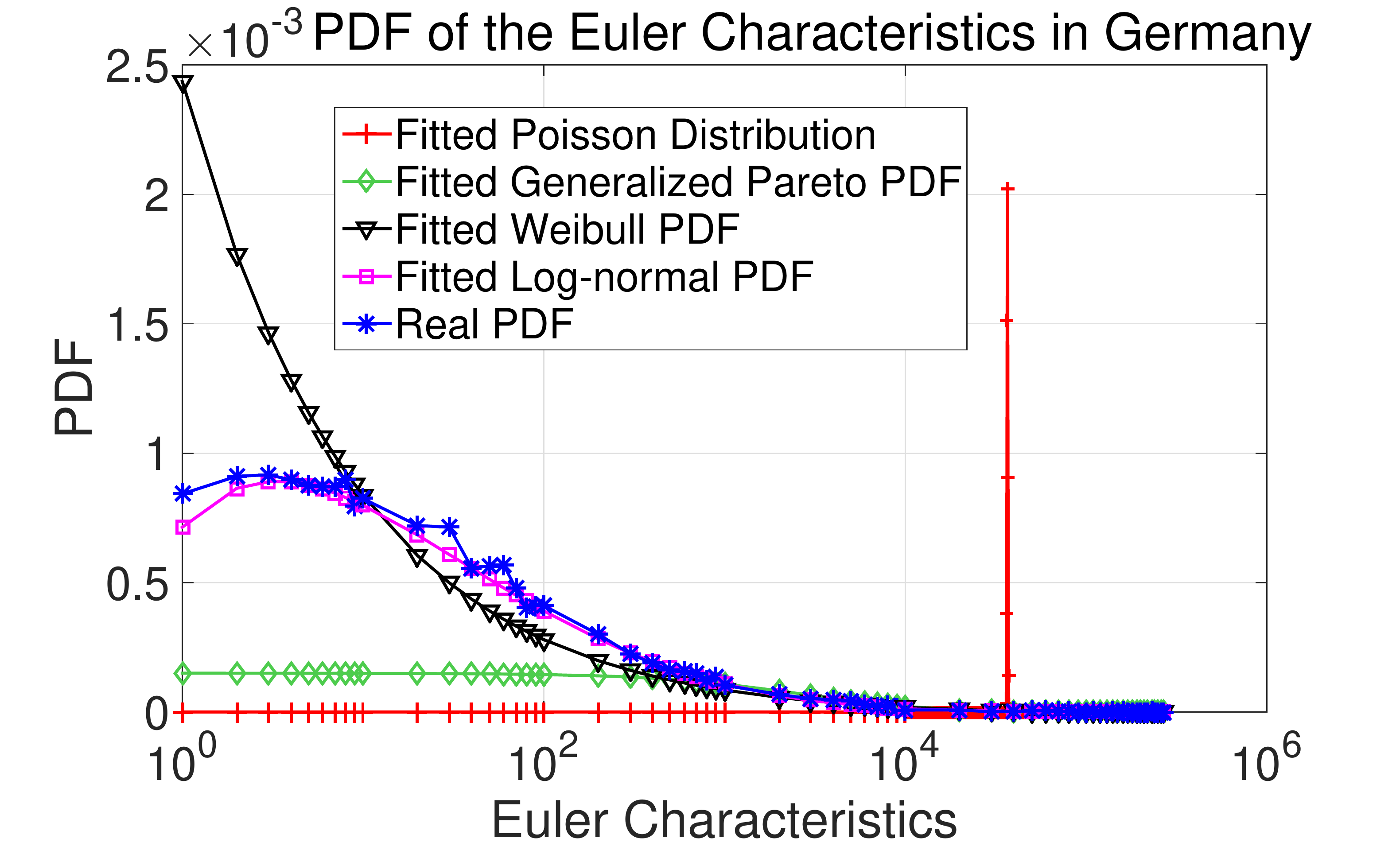}
		}
	}
	
	\subfigure[France $ \qquad\qquad\qquad\qquad\qquad\qquad $(d) Italy$ \qquad\qquad\qquad\qquad\qquad\qquad $ (f) Netherlands]{
		\makebox[5cm][c]{
			\includegraphics[scale=0.17]{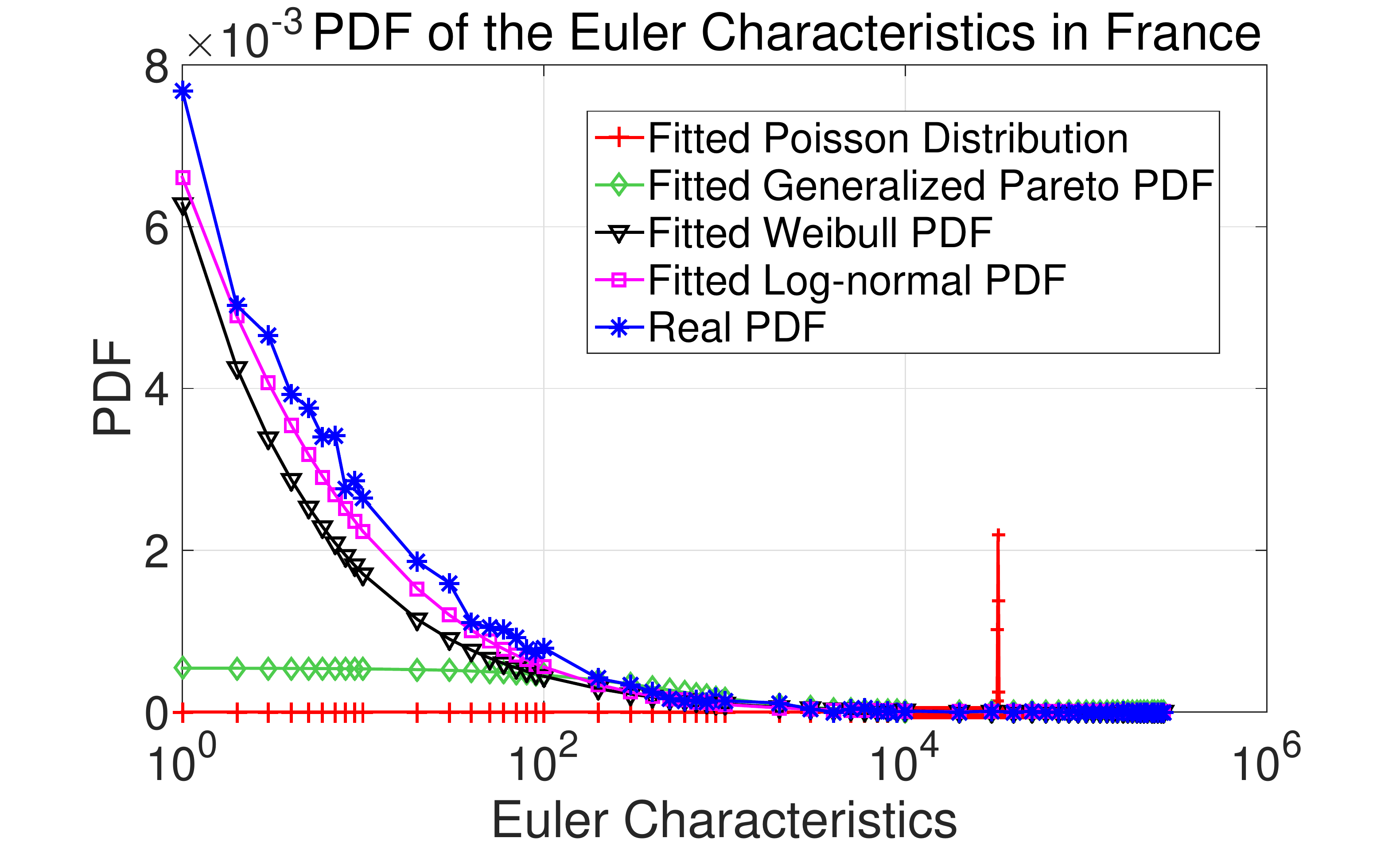}
		}
		
		\makebox[5cm][c]{
			\includegraphics[scale=0.17]{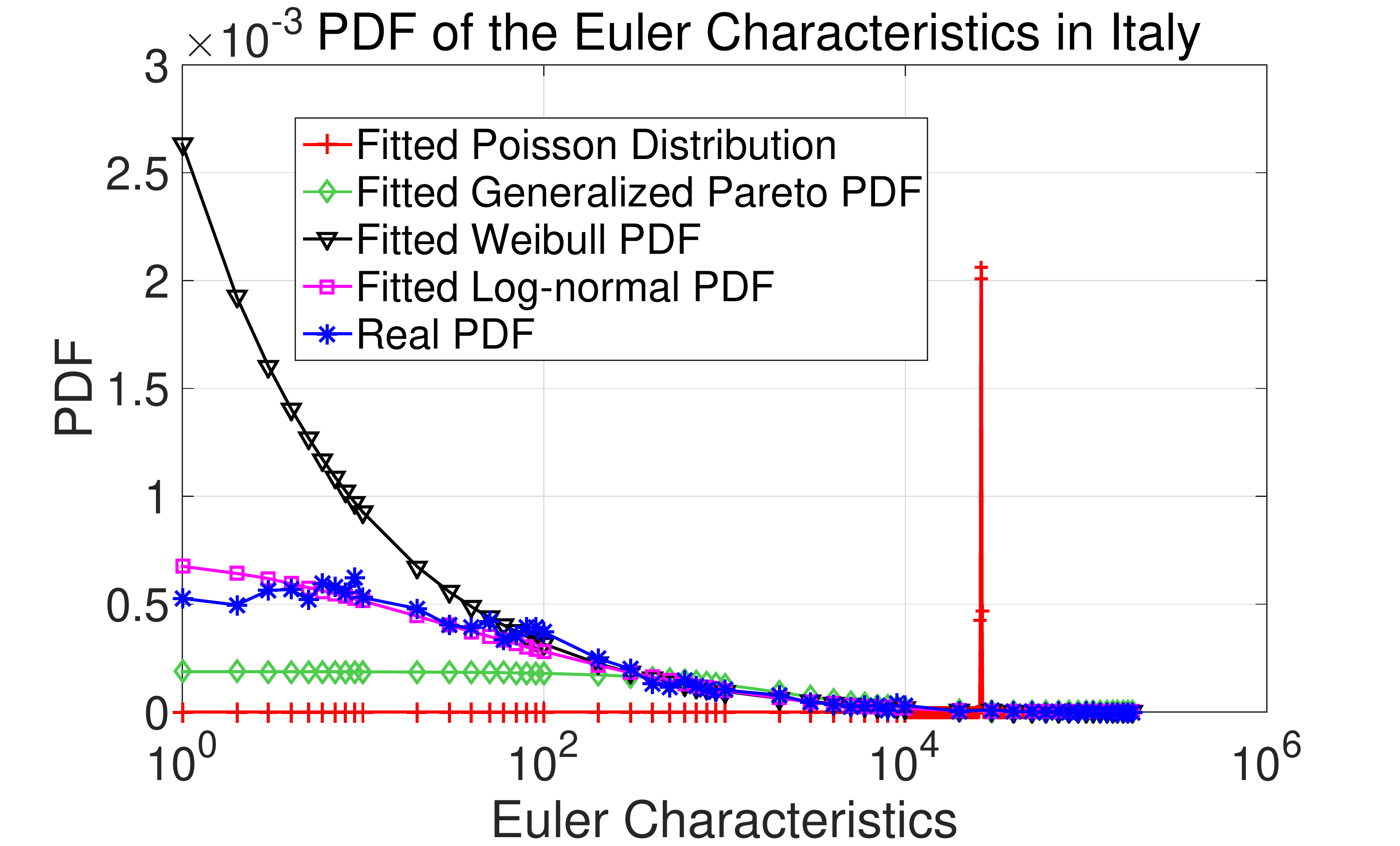}
		}
		
		\makebox[5cm][c]{
			\includegraphics[scale=0.17]{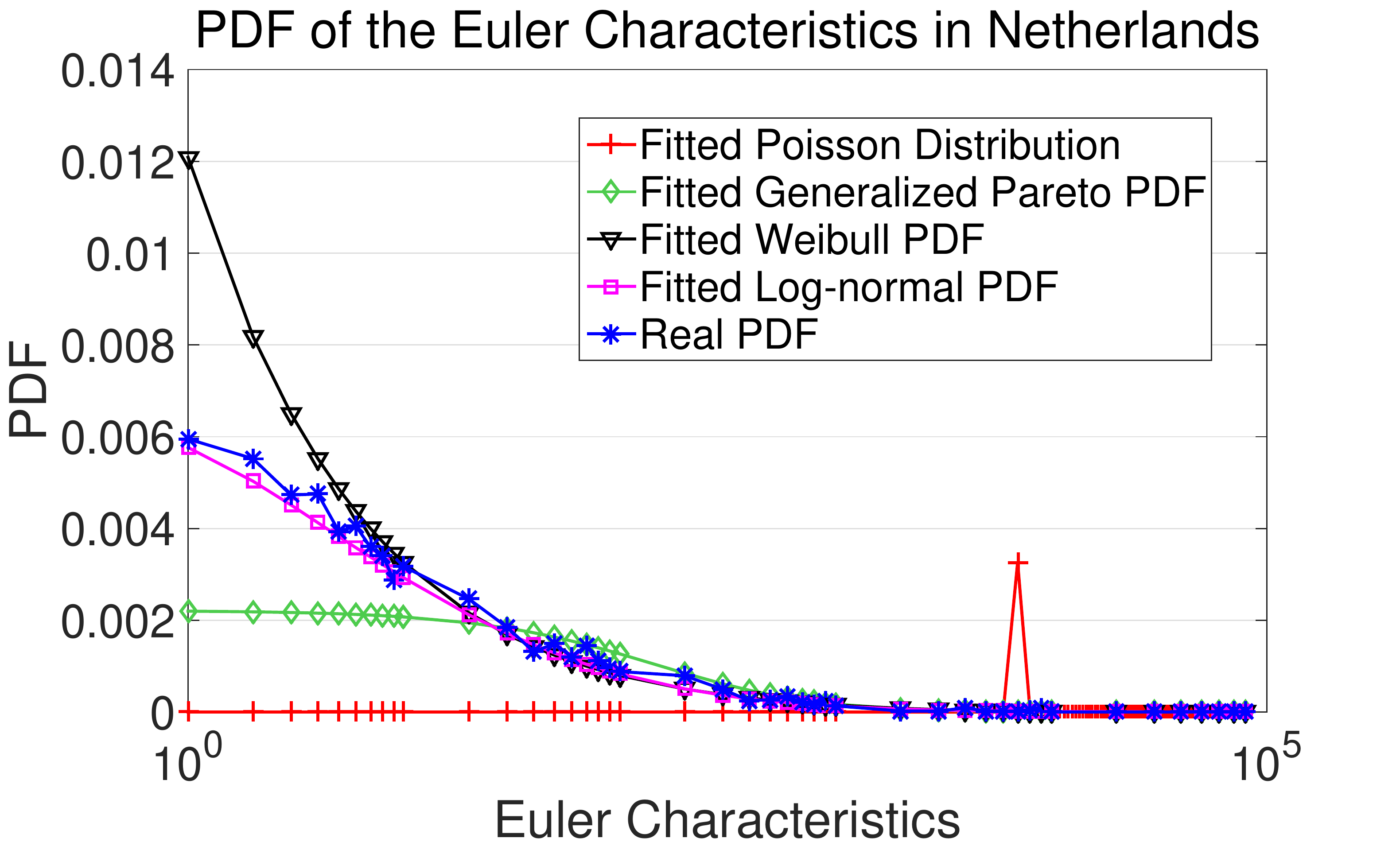}
		}
	}
	\caption{The comparison between the practical PDF and the fitted ones for European countries.}
\end{figure*}

\begin{figure*}[ht]
	\centering
	
	\subfigure[Singapore $ \qquad\qquad\qquad\qquad\qquad\quad $(c)South Korea$ \qquad\qquad\qquad\qquad\qquad\quad $ (e) Japan]{
		\makebox[5cm][c]{
			\includegraphics[scale=0.17]{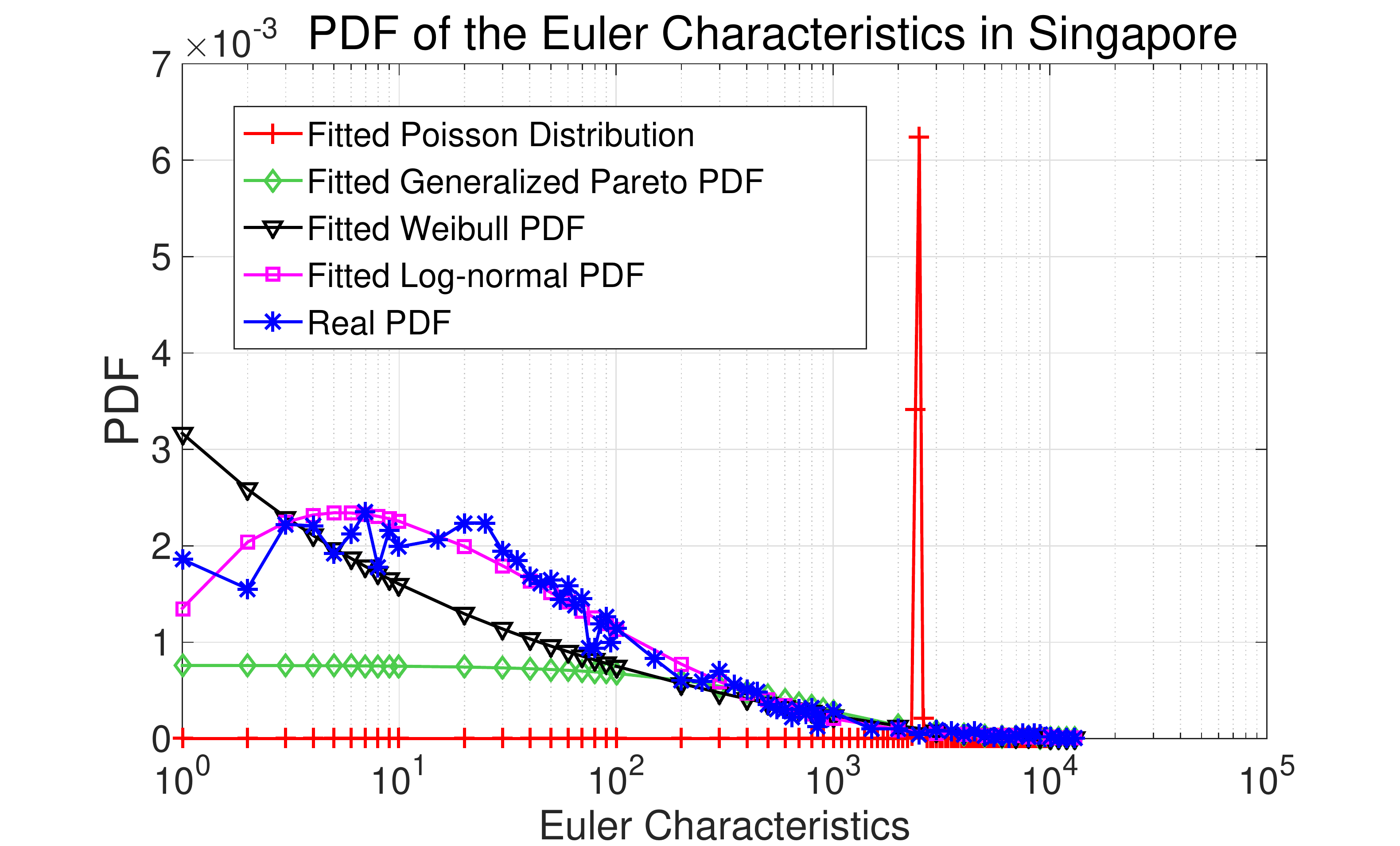}
		}
		
		\makebox[5cm][c]{
			\includegraphics[scale=0.17]{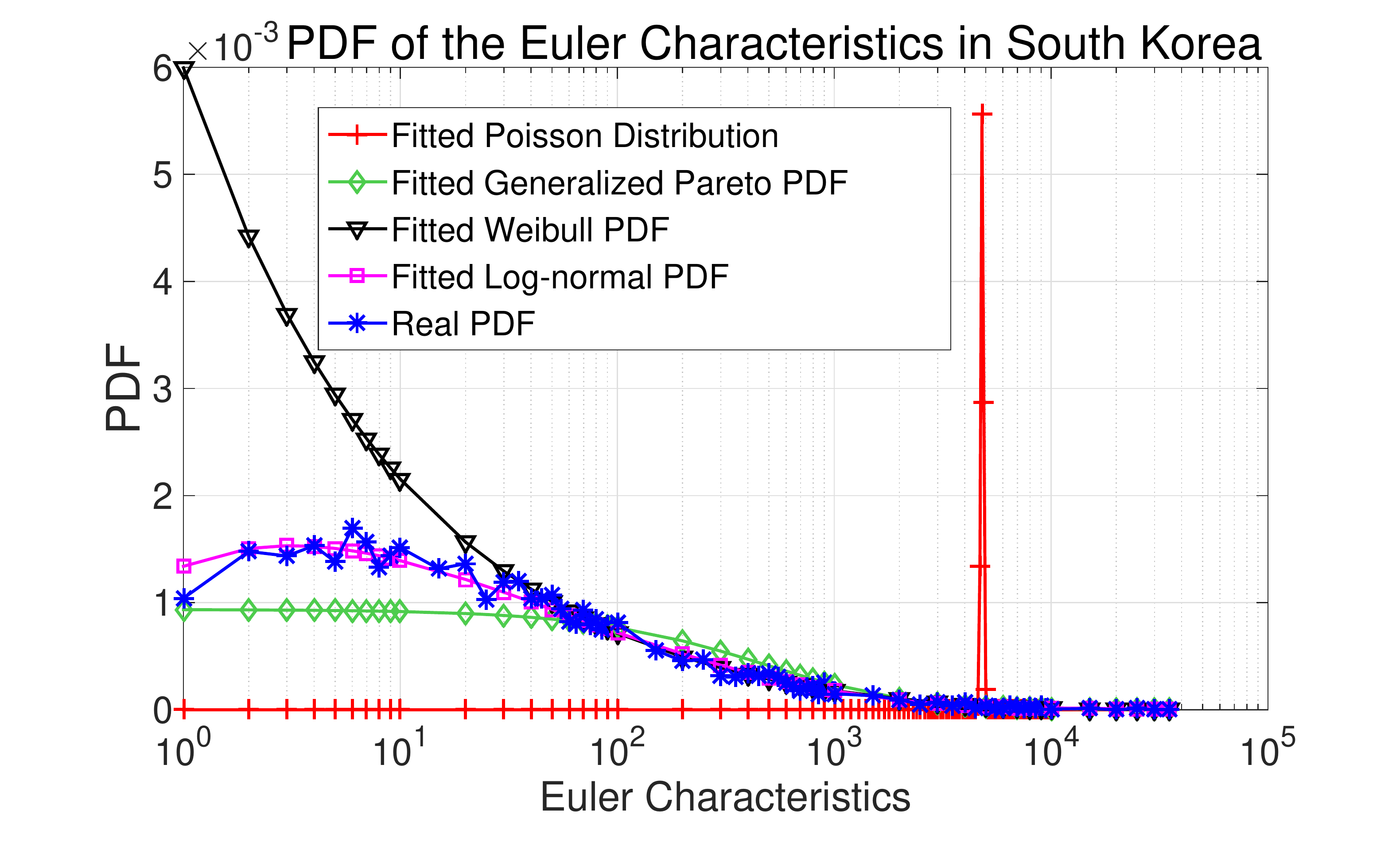}
		}
		
		\makebox[5cm][c]{
			\includegraphics[scale=0.17]{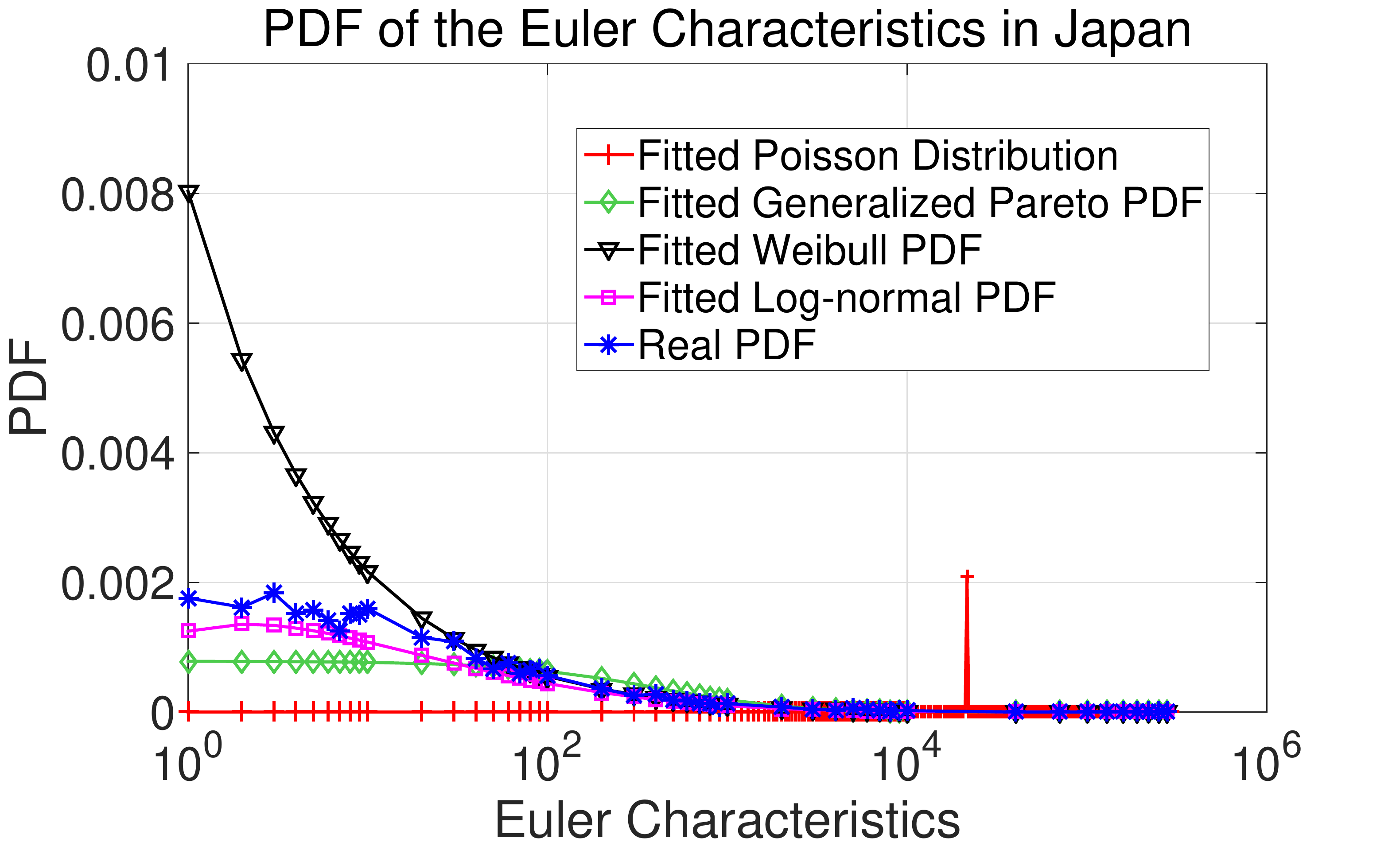}
		}
	}
	
	\subfigure[China $ \qquad\qquad\qquad\qquad\qquad\qquad $(d) India$ \qquad\qquad\qquad\qquad\qquad\qquad $ (f) Tailand]{
		\makebox[5cm][c]{
			\includegraphics[scale=0.17]{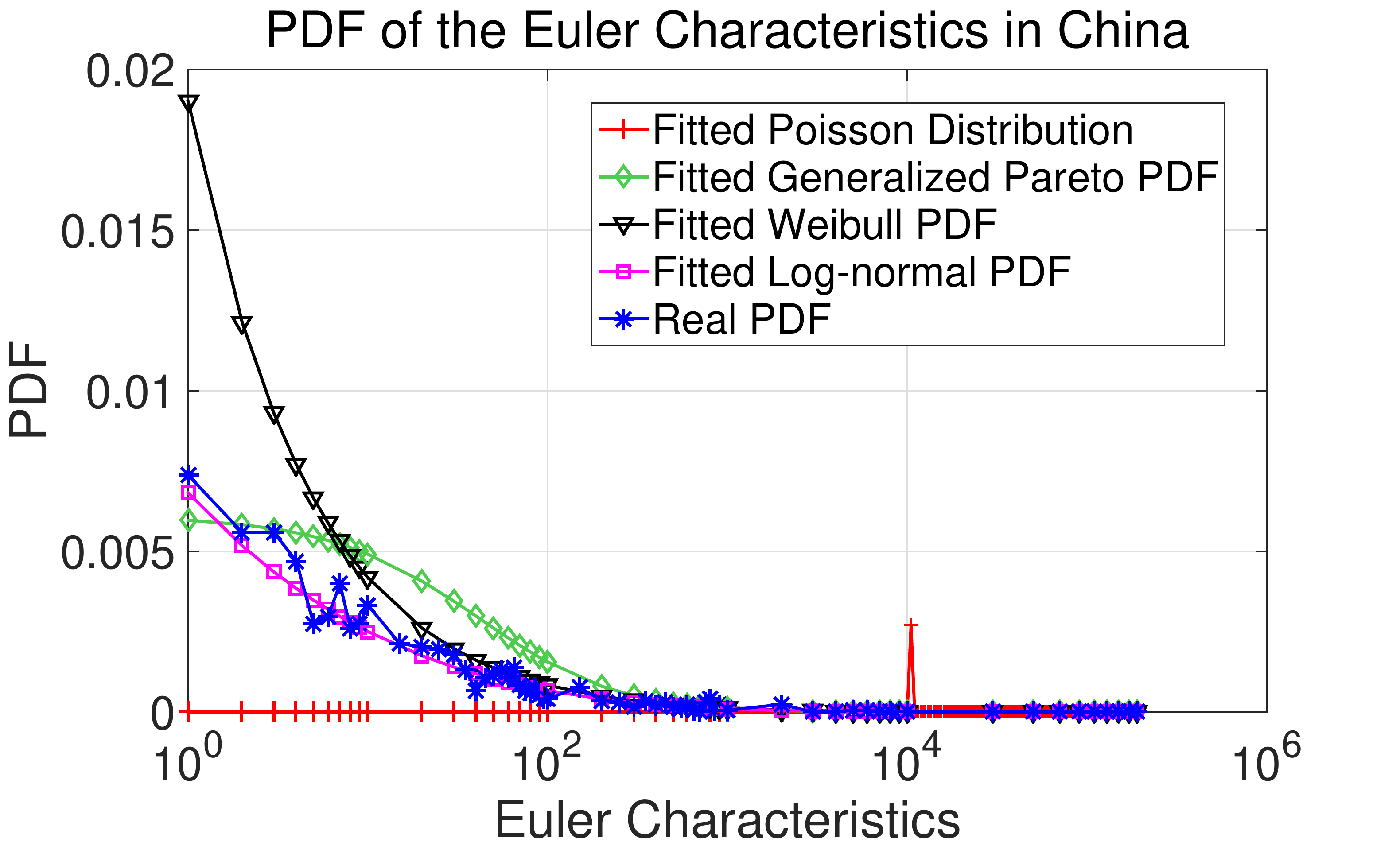}
		}
		
		\makebox[5cm][c]{
			\includegraphics[scale=0.17]{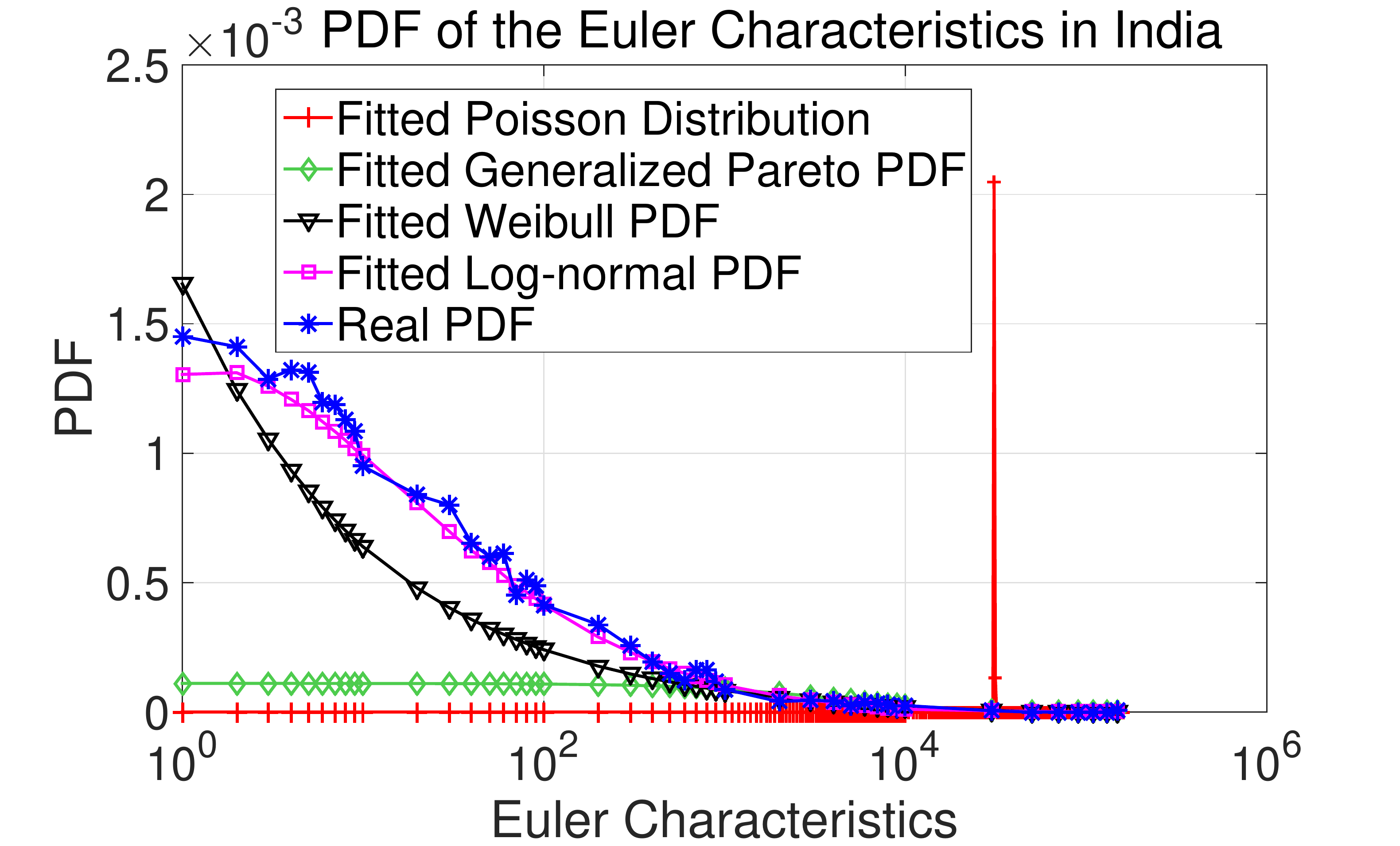}
		}
		
		\makebox[5cm][c]{
			\includegraphics[scale=0.17]{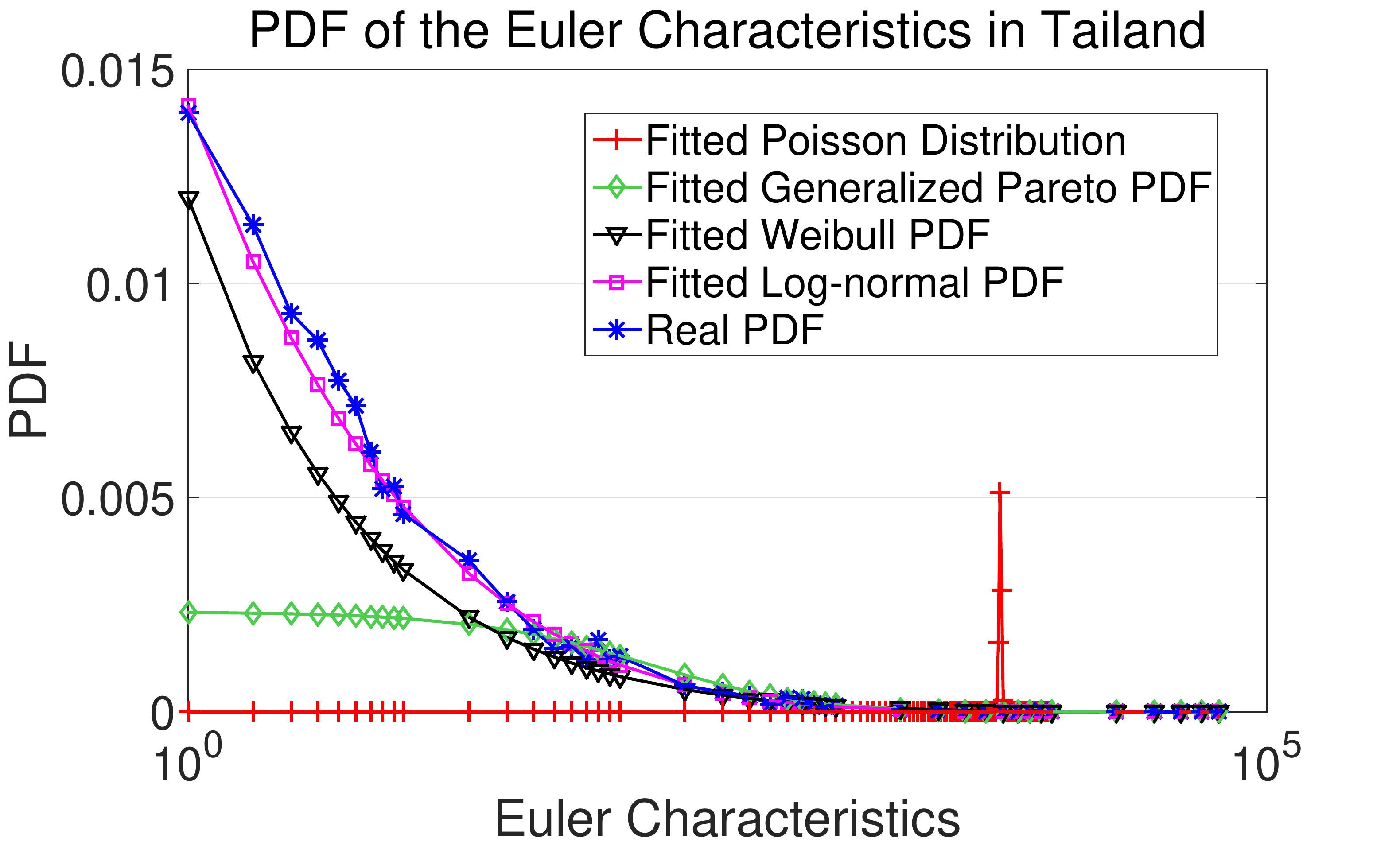}
		}
	}
	
	\caption{The comparison between the practical PDF and the fitted ones for Asian countries.}
\end{figure*}

Besides, the numbers of the ripples or peaks for these countries are not entirety the same, which is 3 for France and Germany, and 2 for other 10 countries. As indicated before, this parameter is associated with the hierarchical levels in the BS distributions. 

In addition, the ripples and peaks appear at completely different values of the scale parameter $\alpha$ for each country, which are listed in Table II. It is noteworthy that the ripples suggest the rapid gradient switch within a narrow range of $\alpha$, and the position of a ripple can be estimated according to the intersection of two straight lines with distinctive slopes respectively, as shown in the first and third columns in Fig. 7 and Fig. 8. The positions are probably related to the area of the countries and the number of BSs per unit area, i.e., the density of BSs.

\subsection{Fractal Features Based on the Hurst Coefficients}
Firstly proposed by H. E. Hurst, the Hurst exponent has recently gained its popularity in plenty of fields, especially in the finance fields as a result of Peter's work \cite{Noll1993Chaos,Peters1994Fractal}. As a measurement for fractality of data series, the Hurst exponent usually falls in the range from 0 to 1. In concrete, Hurst coefficient of 0.5 implies a completely random data series, while the indication of fractal features gets stronger as the value comes closer to 1 \cite{Gneiting2001Stochastic}.

Among the various methods applied in the calculation of the Hurst coefficient, we take advantage of the rescaled range analysis (R/S), one of the most classical methods, in our works. Due to the space limitation, the framework of computing the Hurst coefficient is given in the Algorithm 1 below, and interested readers could refer to \cite{Fern2014An} for the details of R/S method.

\begin{algorithm*}[htbp]
	\caption{Framework of computing the Hurst coefficient.}
	\label{alg:A}
	\begin{algorithmic}
		\STATE {divide a data series $ X={X_{1},X_{2},...,X_{N}} $ of length $ N $ into $ A $ sub-series of the same length $ n $}
\FOR{each $a \in [1,A]$}
\STATE  {\[ \mu_a  = \frac{1}{n}\sum\limits_{i = (a-1)n+1}^{an} {{X_i}}, \] $//$ The mean value $ \mu_a $ of the $ a $-th sub-series.}
\STATE  {\[ {Y_i} = {X_i} - \mu_{a} ,{\rm{   }}i = (a-1)n+1,(a-1)n+2,...,an. \] $//$ The mean adjusted sub-series $ Y $.}
\STATE  {\[ {Z_t} = \sum\limits_{i = (a-1)n+1}^t {{Y_i}} ,{\rm{   }}t = (a-1)n+1,(a-1)n+2,...,an. \] $//$ The cumulative deviate series $ Z $.}
\STATE  {\[ {R_a} = \max \{ {Z_{(a-1)n+1}},{Z_{(a-1)n+2}},...,{Z_{an}}\}  - \min \{ {Z_{(a-1)n+1}},{Z_{(a-1)n+2}},...,{Z_{an}}\}. \] $//$ The accumulated deviation $ R_{a} $ for the $ a $-th sub-series.}
\STATE  {\[ {S_a} = \sqrt {\frac{1}{n}\sum\limits_{i = (a-1)n+1}^{an} {({X_i}  - \mu_{a} )^2}}. \] $//$The standard deviation series $ S_{a} $ for the $ a $-th sub-series.}
\ENDFOR
\STATE  {\[ {(R/S)_n} = \frac{1}{A}\sum\limits_{a = 1}^A {{R_a}} /{S_a}. \]}

	\end{algorithmic}
\end{algorithm*}

Actually, the length $ n $ of sub-series is variable, and it has been found that $ {(R/S)_n} $ scales by power-law as $ n $ grows \cite{Fern2014An}:
\begin{equation}\label{6}
{(R/S)_n} = c \cdot {n^H},
\end{equation}

\noindent where $ c $ is a constant, and the Hurst exponent $ H $ can be estimated by the slope of the least square regression on logarithms of both sides in Eq. \eqref{6}.

In order to testify the fractal property in the cellular networks, the Hurst coefficients are computed for all the BSs in the cellular networks of all the selected countries. In our works, an arbitrary center point is chosen randomly among all the location data of the BSs and a radius is specified firstly, so that a circle is formed given the center BS and the radius. Then the distances between all the BSs within the circle and the center BS are calculated. In other words, the data series consist of a sequence of the distances between each BS and the center BS. Next, the Hurst coefficient is computed according to the data series in the way mentioned above. The similar process is performed a hundred times given different center points and radii, and the average Hurst coefficient is obtained finally. As indicated in Table III, the fractal features are completely confirmed since all the Hurst coefficients are very close to 1.
\begin{table}[H]
	\centering
	\includegraphics[scale=0.35]{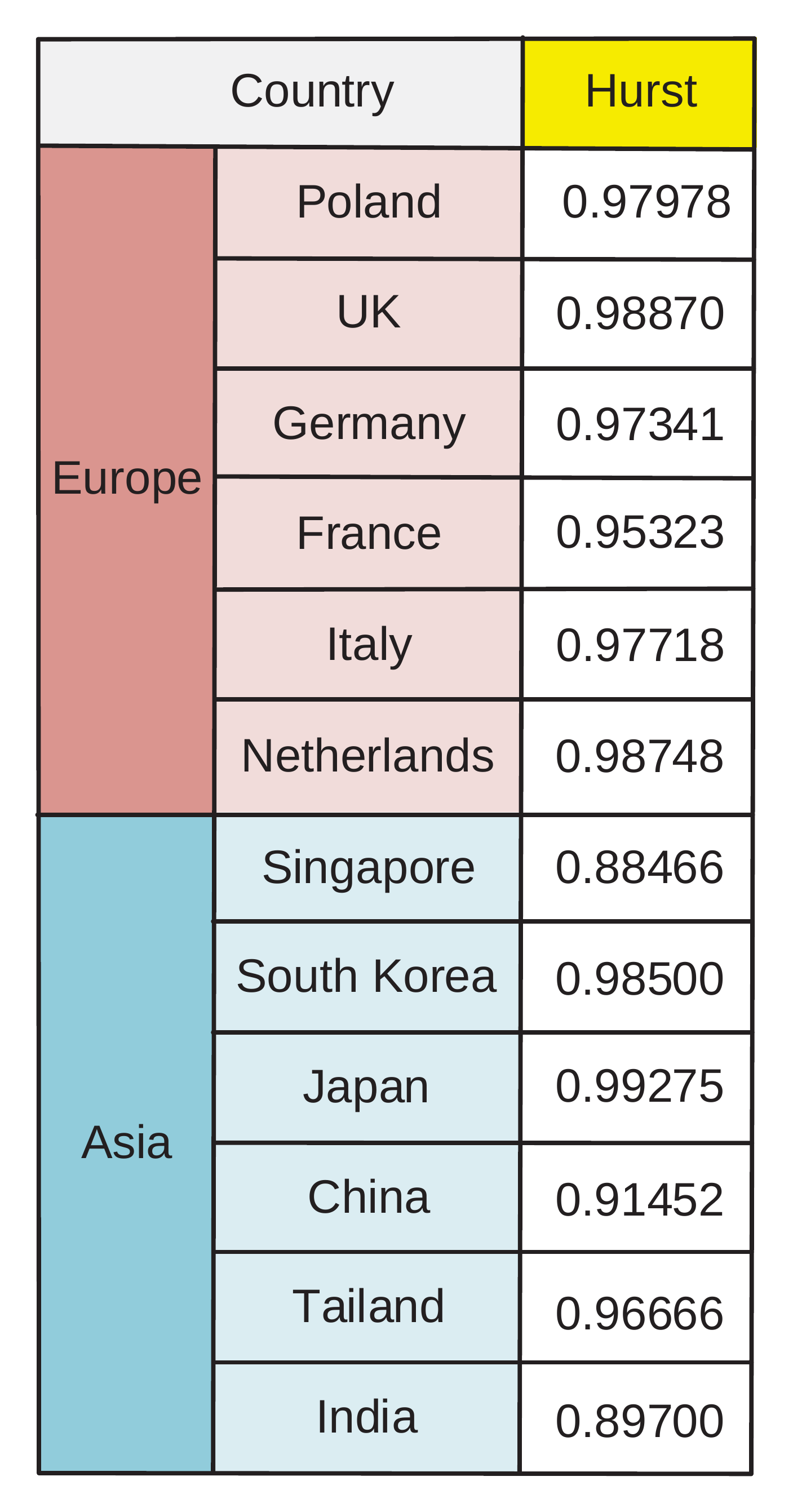}
	\caption{Hurst coefficients for all the 12 countries.}
\end{table}

\section{Consistent Log-normal Distribution of the Euler Characteristics}
In our works, based on Euler-Poincare Formula, the Euler Characteristics are obtained by subtracting $\beta_{1}$ from $\beta_{0}$. Obvious heavy-tail features can be observed from the curves of real probability density functions (PDFs), so several representative heavy-tail distributions are chosen as the candidate distributions for the fitness of real PDF. The candidate distributions and their PDF expressions are listed in Table IV.
\begin{table}[H]
	\centering
	\includegraphics[scale=0.13]{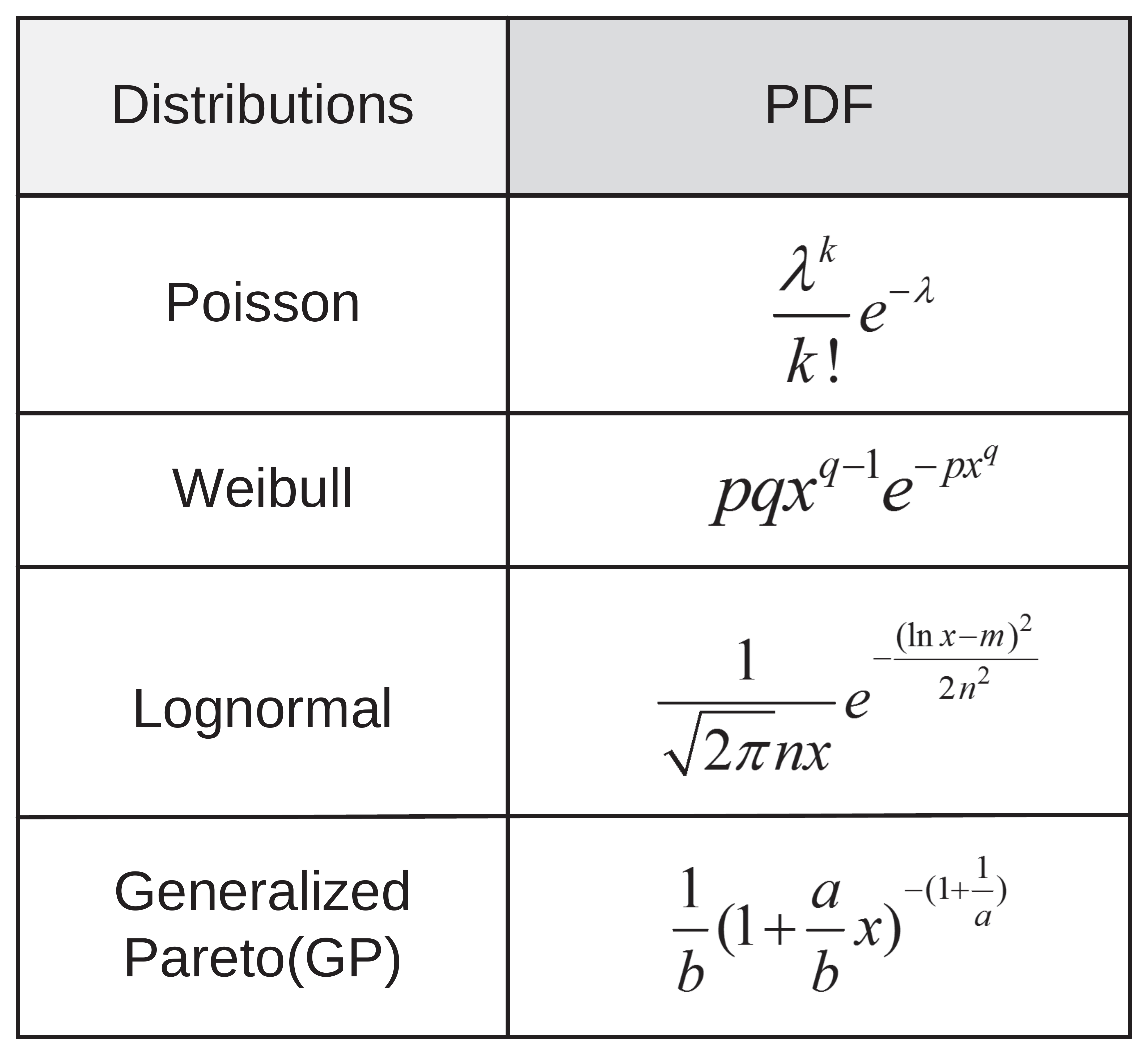}
	\caption{The candidate distributions and their PDF expressions.}
\end{table}

\begin{table}[H]
	\centering
	\includegraphics[scale=0.1]{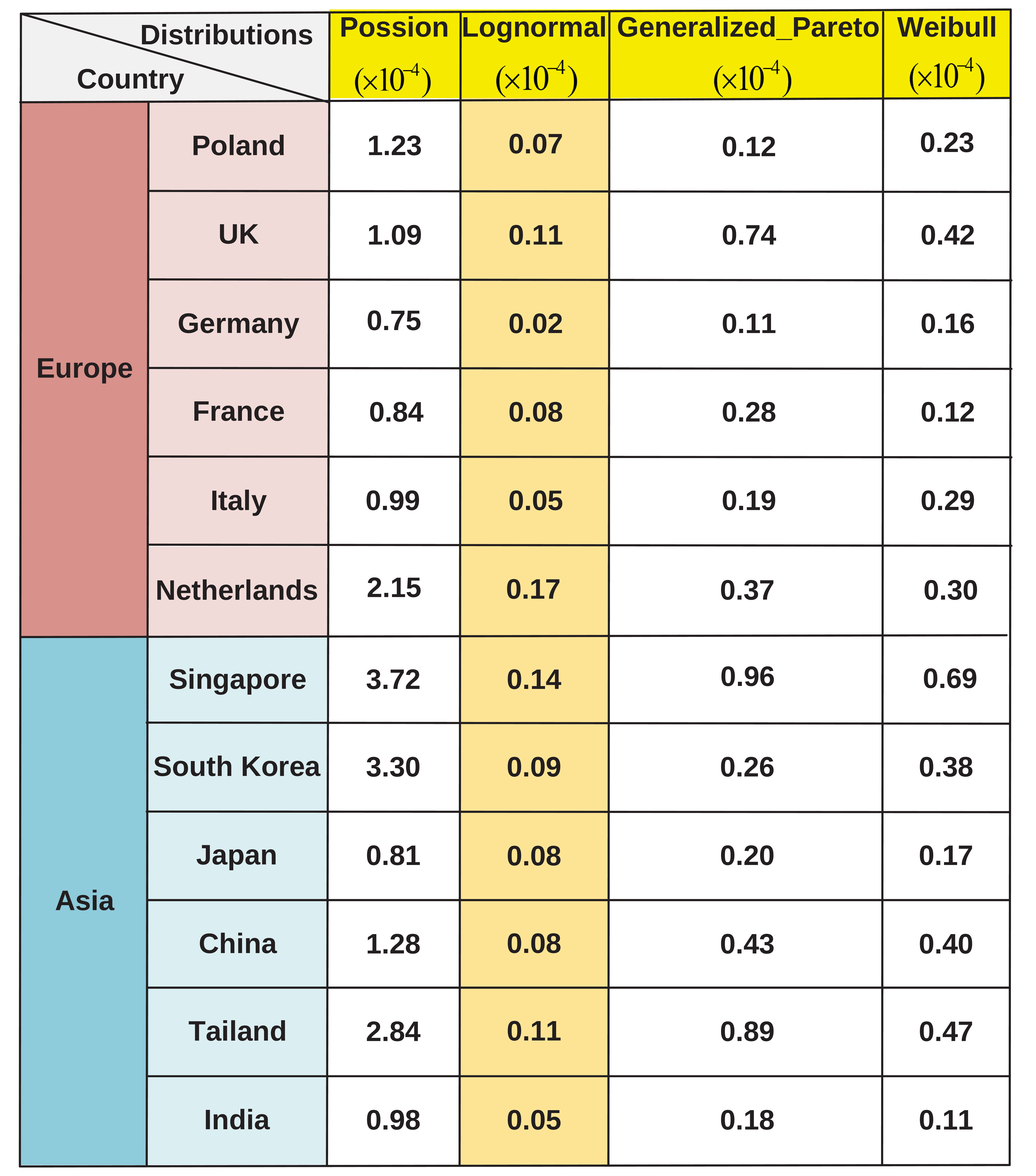}
	\caption{The RMSE between each candidate distribution and the practical one.}
\end{table}

During the process of data fitting, the Euler characteristics are assumed to comply with a specific statistical distribution to estimate the parameters, then an estimated PDF curve is generated to fit the given data. The comparison between the practical PDF and the fitted ones is displayed as Fig. 9 for European countries and Fig. 10 for Asian ones, respectively. As observed from the fitted curves in Fig. 9 and Fig. 10, for each European or Asian country, the log-normal distribution is clearly the closest one to the real PDF.

For the purpose of verifying the best match for the real PDFs, the root mean square error (RMSE) between each candidate distribution and the practical one is computed and listed in Table V. As shown in Table V, the RMSE between the log-normal distribution and the real PDF is also the smallest one for every country, which is almost one order of magnitude smaller than the other terms in the same row. In other words, the log-normal distribution is the best match for the PDF of the Euler characteristics for both European and Asian countries, from the perspective of either intuition or rigorous numerical analysis.

Therefore, a possibly surprising but well-founded conclusion can be drawn here: regardless of the geographical differences as well as the culture and historical factors, the Euler characteristics of either Asian or European countries completely comply with the log-normal distribution.

\section{Conclusion and Future Works}
In this paper, the algebraic geometric tools, i.e., $ \alpha $-Shapes, Betti numbers, and Euler characteristics, have been exploited to discover the intrinsic topological characteristics from the real BS location data of various Asian and European countries. First of all, fractal phenomenon has been confirmed within the BS configurations for either Asian or European countries in terms of both Betti numbers and Hurst coefficients; Secondly, the log-normal distribution has been proven to provide the best fitness to the PDFs of the Euler characteristics among typical distribution candidates in regard to the practical BS deployments in the cellular networks.

Nevertheless, despite all the topological findings above, there are still some challenging issues to be worked out in the future. For instance, what are the definitive factors for the $ \alpha $ values of the ripples and peaks in the Betti curves? What is the intrinsic meaning of the number of levels indicated by the number of ripples or peaks? And how to apply the fractal features and log-normal distribution of the Euler characteristics to the design of BSs deployments? We will investigate all of these problems as our future works.

\bibliographystyle{IEEEtran}
\bibliography{IEEEfull,reference}

\end{document}